\documentclass[preprint,12pt]{elsarticle}

\usepackage{amssymb}
\usepackage{color}
\usepackage{multirow}

\journal{Astroparticle Physics}

\newcommand{\be}{\begin{equation}}
\newcommand{\ee}{\end{equation}}
\newcommand{\beq}{\begin{eqnarray}}
\newcommand{\eeq}{\end{eqnarray}}

\def\nue{\mathrel{{\nu_e}}}
\def\numu{\mathrel{{\nu_\mu}}}
\def\nutau{\mathrel{{\nu_\tau}}}
\def\nux{\mathrel{{\nu_x}}}

\def\barnue{\mathrel{{\bar \nu}_e}}
\def\barnumu{\mathrel{{\bar \nu}_\mu}}
\def\barnutau{\mathrel{{\bar \nu}_\tau}}
\def\barnux{\mathrel{{\bar \nu}_x}}

\def \lta {\mathrel{\vcenter{\hbox{$<$}\nointerlineskip\hbox{$\sim$}}}}
\def \gta {\mathrel{\vcenter{\hbox{$>$}\nointerlineskip\hbox{$\sim$}}}}

\def\t13{\mathrel{{\theta_{13}}}}
\def\y12{\mathrel{{\tan^2 \theta_{12}}}}
\def\c2{\mathrel{{\chi^2 }}}

\newcommand{\df}{DSNB}
\newcommand{\snr}{SNR}
\newcommand{\sfr}{SFR}
\newcommand{\sk}{SK}
\newcommand{\comm}[1]{{}}

\newcommand{\snew}[1]{{ \textmd{#1} } }

\newcommand{\n}{neutrino}
\newcommand{\ns}{neutrinos}

\newcommand{\sn}{supernova}
\newcommand{\sne}{supernovae}

\newcommand{\oss}{oscillations}

\newcommand{\bh}{black hole-forming}
\newcommand{\nts}{neutron star-forming}

\newcommand{\f}{flux}

\newcommand{\ck}{Cherenkov}
\newcommand{\lar}{LAr}

\begin{document}

\begin{frontmatter}

\title{Diffuse supernova neutrinos at underground laboratories}
\author{Cecilia Lunardini}
\address{Arizona State University, Tempe, AZ 85287-1504}
\begin{abstract}  
I review the physics of the Diffuse Supernova Neutrino flux (or Background, \df), in the context of future searches at the next generation of \n\ 
observatories.  The theory of the \df\ is discussed in its fundamental elements, namely the cosmological rate of \sne, \n\ production inside a core collapse \sn,  redshift, and flavor oscillation effects.  The current upper limits are also reviewed, and results are shown for the  rates and energy distributions of the events  expected at 
future liquid argon and liquid scintillator detectors of ${\mathcal O}(10)$ kt mass, and water Cherenkov detectors up to a $0.5$ Mt mass.
Perspectives are given on the significance of future observations of the \df, both at the discovery and precision phases, for the investigation of the physics of \sne\ and of the properties of the \n.
\end{abstract}                                                                 


\end{frontmatter}



\section{Introduction}

\snew{After a first phase of exploration, focused on solar and atmospheric \ns\ studies, \n\ physics has now entered a second phase of greater precision studies.  A new generation of \n\ beam experiments is being developed to achieve a full reconstruction of the \n\ mass spectrum and mixing matrix.  These experiments, primarily designed for oscillation physics, will also serve as powerful \n\ observatories: thanks to their larger detector masses and improved technologies, they will surpass their predecessors in the ability to detect and study \n\ sources of increasing distance from Earth,  increasing energy, and  increasing physical complexity. }

The still mysterious core collapse supernovae are among these sources.  After the handful of \n\ data from SN1987A, the scientific community is still waiting for the next detection of \sn\ \ns, to have the opportunity to  learn about the physics of core collapse, to test \n\ properties, and to answer a large number of questions regarding new particles and new forces of nature.
Considering that \sne\ in our galaxy and its satellites are rare (1-3 per century, see e.g. \cite{Arnaud:2003zr,Ando:2005ka}), it is likely that the opportunity will first be offered by the {\it diffuse  }  \sn\ \n\ flux (commonly called ``diffuse supernova neutrino background", \df). 
This flux receives contributions from all the \sne\ in the universe and therefore is practically constant in time\footnote{ Fluctuations in time could be seen due to individual supernovae at several megaparsecs of distance \cite{Ando:2005ka}.}, requiring only the right experimental sensitivity to be seen.  Once observed, it will turn the field of \sn\ \ns\ from the realm of rare events to  the
territory of a moderately paced and steady progress.

In addition to testing the variety of physics already probed by SN1987A -- like \n\ masses and mixings, \n\ spectra formation in the star, and a number of exotica --  the diffuse flux will offer other, complementary, information.
 Most importantly, the diffuse flux images the whole supernova population of the universe, comprised of progenitor stars of different mass and distance. Thanks to the fast rising of the \sn\ rate with the redshift, a substantial fraction of the \df\ at Earth originates at cosmological distances.
This opens the exciting possibility to do cosmology with \ns, and test not only the \sn\ rate, but also the rate of star formation, of which \sne\ are tracers. 

Since the original idea that diffuse \sn\ \ns\ might be detectable \cite{Bisnovatyi-Kogan:1982rd,Krauss:1983zn},  the physics of the \df\ has matured considerably.   After early upper limits that exceeded the predictions by orders of magnitude \cite{Zhang:1988tv,Aglietta:1992yk}, a turning point happened in 2003 when SuperKamiokande \cite{Malek:2002ns} 
placed   a bound that touched the interval of existing theoretical predictions, $\Phi \sim 0.1 - 1~{\rm cm^{-2} s^{-1}} $ above 19.3 of neutrino energy (see e.g.: \cite{woosleyetal,Totani:1995rg,Totani:1995dw,Malaney:1996ar,Hartmann:1997qe,Kaplinghat:1999xi}),
%
thus rising the hopes that the \df\ might be detected soon.  
 That first SuperKamiokande result, and its later updates \cite{Bays:2011si,Zhang:2013tua}, have
 motivated more detailed theoretical predictions of the \df, which now include several \n\ oscillation  effects \cite{Ando:2002zj,Fogli:2004ff,Ando:2004hc,Cocco:2004ac,Lunardini:2005jf,Chakraborty:2008zp,Galais:2009wi,Chakraboty:2010sz,Lunardini:2012ne}, \sn\ rate functions motivated by different data sets  \cite{Ando:2002ky,Strigari:2003ig,Ando:2004sb,Ando:2004hc,Iocco:2004wd,Daigne:2005xi,Lunardini:2005jf,Mathews:2014qba}, \n\ spectra from several numerical calculations \cite{Ando:2004hc,Horiuchi:2008jz,Lunardini:2012ne,Nakazato:2013maa} and inspired by SN1987A as well  \cite{Fukugita:2002qw,Lunardini:2005jf,Yuksel:2007mn,Horiuchi:2008jz,Vissani:2011kx}, and even possible new non-standard physics \cite{Ando:2003ie,Fogli:2004gy,Goldberg:2005yw,Volpe:2007qx,Farzan:2014gza} and new \sn\ types \cite{Daigne:2005xi,Lunardini:2009ya,Lien:2010yb,Yuksel:2012zy,Mathews:2014qba}.    
Studies show that the current  bound implies conditional constraints on the \sn\ rate \cite{Strigari:2005hu,Lien:2010yb} and on the \n\ flux parameters \cite{Yuksel:2005ae,Bays:2011si,Yuksel:2012zy}, and discuss  what will be learned from a future detection on the \n\ spectra \cite{Lunardini:2007zz}.  

While developing the phenomenology of the \df, the \n\ community looks ahead to the next generation of large scale detectors \cite{Abe:2011ts,Agostino:2012fd,Wurm:2011zn,Stahl:2012exa,Adams:2013qkq,An:2015jdp,Kim:2014rfa}
of which some will expand existing projects, while others will be in completely new  multi-disciplinary facilities, like the 
\snew{
US-based  Deep Underground Neutrino Experiment (DUNE) \cite{Stahl:2012exa,Adams:2013qkq,p5report,cdr}. 
}

 For the new \n\ experiments, observing the \df\ is an important item of the agenda, to the point that technical upgrades are sometimes driven by this specific goal \cite{Beacom:2003nk}.    For all detection technologies backgrounds are the main limiting factors, as they often restrict the sensitive energy window considerably (to the $\sim$20-40 MeV interval, in the case of a water \ck\ detector).  Even within the energy window, backgrounds limit the benefits of the larger detector mass. 

Interestingly, searches for the \df\ show how, in the new chapter of \n\ astrophysics, what were once sought after signals -- such as solar and atmospheric \ns\ --  will become well known backgrounds that will have to be reduced or subtracted. This shift in the focus  might have interesting implications on what characteristics might define an ideal detector several decades from now.

In this time of intense activity on the diffuse  \sn\ \n\ flux, this review may offer a timely summary as well as a useful perspective on this new direction of  research,  within the activity of scoping of the next generation of \n\ \snew{observatories}.   The paper opens with a section of essential facts (sec. \ref{essentials}), which are then developed in sec. \ref{theory} for the theory and in sec. \ref{detection} for detection aspects.  A section of discussion of the physics potential of the \df\ follows in closing (sec. \ref{perspectives}).  

\emph{\underline{ Note}:} Many of the numerical results (figures and tables) that appear here were prepared specifically for this review -- and not taken from previous literature --  for the sake of consistency in the graphics styles and in the sets of input parameters.    The original works where analogous results appear will be referenced as accurately as possible, with apologies in advance for involuntary omissions.

\section{Diffuse \sn\ \ns : the essentials}
\label{essentials}

This section gives a minimal introduction to the subject of diffuse \sn\ \ns. It might be useful to the reader who needs only the essential information, and to others as a summary of the reminder of the review.

Stars with masses larger than $\sim 8 M_\odot$ (with $M_\odot=1.99 \cdot 10^{30}$ Kg the mass of the Sun) end their lives with the gravitational collapse of their core, followed first by  \n\ emission  over a time scale of about 10 s, and then by a shock-driven, very luminous, explosion called a supernova (SN).  These core collapse supernovae\footnote{Core collapse \sne\ are astronomically classified as type II, Ib and Ic.  Type Ia \sne\ are of entirely different nature and do not involve a collapse of the stellar core.} are relatively rare phenomena: their rate in the universe today (redshift $z=0$) is $R_{SN}(0)\sim  10^{-4}~{\rm yr^{-1}~Mpc^{-3}} $. Interestingly, the \sn\ rate is a growing function of the redshift,  $z$, signifying that \sne\ were more frequent in the past (sec. \ref{snrate}).

The matter inside a \sn\ is dense enough (reaching nuclear density in the core, 
\snew{
$\rho \simeq  3 \cdot10^{14}~{\rm g~ cm^{-3}}$) 
}
to host a thermal population of \ns\ of all species ($\nue, \barnue, \numu, \nutau, \barnumu, \barnutau$) which then diffuse out and reach the Earth, carrying information on the stellar temperature in their nearly thermal energy spectrum, which peaks at $\sim 10-20$ MeV.  It is expected that $\nue$ and $\barnue$ have colder spectra than the other species, as they are more strongly coupled to matter (sec. \ref{snspectra}). 
Neutrinos  dominate the energetics of a \sn: they carry away about 99\% of the gravitational binding energy released in the collapse, $E_b  \simeq 3 \cdot 10^{53}~{\rm ergs}$ ($=3 \cdot 10^{46}$ J), which is roughly equipartitioned between the six \n\ species.

On the way between their production point and a detector on Earth, the \ns\ undergo redshift of energy and flavor conversion (oscillations), so that the flux of \ns\ (antineutrinos) of a given flavor in a detector is a linear combination of the fluxes of neutrinos (antineutrinos) originally produced in different flavors (sec. \ref{flavconv}).
If all \sne\ are outside our immediate galactic neighborhood (farther than few megaparsecs), the flux we receive from each of them is practically infinitesimal, but the total, diffuse, flux from all \sne\ combined is in principle observable.  In terms of the (comoving) \sn\ rate, $R_{
SN}(z)$,  the diffuse  flux of $\barnue$ in a detector at Earth, differential in  energy,
surface and time, is given by:
 \be
\Phi_{\bar e}(E)=\frac{c}{H_0}\int_0^{z_{ max}} R_{ SN}(z)
F_{\bar e}(E^\prime)
 \frac{{d} z}
{\sqrt{\Omega_{ m}(1+z)^3+\Omega_\Lambda}}~ 
  \label{flux}
\ee
 (see
 e.g. \cite{Ando:2004hc}), where $F_{\bar e}(E^\prime)$ is the contribution of an individual
supernova, inclusive of neutrino oscillations and of the redshift of
energy, $E' = E(1+z)$, and differential in $E^\prime$.   $\Omega_{ m}$ and $\Omega_\Lambda$ are the fractions of the
cosmic energy density in matter and dark energy respectively;  $c$ is
the speed of light and $H_0$ is the Hubble constant.  $z_{max}$ is the maximum redshift for which there is substantial star formation, $z_{max} \sim 5$ (sec. \ref{snrate}).

\begin{figure}[htbp]
\includegraphics[width=0.75 \textwidth]{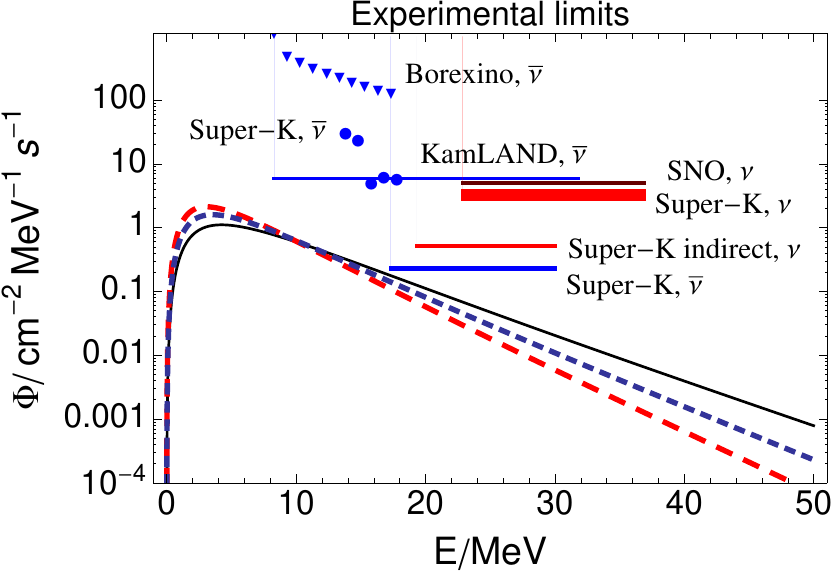}\hspace{2pc}%
\caption{\label{limitssummary}  Experimental limits for the $\nue$ and $\barnue$ components of the \df, (Table 
\ref{ex}) 
compared with theoretical predictions for three different examples of \n\ spectra. 
\snew{
The upper to lower curves at 20 MeV refer to the H, W and C spectrum (Table \ref{SNmodeltable}), 
}
and complete flavor permutation ($p=\bar p=0$), for which fluxes are maximal for the energies of interest.  To make the comparison meaningful,  limits on the energy-integrated fluxes have been divided by the size of the energy window of sensitivity of the experiment (see sec. \ref{enwindow}).  For the \sk\ limits, the widths of the lines represent 
how each bound varies with the variation of the \n\ energy spectrum. See sec. \ref{upperlimits} for details. }
\end{figure}

Estimates (fig. \ref{limitssummary}) show that for realistic \n\ spectra and flux normalizations, the \df\ peaks around $5 -7$ MeV of energy, where it can be as large as $\Phi \sim 5 ~{\rm cm^{-2} s^{-1} MeV^{-1}}$  for each \n\ species. It decays exponentially with energy above the peak.  The flux in each \n\ type is typically in the range $12 - 20~{\rm cm^{-2} s^{-1} }$ if integrated over all energies, and $\sim 0.1 - 0.8~{\rm cm^{-2} s^{-1} }$ in the energy window of current experimental interest: $E \sim 18 - 35$ MeV. This window  is determined by backgrounds such as spallation and solar neutrinos at low energy, and atmospheric neutrinos at high energy (sec. \ref{enwindow}). 

\begin{figure}[htbp]
 \begin{center}
\includegraphics[width=0.75 \textwidth]{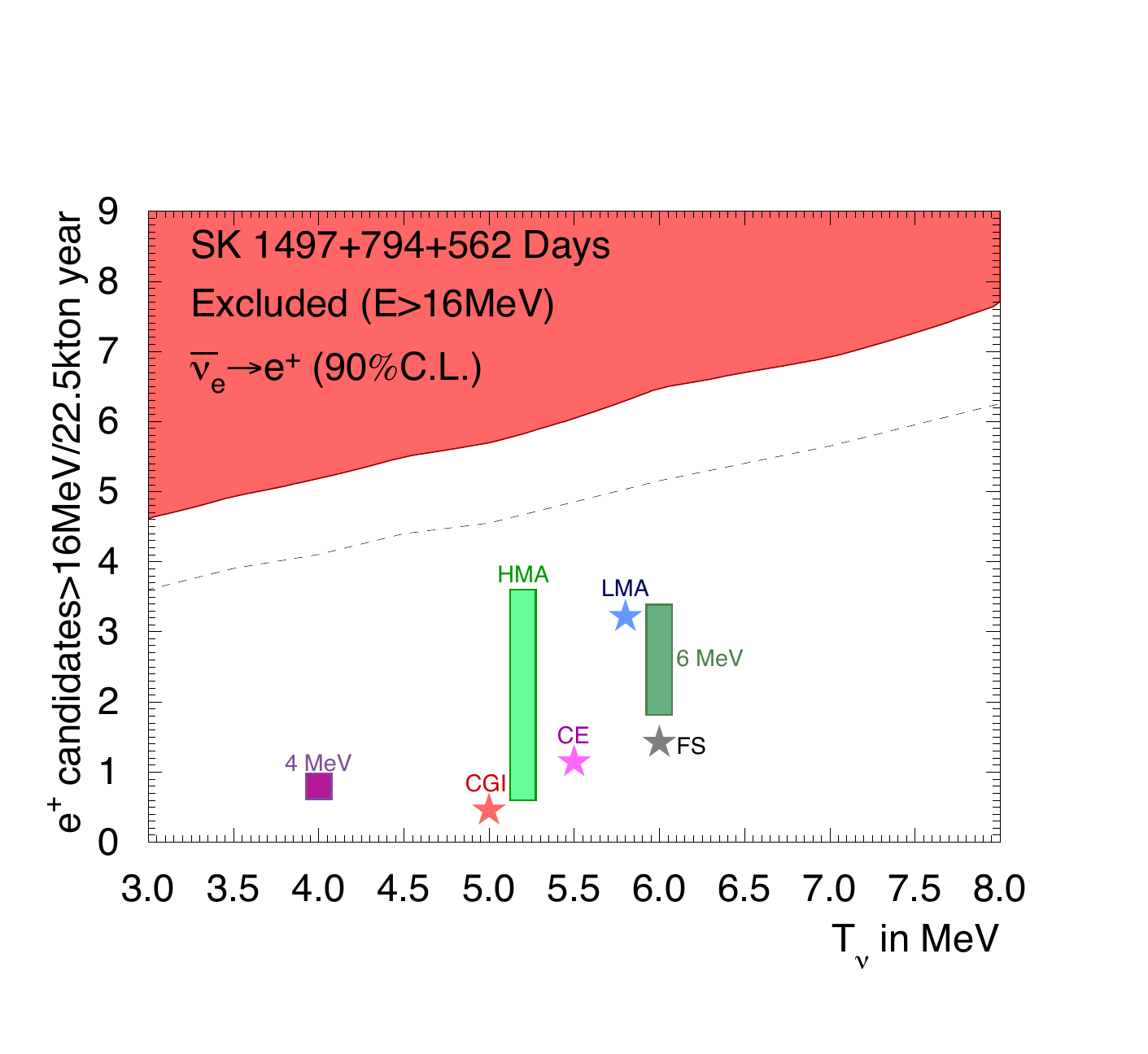}\hspace{2pc}%
\end{center}
\caption{\label{comparefig} 
From \cite{Bays:2011si}: The SuperKamiokande 90\% C.L. two-dimensional exclusion contour  (shaded, red region) in the parameter space of event rate versus \n\ temperature (a Fermi Dirac spectrum is assumed here for  the \ns\ from an individual \sn).  For comparison, a number of theoretical predictions are shown. The 6 and 4 MeV cases are from \cite{Horiuchi:2008jz}, and refer to Fermi-Dirac spectra. The other models use a variety of prescriptions, and are: the Cosmic Gas Infall model (CGI) \cite{Malaney:1996ar}, the Heavy Metal Abundance model (HMA) \cite{Kaplinghat:1999xi}, the Chemical Evolution model (CE) \cite{Hartmann:1997qe}, the Large Mixing Angle model \cite{Ando:2002ky}, and  Failed Supernova (FS) model \cite{Lunardini:2009ya}.   For these models, the histograms represent the theoretical uncertainties. The star symbol is used for predictions with no published uncertainty. The dashed line is the one-dimensional limit obtained for each (fixed) \n\ temperature.  The threshold of this analysis is 16 MeV of positron energy (17.3 MeV of \n\ energy).
}
\end{figure}

Experimentally, many upper bounds on the \df\ exist (fig. \ref{limitssummary}; sec. \ref{upperlimits}). 
The strongest is on the $\barnue$ component, from the positron search at SuperKamiokande (\sk) 
 \cite{Bays:2011si}:
\be
\Phi_{\barnue}(E>17.3~ {\rm MeV})<2.8- 3.0~{\rm cm^{-2} s^{-1}~~~~at~ 90\% C.L.}~.
\label{sklim}
\ee
 This limit is a factor of $\sim 2 - 10$ away from theoretical predictions (fig. \ref{comparefig}), and thus suggests the detectability of the \df\ at current, or, more likely,  near future detectors.  
\snew{
Currently, the most realistic projects for new \n\ detectors include a $\sim$ 0.5 Mt water \ck\ detector \cite{Abe:2011ts}, two $\sim$20 kt liquid scintillator designs \cite{An:2015jdp,Kim:2014rfa} and a $\sim40 $ kt liquid argon experiment \cite{cdr} (secs. \ref{waterc}-\ref{scintgad}). 
}
These are largely complementary in their capability for the \df: the Mt water \ck\ design will have the largest statistics ($\sim 25 - 120 $ events for a $2.5~{\rm Mt \cdot yr}$ exposure, sec. \ref{waterc}), while a solution of water and Gadolinium as well as  liquid scintillator allow to reduce background, and liquid argon has a unique sensitivity to electron \ns.

The first phase of the \df\ detection will probably be a test of the theoretically predicted event rates, and will constrain the multi-dimensional space of the parameters that govern the \df.  A more mature phase will require high energy resolution and lower energy thresholds (compared to \sk) as key elements.  It would allow to reconstruct the energy spectrum of at least some flavor components of the \df\ and therefore infer information  -- even though in a model dependent way --  on the spectrum of the \ns\ that emerge from an individual \sn\ (sec. \ref{spectesting}).  Tests of the cosmological \sn\ rate, $R_{SN}(z)$, will be complementary to, but probably not as sensitive as,  astrophysical ones (sec. \ref{perspectives}). Neutrino oscillation effects will be probed at a basic level, limited by the high energy thresholds and by the integration over $z$ washing out interesting spectral features (sec. \ref{osctesting}). Several exotica will be probed, improving on the many strong constraints already placed by SN1987A, especially thanks to the longer propagation distances involved.
%


\section{The theory: diffuse \ns\ from \sne}
\label{theory}

\subsection{Core collapse \sne\ and their cosmological rates}
\label{snrate}

How common are \sne\ in the universe?
The question of the rate of stellar death by core collapse is interestingly related to the question of stellar birth.  
Indeed, supernova progenitors have a typical life span of  $\sim 10^7$ years, much smaller than 
the characteristic time of formation of a star. 
 Thus, it is expected that the supernova rate  (\snr) as a function of the redshift $z$, should be proportional to the cosmological star formation rate  (\sfr), defined as the mass that forms stars per unit of (comoving) volume per unit time.   Assuming that stars are are distributed in mass according to the Salpeter Initial Mass Function, $\phi(m) \propto m^{-2.35}$ \cite{Salpeter:1955it}, this proportionality between the supernova \snew{number} rate  $R_{SN}(z)$ and the star formation rate $R_{SF}(z)$ \snew{(which is a mass rate, note the different units) } reads:
\be
R_{SN}(z)= \frac{\int^{125 M_\odot}_{8 M_\odot} dm \phi(m)}{\int^{125 M_\odot}_{0.5 M_\odot} dm m \phi(m)} R_{SF}(z)  \simeq \snew{0.014}~ M^{-1}_{\odot} R_{SF}(z)~,
\label{rsn}
\ee
where a cutoff of $0.5 M_\odot$ has been assumed and  $125 M_\odot$ is a tentative upper limit for the occurrence of normal core collapse \sne\ (as opposed to pair instability ones or black-hole forming events) \footnote{
\snew{Although generally motivated by stellar evolution studies (see, e.g. \cite{Woosley:2002zz}), the integration limits in eq. (\ref{rsn}), are not robustly known.  The resulting uncertainty on the supernova rate is subdominant compared to  other major  uncertainties of a factor $\sim 2$ that are discussed in this section, see fig. \ref{snrplot} and \ref{snrcontours}. } }. 

\begin{figure}[htbp]
\begin{center}
      \includegraphics[width=0.75\textwidth]{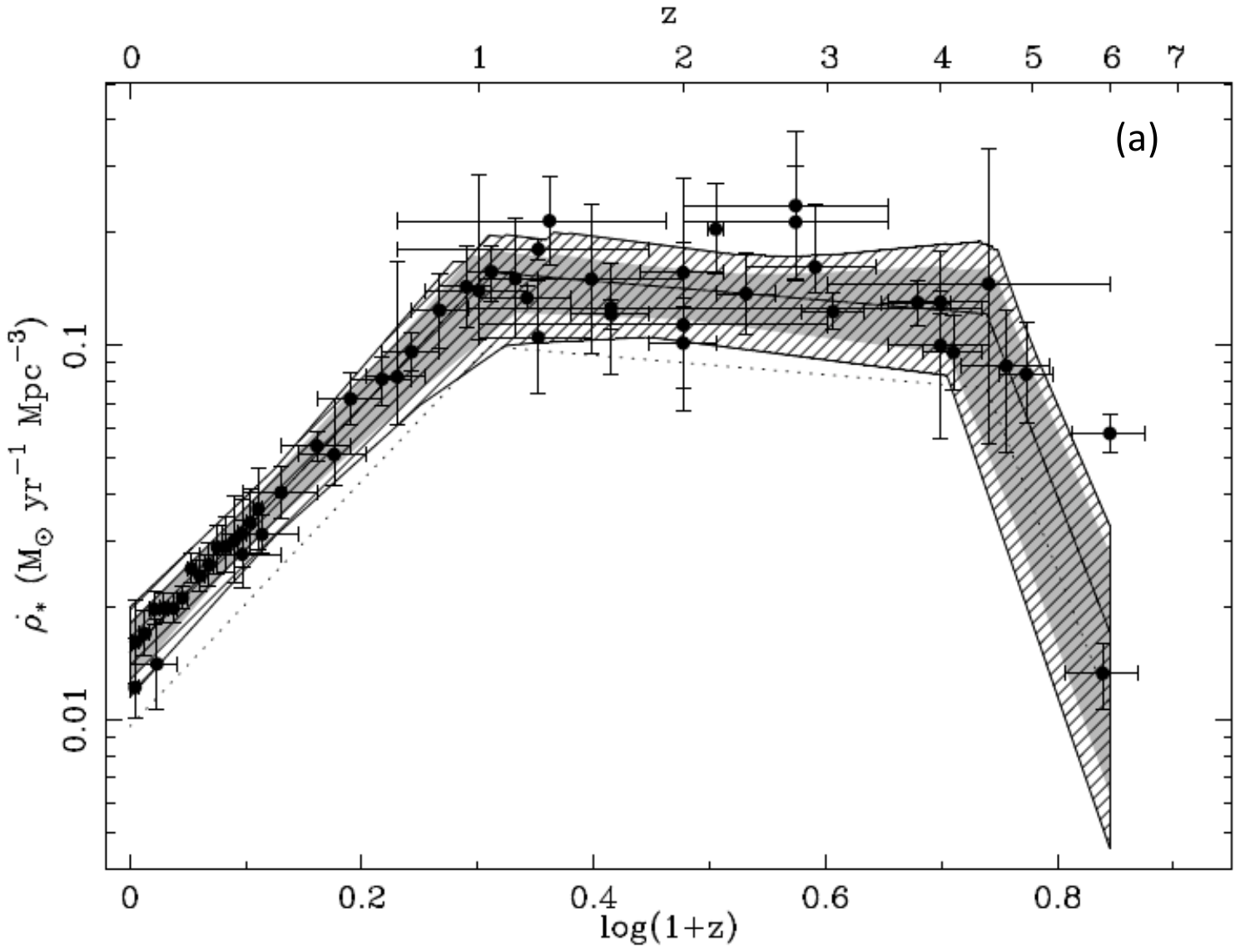}\\
      \includegraphics[width=0.75\textwidth]{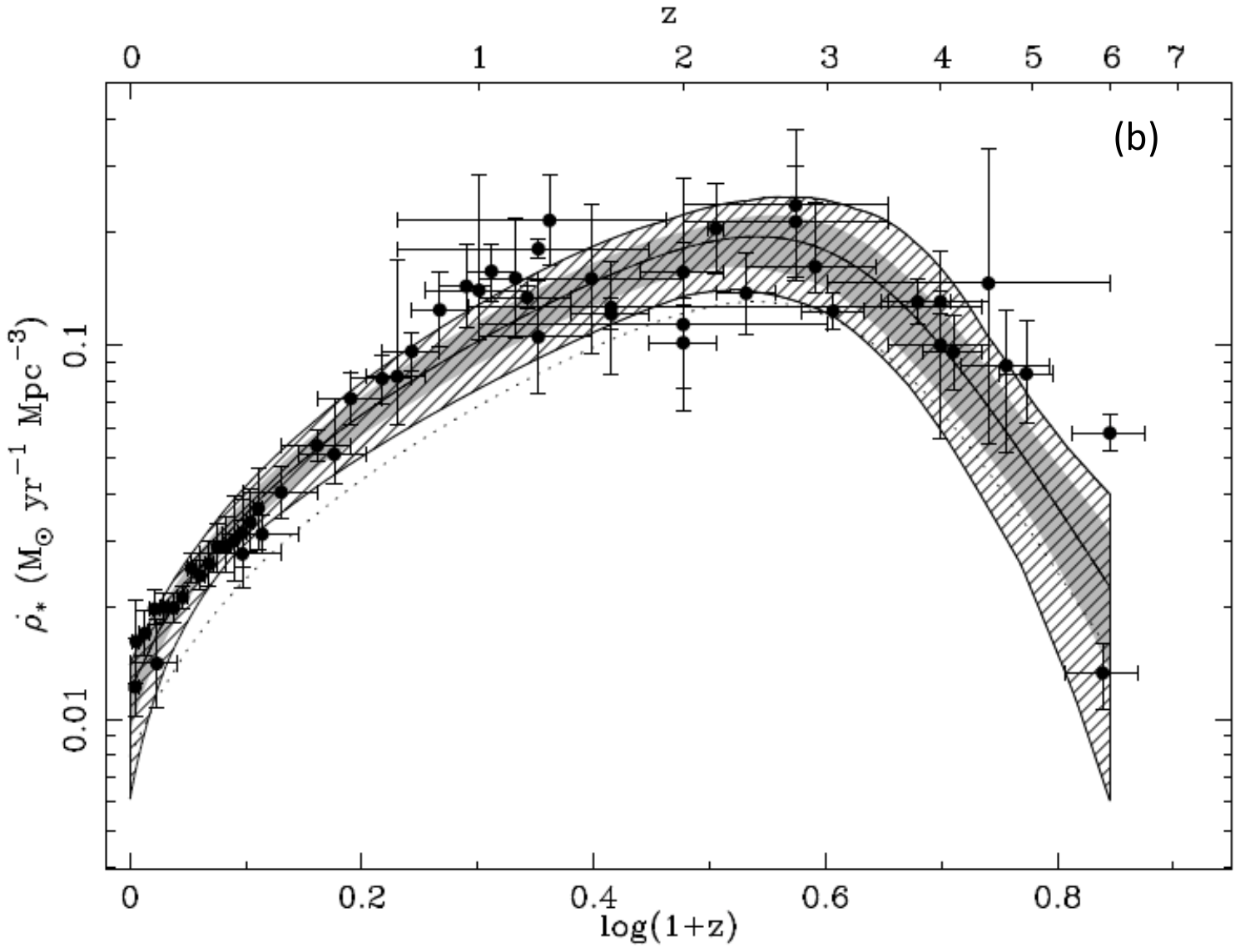}
      \end{center}
    \caption{Measurements of the cosmological star formation rate and functional fits, from \cite{Hopkins:2006bw}.  The grey shaded and hatched regions are the 1$\sigma$ and 3$\sigma $ confidence regions obtained with (a) the piecewise parameterization and (b) the Cole et al. parameterization, Eq. (\ref{sfrcurves}). 
The \sk\ limit on the \df, Eq. (\ref{sklim}), constrains the normalization to values lower by a factor 0.74 - 1 depending on the \n\ spectrum; the rescaled \sfr\ is indicated by the dotted  curves, see \cite{Hopkins:2006bw}  for details.}
    \label{beacomplots}
\end{figure}
 The  \sfr\ is fairly well known, even though with some uncertainties, especially on its normalization.  Different tracers of star formation, such as  ultraviolet
and far-infrared radiation, indicate that the \sfr\ grows with $z$ (see e.g., \cite{Hopkins:2006bw} for a review). Fig. \ref{beacomplots} shows the data and two commonly used functional fits, a piecewise one and a continuous one from  \cite{Cole:2000ea}:  
\beq
{R_{SF}(z)}\propto\Bigg\{ \begin{array}{lc}{(1+z)^{\beta}} & z<1 \\
{(1+z)^{\alpha}} & 1<z<4.5\\
{(1+z)^{\gamma}} & 4.5<z\\
\end{array} \nonumber \\
{R_{SF}(z)}\propto \frac{a + b z}{1 + (z/c)^d}~,
\label{sfrcurves}
\eeq
where $\alpha,\beta,\gamma, a,b,c,d$ are fit parameters.  
In the next sections the piecewise form will be used, for its transparency and easier comparison with other literature. Fig.  \ref{beacomplots}, as well as the detailed statistical analysis  in \cite{Hopkins:2006bw}, show that the \sfr\ is about $R_{SF}(0)=0.015 M_\odot {\rm  yr^{-1} Mpc^{-3}}$ today,  grows with power $\beta \simeq 3$ ($\beta = 3.28$ best fit) up to $z\simeq 1$, flattens at larger redshift  ($\alpha=-0.26$ best fit) and decreases at $z\gta  4.5$ ($\gamma = -7.8$ best fit).  

\begin{figure}[htbp]
\newpage
\centering
\includegraphics[width=0.6\textwidth]{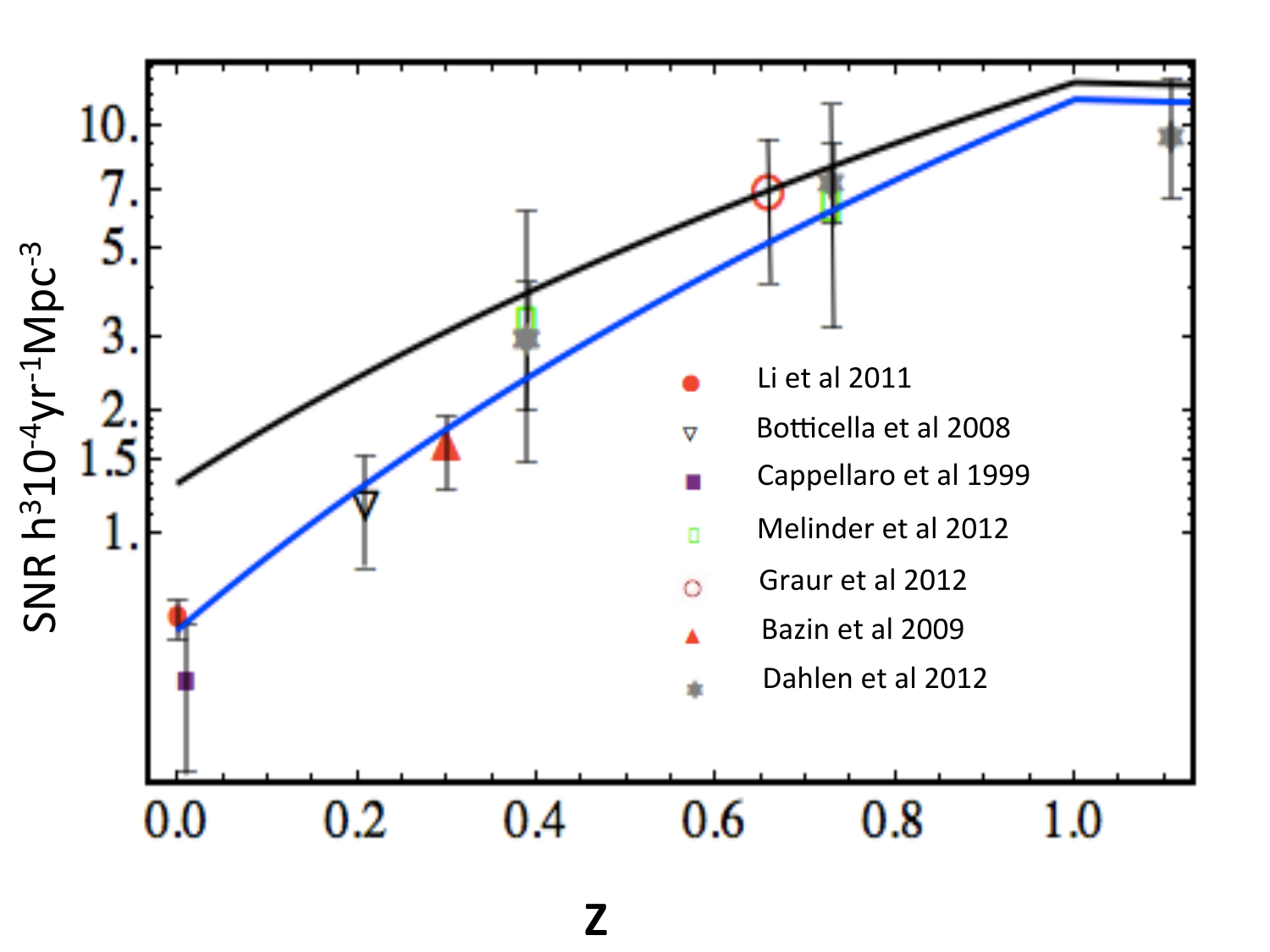}
      \caption{  
      Direct \sn\ rate measurements as of 2013 \cite{Botticella:2007er,Cappellaro:1999qy,Li:2010kd,Melinder:2012nv,Graur:2011cv,Bazin:2009mp,Dahlen:2012cm};   from \cite{yangthesis}, with their best fit function (lower curve). The flattening of the function at $z \simeq 1$ is motivated by \sfr\ observations (fig. \ref{beacomplots}) and not by the \sn\ data themselves.  The upper curve represents Eq. (\ref{snrpractical}) and is the \snr\ obtained from a \sfr\ data fit via Eq. (\ref{rsn}) \cite{Hopkins:2006bw}.     
      }
    \label{snrplot}
\end{figure}
%
\begin{figure}[htbp]
  \centering
\includegraphics[width=0.55\textwidth]{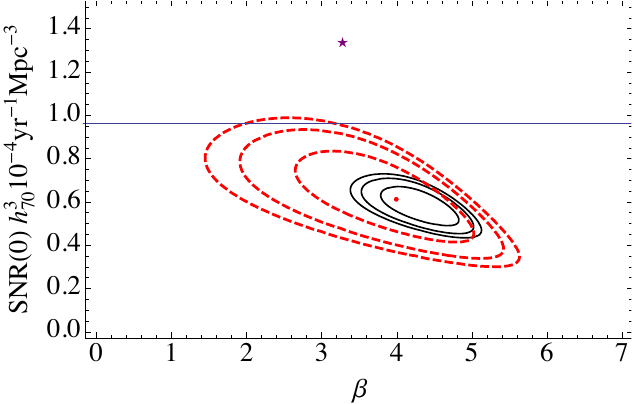}
\caption{
Best fit point and isocontours of $\chi^2$ in the space of the parameters describing the \snr\ function, $R_{SN}(z)$ \cite{yangthesis}. These are the intercept, $R_{SN}(0)$ (in units of $10^{-4}~{\rm yr^{-1}~Mpc^{-3}}$) and the power, $\beta$ (for the piecewise parameterization in Eq. (\ref{sfrcurves})).  Results are presented for two different analyses, with statistical errors only (solid, black curves), and with statistical and correlated systematic errors (dashed regions, in red). For each, the contours refer to 68.3, 90, 95.4\% C.L..   The star represents the star formation rate-favored  parameters, $(\beta, R_{SN}(0))=(3.28, 1.33)$, \cite{Hopkins:2006bw}.  The horizontal line is a lower bound from Smarrtt et al. \cite{Smartt:2008zd}. 
}
\label{snrcontours}
\end{figure}
The corresponding  best fit \snr, obtained via Eq. (\ref{rsn}), can be written as:
\be
{R_{SN}(z)} =R_{-4} 10^{-4}~{\rm yr^{-1} Mpc^{-3}}  \Bigg\{ \begin{array}{lc}{  (1+z)^{\beta}} & z<1 \\
{2^{\beta - \alpha}(1+z)^{\alpha}} & 1<z<4.5\\
{2^{\beta - \alpha}~ 5.5^{\alpha -\gamma} (1+z)^{\gamma}} & 4.5<z\\
\end{array}
\label{snrpractical}
\ee
with $R_{-4} \sim 1$ describing the uncertainty in the normalization.   It is plotted in fig. \ref{snrplot} (left panel, upper dashed line).  

Fig. \ref{snrplot}  also shows
the  {\it direct} measurements of the \snr\ as of 2013  \cite{Botticella:2007er,Cappellaro:1999qy,Li:2010kd,Melinder:2012nv,Graur:2011cv,Bazin:2009mp,Dahlen:2012cm} and their piecewise fitting curve \cite{yangthesis}.  
For the same fit, fig. \ref{snrcontours} gives 
the allowed region  of the parameters $R_{SN}(0)$ and $\beta$; the  correlation between the two quantities is evident in the figure.
   The comparison between the \snr\ obtained directly and that derived from the \sfr\ (upper curve in fig. \ref{snrplot}) shows  a discrepancy at low redshift, which can be interpreted as a measure of the systematic errors involved. At least part of this discrepancy is resolved when including faint or completely dark \sne\ \cite{Horiuchi:2011zz,Mathews:2014qba}. 

\subsection{Neutrinos from \sne}
\label{snspectra}

Core collapse supernovae are the only site in the universe today 
 where the matter density is large enough  to have the buildup of a
thermal gas of neutrinos.  Thanks to their lack of electromagnetic
interaction, these neutrinos can diffuse out of the star over a time
scale of few seconds, much shorter than the diffusion time of photons. This makes
the neutrinos the principal channel of emission of  the ${\mathcal
O} (10^{53})$ ergs of gravitational energy that is liberated in the
collapse.
 The energy spectrum  of neutrinos of each flavor is expected
to be thermal near the surface of decoupling from matter (``neutrinosphere"), but then it changes due to propagation effects. 
One of these effects is scattering. 
Numerical modeling indicates that, after scattering right outside the decoupling region, neutrinos  of a given flavor $w$ ($w=e,\mu,\tau$) have energy spectrum:  
\be
F^0_w= \frac{{d} N_w}{{d} E}\simeq \frac{(1+\alpha_w)^{1+\alpha_w}L_w}
  {\Gamma (1+\alpha_w){E_{0w}}^2}
  \left(\frac{E}{{E_{0w}}}\right)^{\alpha_w}
  e^{-(1+\alpha_w)E/{E_{0w}}}, 
  \label{nuspec}
\ee
 \cite{Keil:2002in}, where $E$ is the neutrino energy, $L_w$ is the energy emitted in the species $w$ and $E_{0w}$ is the average energy of the spectrum.  The quantity $\alpha_w$ is a numerical parameter, $\alpha_w \sim 2 - 5$ \cite{Keil:2002in}, describing the shape of the spectrum.  
Considering typical temperatures of matter near the collapsed core, one expects average energies in the 10-20 MeV range. 
The non-electron neutrino flavors, $\numu$, $\nutau$, $\barnumu$ and $\barnutau$ (each of them denoted as $\nu_x$ from here on\footnote{For the energies of interest here, the produced fluxes of $\numu$, $\nutau$ are equal with very good approximation. They are also nearly equal to the fluxes of $\barnumu$ and $\barnutau$, up to corrections due to weak magnetism \cite{Horowitz:2001xf}.}), interact with matter more weakly than $\nue$ and $\barnue$ (via neutral current processes only), and therefore decouple from matter in a denser and hotter region.  Additionally, an asymmetry exist in the coupling with matter of $\nue$ and $\barnue$: these are kept in thermal equilibrium by charged current interaction on neutrons and protons respectively, and  the overabundance of neutrons relative to protons implies a lower decoupling temperature for $\nue$. Therefore, we expect a hierarchy of average energies: 
\be
 E_{0  e} < E_{0 \bar e} < E_{0 x}~.
\label{ehier}
\ee


\begin{table}[htb]
\begin{center}
\begin{tabular}{| c |c |c |c |c |c |c |}
\hline
\hline
Case & $E_{0 e}$ (MeV)  &  $E_{0 \bar e}$ (MeV) &  $E_{0 x}$ (MeV) & $\alpha_e$ & $\alpha_{\bar e} $ & $\alpha_x$ \\
\hline
\hline
Cold (C) &  9 &  11 & 13 &  3 & 3 & 2 \\
\hline
Warm (W) & 11 & 14 & 15 &   3 & 3 & 2 \\
\hline
Hot (H) &  12 &  15 & 18 &  3 & 3 & 2 \\ 
\hline
\hline

\end{tabular}
\caption{\snew{ The set of spectral parameters used here. Their labels, Hot, Warm and Cold, are purely conventional. The total energies emitted in each flavor are $L_e= L_{\bar e}= L_x=5\times 10^{52} $ ergs in all cases.  } } 
\label{SNmodeltable}
\end{center}
\end{table}
While the basic spectral features and the total energy emitted in \ns\ are generic predictions, a detailed study of \n\ emission in a \sn\ requires complex numerical calculations, that have been performed  by several groups in the quest of reproducing the observed explosion that follows the collapse
(see e.g. \cite{Cardall:2005ib,Papish:2014tya} for a review).

\snew{Here the form (\ref{nuspec}) will be used for the spectrum of each \n\ flavor before oscillations, integrated over the $\sim 10$ s duration of the \n\ burst. 
For illustration,  three different sets of spectral parameters are used, conventionally named Hot (H), Warm (W) and Cold (C). They are summarized in Table \ref{SNmodeltable}. They represent examples, loosely covering the range of parameters that have emerged from recent numerical simulations  (e.g., \cite{Marek:2007gr,Ott:2008jb,Huedepohl:2009wh,Fischer:2009af,Sumiyoshi:2006id,O'Connor:2012am,Nakazato:2012qf}).   The C case is quantitatively close to the results of  the Basel group \cite{Fischer:2009af} (after time-integration), that were calculated over several seconds after the collapse. The W case is meant to represent the situation in which the 
spectra of $\barnue$ and $\nux$ quickly become similar immediately after $\sim 1$ s accretion phase (see e.g., \cite{Huedepohl:2009wh}). The H case account for the possibility of more energetic \n\ spectra, that might be realized in a star with a longer accretion phase (see e.g., \cite{Nakazato:2012qf}).  
}

\begin{figure}[htbp]
  \centering
 \includegraphics[width=0.47\textwidth]{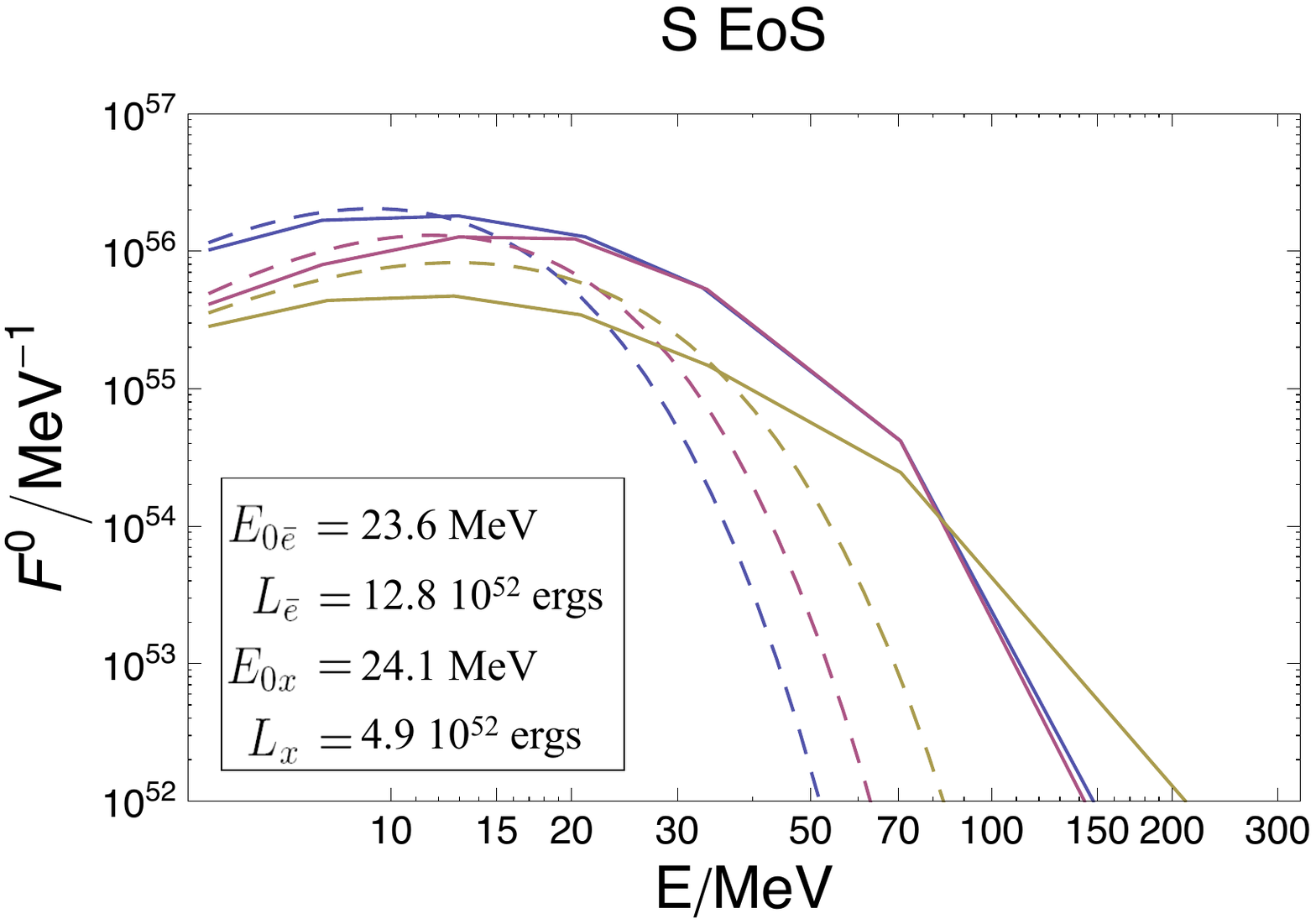}
 \includegraphics[width=0.47\textwidth]{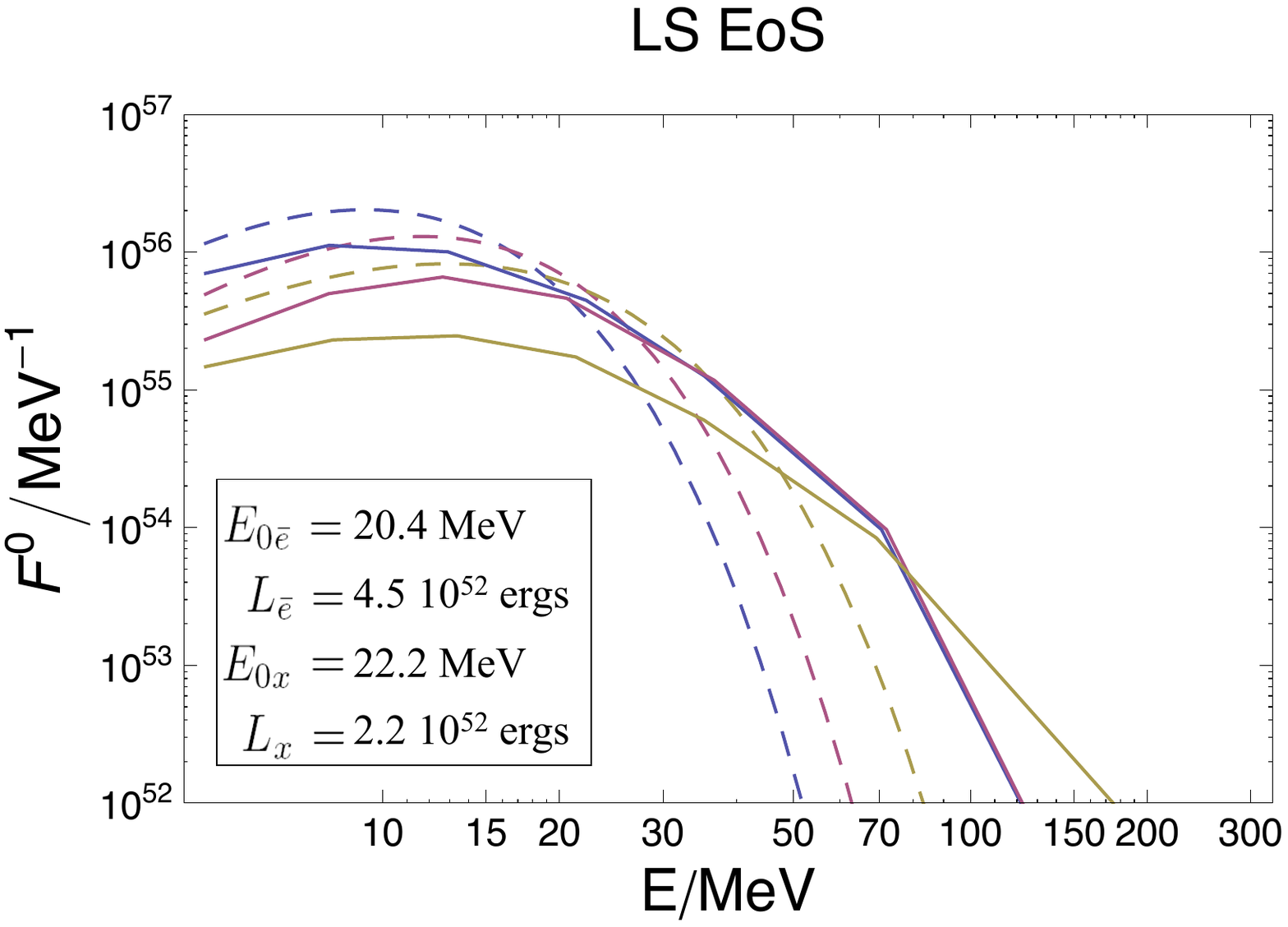}
   \caption{From \cite{Lunardini:2009ya}: neutrino fluxes at production inside the star for direct \bh\ collapse (solid lines), taken from the numerical calculation of \cite{Nakazato:2008vj}. Curves from upper to lower at 5 MeV correspond to $\nue$, $\barnue$, $\nux$.  Spectra are shown for the Shen et al. (left panel) and Lattimer-Swesty (right) equation of state.  For each, the \n\ average energies and the total energy emitted per flavor are given (inserts).  See text for details. The dashed lines represent typical spectra  for \nts\  collapse (Eq. (\ref{flux})), with the parameters  $E_{0 \bar e} = 15$ MeV, $E_{0
x} = 18$ MeV, $L_{\bar e}=L_x=5 \cdot 10^{52}$ ergs, $\alpha_{\bar
e}=3.5$ and $\alpha_x=2.5$. }
\label{BHspectra}
\end{figure}
The \snew{examples of spectra above} refer to the most common scenario for a \sn: a collapse that leads to the formation of a neutron star (``neutron star-forming collapses", NSFCs).
 Detailed studies have appeared 
 \cite{Liebendoerfer:2002xn,Sumiyoshi:2006id,Sumiyoshi:2007pp,Fischer:2008rh,Sumiyoshi:2008zw,Nakazato:2008vj}. 
 on the rarer case of \emph{ direct} collapse into a black hole
 without explosion, i.e., a {\it failed \sn}.  This is the fate of stars with $M \gta 25 - 40 M_\odot$, which comprise a 9 - 22\%
fraction of all \sn\ progenitors \cite{Lunardini:2009ya}.
 It was shown that the
 neutrino emission from these direct \bh\ collapses (DBHFCs) is somewhat more luminous and decidedly more
 energetic than for \nts\ collapses -- depending on the equation of state (EoS) of nuclear matter -- due to the rapid contraction of the newly formed
protoneutron star preceding the black hole formation. This suggests that the hotter contribution of \bh\ collapses 
 to the \df\ might exceed that of \nts\ ones in part of
 the energy spectrum \cite{Lunardini:2009ya,Lien:2010yb,Keehn:2010pn,Yuksel:2012zy}. 

Fig. \ref{BHspectra} shows  the \n\  fluxes emitted in a direct \bh\ collapse, from 
\cite{Nakazato:2008vj}. They  were obtained for the $40 M_{\odot}$ progenitor in  \cite{woosley} with the stiffer  Shen et al. (S) EoS 
\cite{shenetal} (incompressibility $K=281$ MeV) and the softer  Lattimer-Swesty (LS) one  \cite{lattimer91generalized} (with $K=180$ MeV, see  \cite{Nakazato:2008vj}).  
 It appears that average energies are $E_0 \sim 20-24$ MeV for all \n\
 flavors, with a stiffer EoS  corresponding to more energetic \n\ spectra. Notice how  the  $\nue$ and $\barnue$ components are especially luminous due to the high rate of capture of
electrons and positrons on nuclei.

\subsection{Neutrino flavor conversion}
\label{flavconv}

For a long time a mere hypothesis, \n\ flavor conversion is now a reality well established experimentally  \cite{Ahmad:2001an,Eguchi:2002dm}. The leading mechanism driving conversion is \n\ oscillations due to the \n\ being massive and having non-zero mixing between the \n\ mass eigenstates, $\nu_i$ ($i=1,2,3$) with masses $m_i$, and the  flavor eigenstates, $\nu_\alpha$
($\alpha=e,\mu,\tau$): $\nu_\alpha=\sum_{i}
U_{\alpha i} \nu_i$.  The standard parameterization of the mixing matrix (see e.g. \cite{Krastev:1988yu}) is in terms of three mixing angles: $\theta_{12},\theta_{23},\theta_{13}$, and a complex phase, $\delta$. Oscillation effects depend on these parameters, 
and on the mass squared differences, $\Delta m^2_{ij}=m^2_i - m^2_j$, as well as on the \n\ energy and on the density and composition of the medium of propagation due to forward scattering on the matter constituents (refraction).

After an intense phase of \n\ experiments on solar, atmospheric, reactor and accelerator \ns, measurements are available for 
$\theta_{12}, \theta_{13},\theta_{23}, \Delta m^2_{21}$ and $| \Delta m^2_{31}|$ (see e.g., \cite{Beringer:1900zz} and references therein): 
\beq
|\Delta m^2_{31}| =  (2.43^{+0.06}_{-0.10}) \cdot 10^{-3}{\rm eV}^2~, \hskip 0.6truecm   \Delta m^2_{21} = (7.54^{+0.26}_{-0.22}) \cdot 10^{-5}{\rm eV}^2~, \nonumber\\
 \sin^2 \theta_{12}=0.307^{+0.018}_{
-0.016}~,\hskip 0.6truecm
\sin^2 \theta_{23} = 0.386^{+0.0244}_{-0.021}~, \hskip 0.6truecm
\sin^2 \theta_{13} = 0.0241 \pm 0.0025~.  \nonumber\\
\label{values}
\eeq
The phase $\delta$ currently remains unmeasured; it has a 
subdominant or negligible effect on \sn\ neutrinos \cite{Akhmedov:2002zj,Balantekin:2007es}, and therefore it will be set to zero from now on.  
Current data are not sufficient to distinguish the sign of $\Delta m^2_{31}$: The two possibilities, $\Delta m^2_{32} \approx \Delta m^2_{31} > 0$ and
$\Delta m^2_{32} \approx \Delta m^2_{31} < 0$, are
referred to as {\it normal}  and {\it inverted} mass hierarchy/ordering\footnote{The parameter values in eq. (\ref{values}) refer to the normal mass hierarchy; the corresponding numbers for inverted hierarchy are only minimally different, and consistent within the error with those in eq. (\ref{values}). See \cite{Beringer:1900zz} for details. }.  

With these information, it has been possible to study flavor conversion of \sn\ \ns\ in  detail, uncovering a very rich pattern.  One can distinguish four spatially separated stages of conversion, that contribute to the final $\nue$ and $\barnue$ survival probabilities (Table \ref{probtable}):

\begin{itemize}

\item  Within $\sim 200$ Km radius in the star, collective flavor oscillations occur due to \n-\n\ coherent scattering \cite{Duan:2006an,Duan:2006jv}\footnote{See e.g., \cite{Duan:2009cd} for a comprehensive review of the vast literature on this topic.} when the density of \ns\ exceeds that of electrons \cite{EstebanPretel:2008ni}. This should be the case  $\sim 1-5$ s after the core bounce.  
The most common effect observed in numerical simulations
 is a swap of the spectra of the  electron and non-electron \ns\ and antineutrinos for the inverted mass hierarchy and  above a certain critical energy, $E_c$. $E_c$ is small 
(below typical detection thresholds)
for antineutrinos, while for \ns\ it depends on the fluxes of $\nue$, and $\barnue$ and $\nux$ ($\nux=\nu_\mu,\nu_\tau,\barnumu,\barnutau$) as: 
\be
\int_{E_c}^{\infty} (F^0_{e}-F^0_{x})=\int_{0}^{\infty} (F^0_{\bar e}-F^0_{ x})~,
\label{ecrit}
\ee
with $E_c \simeq 3-10 $ MeV as typical values \cite{Duan:2007bt}.
Here I denote as  $P_c$ and $\bar P_c$, the step-like $\nue$ and $\barnue$ survival probabilities after collective effects. They vary over the duration of the \n\ burst, due to possible suppression of collective effects in the accretion phase \cite{Chakraborty:2011nf,Chakraborty:2011gd,Dasgupta:2011jf}, and to the time variation of the fluxes in eq. (\ref{ecrit}).

\item At larger radii, where \n-\n\ coherent scattering is negligible, conversion is driven by coherent scattering on electrons, according to the Mikheev-Smirnov-Wolfenstein effect (MSW) \cite{Wolfenstein:1977ue,Mikheev:1986gs}, that was first elaborated in the context of solar \n\ and then applied to \sn\ \ns\  \cite{Mikheev:1986if}.
In essence, the MSW effect is the idea  that a small
neutrino mixing   induces more than
50\% change of flavor if neutrinos propagate in matter and: \\
(i) the matter density along the \n\ trajectory varies slowly enough that there are no quantum transitions between different eigenstates of the Hamiltonian ({\it adiabatic propagation}).  Under this condition, a neutrino that is produced in given eigenstate of the Hamiltonian in matter, $\nu_{i,m}$ will remain in such state while the flavor composition of the state itself varies as an effect of the varying matter density along the \n\ trajectory.  At emergence from the star, the \n\ is in the vacuum state $\nu_i$, whose flavor composition is described by the mixing matrix. The process results in a strong change of flavor if \\
(ii) the neutrinos
cross a region where the density is such  ({\it resonance density}) that a cancellation  occurs between the matter and kinetic terms of the Hamiltonian, thus producing a resonant behavior.

 \begin{figure}[htbp]
 \newpage
  \centering
\includegraphics[width=0.65\textwidth]{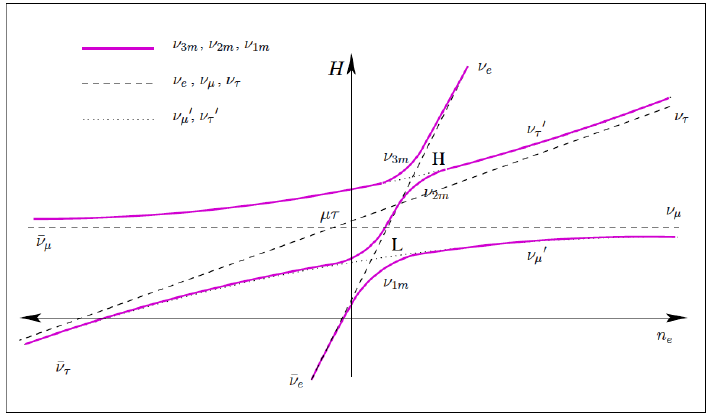}
\includegraphics[width=0.65\textwidth]{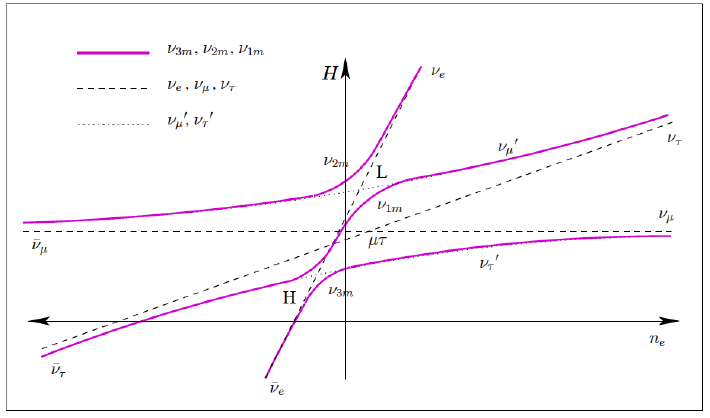}
\caption{
Schematic representation of the neutrino energy levels in matter vs electron number density, $n_e$, for normal (upper pane) and inverted (lower pane) mass hierarchy, from \cite{Akhmedov:2002zj}.  Dashed line: in the
absence of mixing (unphysical), solid line: with mixing (physical case).
The semi-plane with positive (negative) density, $n_e >0$ ($n_e < 0$), describes the conversion of neutrinos (antineutrinos), to account for the fact that the refraction potential has opposite sign for \ns\ and antineutrinos. 
The positions of the high (H) and low (L) density MSW resonances are marked.
See \cite{Akhmedov:2002zj,Lunardini:2003eh} for details.
}
\label{resonances}
\end{figure}
 Two MSW resonances relevant for $\nue$ and $\barnue$ conversion are realized inside a supernova, corresponding to the two independent mass squared splittings (fig. \ref{resonances}). 
The first resonance occurs
at    $\rho \sim 10^3~{\rm g\cdot cm^{-3}}$. 
For a power-law profile 
\be
 \rho(r)=10^{13}~ C  \left(\frac{10 ~{\rm km}}{r} \right)^3~{\rm g\cdot cm^{-3}}  \hskip 1 truecm C \simeq 1-15~
 \label{powerlaw}
 \ee
  \cite{Lunardini:2003eh}, that applies before the shock wave reaches the resonance layer \cite{Schirato:2002tg},  this resonance is adiabatic \cite{Dighe:1999bi}. 
 it depends on the \n\ mass hierarchy: it affects neutrinos (antineutrinos) if the mass hierarchy is normal (inverted).

\item The second resonance is at  $\rho \sim 10~{\rm g\cdot
    cm^{-3}}$ and is described by
  the parameters $\theta_{12}$ and $\Delta m^2_{21}$,  known as the LMA
  (Large Mixing Angle MSW)  solution of the solar neutrino
  problem.  The resonance is adiabatic and it affects neutrinos.
  Antineutrinos undergo non-resonant conversion, thanks to $\theta_{12}$ being large.   

\item A fourth stage of oscillations happens inside the Earth, where the physical conditions resemble those of the second resonance.  While potentially important for \ns\ from a galactic \sn, these oscillations affect the \df\ by less than few per cent \cite{Lunardini:2006sn}, so they will not be discussed here.  

\end{itemize}

The fluxes of $\nue$ and $\barnue$ after oscillations are a mixture of the original fluxes in these flavors and of the original muon and tau component.  Up to a geometric factor due to the distance from the star (omitted here for simplicity), they are described by the permutation parameters $p, \bar p$ as: 
\beq
&&F_{ e} = { p} F_{ e}^0 + (1-{ p}) F_{ x}^0  ~~,
\label{fluxes1}\\
&&F_{\bar e} = {\bar p} F_{\bar e}^0 + (1-{\bar p}) F_{ x}^0 ~.
\label{fluxes2}
\eeq
The oscillated fluxes, $F_{ e},F_{\bar e}$, are more energetic than the original ones due to the higher average energy of $ F_{ x}^0$. This spectral hardening is one of the main signature of \n\ flavor conversion in a \sn. 
The calculation of $p$ and $ \bar p$ is at times complex, and for this we refer to the literature (e.g., \cite{Dighe:1999bi,Lunardini:2003eh,Duan:2007fw,Dasgupta:2008cd}). Here the results are shown in Table \ref{probtable}.

\begin{table}[htbp]
\centering
\begin{tabular}{lclclcl}
\hline
 &  Normal hierarchy &  Inverted hierarchy \\ 
\hline
 $p$  &   $\sin^2 \theta_{12}(1- P_c)$  & $\sin^2 \theta_{12} P_c$  \\
\vspace{0.1cm}
 $\bar p $ &  $\cos^2 \theta_{12}\bar P_c $ &   $\cos^2 \theta_{12} (1-\bar P_c)$ \\
\hline
\end{tabular}
\caption{The electron neutrino (antineutrino) survival probability, $p$ ($\bar p$)  for \n\
emerging from a \sn. 
Here $P_c$ and $\bar P_c$ are the step-like survival probabilities after collective effects.} 
\label{probtable}
\end{table}

It appears that, considering $0 \leq P_c, \bar P_c \leq 1$,
 $p$ and $\bar p$ are in the ranges:
\beq 
&&p= 0 - \sin^2\theta_{12}\simeq 0 -  0.32~ , \nonumber \\
&&\bar p= 0 - \cos^2\theta_{12}\simeq 0 -  0.68~.
\label{rangesprob}
\eeq
In general, both quantities vary with the neutrino energy and with time, via the functions $P_c$ and $\bar P_c$, 
and also due to a change in the adiabaticity of the high density resonance due to shockwave effects \cite{Ando:2002zj,Galais:2009wi}.  However, for the purpose of the \df\ calculation, $p$ and $\bar p$ can typically be taken as constant in energy. This is justified if the  critical energy $E_c$, is below the energy window for detection of the diffuse flux. Furthermore, 
 integrating over the duration of the burst and over the \sn\ population smears out the energy dependence of the oscillation effects \cite{Chakraboty:2010sz,Lunardini:2012ne}. 

As a cautionary remark, it should be noted that the phenomenology of oscillations induced by \n-\n\ scattering is still in the phase of initial exploration, and therefore the  results in Table \ref{probtable} may change as studies progress.  For example, recently it has been understood that multiple spectral swaps due to \n-\n\ scattering can occur  \cite{Dasgupta:2009mg}, 
and that  a richer pattern of spectral splits may emerge from including the effect of the so called \n\ ``halo" \cite{Cherry:2012zw,Sarikas:2012vb,Cherry:2013mv} ,and the breaking of the azimuthal asymmetry of the collective oscillations \cite{Raffelt:2013rqa,Mirizzi:2013rla,Mirizzi:2013wda}. 
Therefore the expressions in Table \ref{probtable} may not be entirely applicable to the \df.   Variations between individual \sne\ (e.g., in the matter density profile, see \cite{Nakazato:2008vj}) may further complicate the description of conversion effects for the \df. 
In view of these uncertainties, it is adequate and convenient -- for the sake of generality -- to describe the effects of oscillations on the \df\ in terms of $p$ and $\bar p$ intended as averaged over time and over the \sn\ population.  I neglect their energy dependence for the reasons discussed so far, and vary them in the numerical intervals (\ref{rangesprob}), which are guaranteed to be valid under the sole conditions of adiabaticity of the lower density resonance and  absence of turbulence (see e.g. \cite{Kneller:2010sc}). 

\subsection{Constraints from SN1987A}
\label{SN1987Acons}

\begin{figure}[htbp]
\newpage
  \centering
    \includegraphics[width=0.65\textwidth]{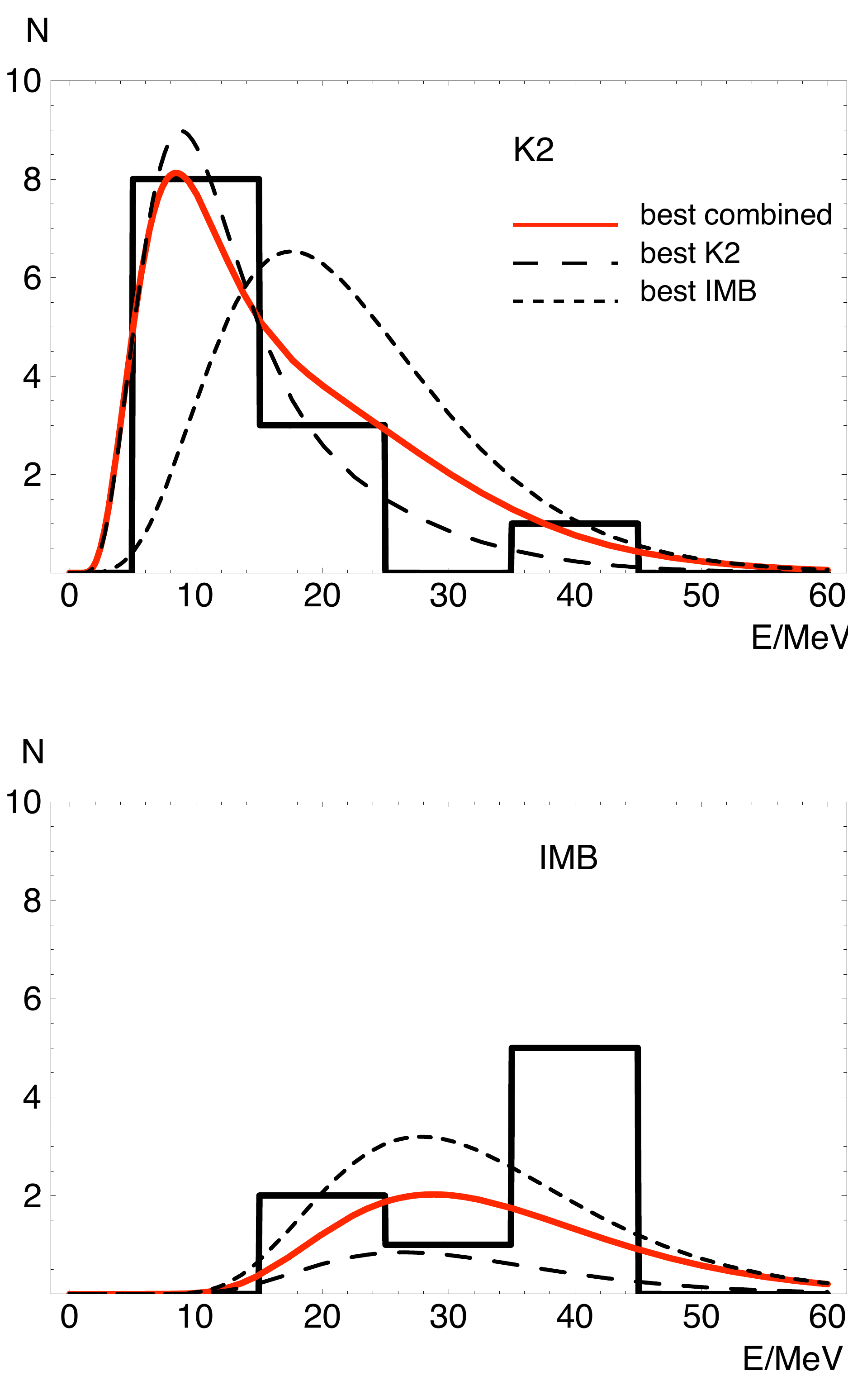}
\caption{The observed  positron energy spectra of events at Kamiokande II (K2) and IMB, compared with the predicted spectra in the points of minimum $\chi^2$ for the two data sets saparately and combined \cite{Lunardini:2005jf}.} 
\label{histo}
\end{figure}
It is interesting to check how predictions compare with the only data we currently have on \sn\ \ns\ -- those from SN1987A -- and what these data alone tell us without theoretical priors \cite{Fukugita:2002qw,Lunardini:2005jf,Vissani:2011kx}. Fig. \ref{histo} shows the data published by the Kamiokande-II  and IMB  experiments   \cite{Hirata:1987hu,Bionta:1987qt} and the best fits from a maximum likelihood analysis of the two datasets separately and combined \cite{Lunardini:2005jf}. 
The analysis was carried out assuming that all the events were from inverse beta decay ($\barnue + p \rightarrow n + e^+$), 
and with five fit parameters: $L_{\bar e},~L_x,~E_{0 \bar e}, ~E_{0 x}$ and $\bar p$ (the latter was constrained to be in the range in eq. (\ref{rangesprob})).
For simplicity, $\alpha_{\bar e}=\alpha_{x}=2.3$ (which give a spectral shape close to
Fermi-Dirac) was held fixed.  The figure shows that the two datasets are compatible, in spite of the tension in their favoring different \n\ energy spectra.  The best fit of all the data together is realized for maximum permutation of fluxes ($\bar p = 0.68$) and:
\beq
&& E_{0 \bar e}= 4.2~{\rm MeV} \hskip 0.7 truecm
 L_{\bar e}  = 4.4 \cdot {\rm 10^{53}~ergs}  \nonumber \\
&& E_{0x}= 14.9  ~{\rm MeV} \hskip 0.7 truecm
 L_{x}=0.8 \cdot {\rm 10^{53}~ergs}  ~,
\label{87abest}
\eeq
where the $\barnue $ parameters are somewhat in tension with the theoretical expectations for the  unusually low average energy and the very large total energy $L_{\bar e}$.  

%
\begin{figure}[htbp]
  \centering
\includegraphics[width=0.9\textwidth]{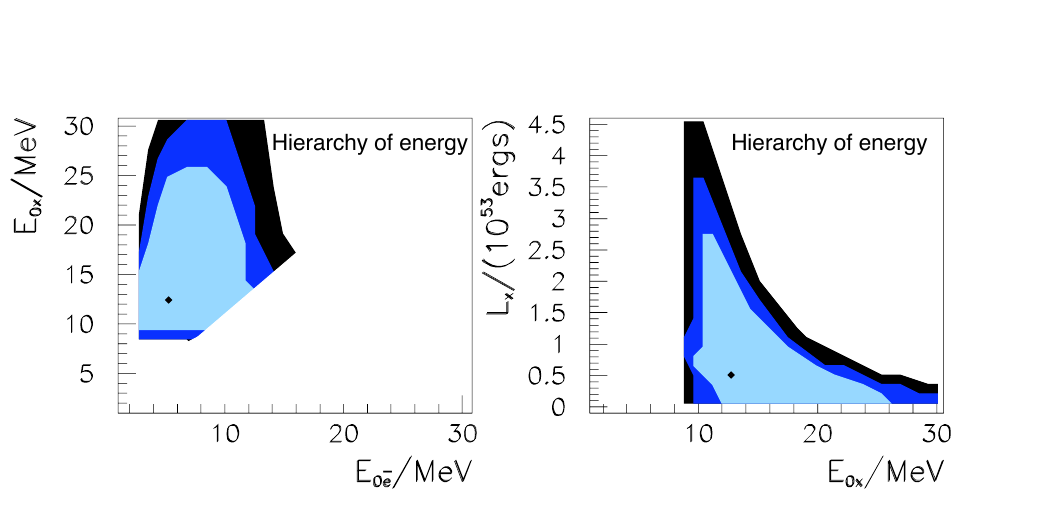}
\caption{From \cite{Lunardini:2005jf}: projections of the 68\%,90\%,99\% C.L. regions allowed by the SN1987A data on the planes  $E_{0 \bar e}-E_{0x}$ and   $E_{0 x}-L_{x}$,  with the hierarchy $E_{0 \bar e}<E_{0 x}$ imposed as a prior. The dots in each panel mark the projections of the points of maximum likelihood.  The entire plane  $E_{0 \bar e}-L_{\bar e}$ (not shown) is allowed at 68\% C.L..  }
\label{1987aplot}
\end{figure}
The region of the parameter space allowed at different confidence levels is large and is shown in fig. \ref{1987aplot}, where the hierarchy  $E_{0\bar e} \leq E_{0x}$ (Eq. (\ref{ehier})) has been imposed.  A comparison between this figure and Table \ref{SNmodeltable} shows that the data are compatible with numerical predictions (only marginally for the H spectrum), but at the same time allow much softer \n\ spectra, which thus remain a possibility on the basis of data alone.  How these softer spectra impact the \df\ is discussed in sec. \ref{diffearth}.

\subsection{Diffuse flux at Earth}

\subsubsection{Generalities}

Armed with the ingredients discussed so far, one can calculate the \df\ expected at Earth, using Eq. (\ref{flux}). 
Here I first discuss a scenario in which a number of approximations and simplifying assumptions hold:

\begin{itemize}

\item all the \ns\ are from \nts\ collapses (Table  \ref{SNmodeltable}) 
 
\item the \n\ emission is identical for all \nts\ progenitors, which is a condition for the validity of Eq.  (\ref{flux}).  The generalization to a mixed population (\nts\ and \bh) will be discussed in sec. \ref{diffailed}. 

\item the total energy emitted in \ns\   is fixed and equipartitioned among the flavors: 
$L_e = L_{\bar e}=L_x=0.5 \cdot 10^{53}$ ergs. 

\item the piecewise form of the \snr, Eq. (\ref{snrpractical}), is used, with $R_{-4} = 1$.  \footnote{Results can be rescaled to reproduce those of other literature, e.g. \cite{Ando:2004hc} ($R_{-4} =0.87$), \cite{Lunardini:2005jf} ($R_{-4} =0.67$, best fit) and \cite{Hopkins:2006bw} ($R_{-4} =1.84$). }

\item the survival probabilities $p$, $\bar p$  are constant in energy, which is valid in several representative cases (sec. \ref{flavconv}). 

\item due to backgrounds, the energy thresholds that are considered realistic for the detection of the \df\ are $E_{th}\simeq 11 - 20$ MeV (see sec. \ref{detection}).These will be the focus of the discussion.  

\end{itemize}

Under these conditions, one can write the $ \nue,\barnue$ component of the \df\  as:
\beq
\Phi_e (E)= p \Phi^0_e(E) + (1 - p )  \Phi^0_x(E)~,\nonumber\\
\Phi_{\bar e} (E)= \bar p \Phi^0_{\bar e}(E) + (1 - \bar p )  \Phi^0_x(E)~,\nonumber\\
\label{practical}
\eeq
where
\be
\Phi^0_w(E) =\frac{c}{H_0}\int_0^{z_{ max}} R_{ SN}(z) F^0_w(E^\prime) \frac{{d}
z}{\sqrt{\Omega_{ m}(1+z)^3+\Omega_\Lambda}}~ ,
\label{phi0}
\ee
 is the component of the \df\  of the species $w$ in absence of oscillations and $E' = E(1+z)$ (sec. \ref{flavconv}).
While 
not directly observable, $ \Phi^0_e,\Phi^0_{\bar e},\Phi^0_x$ are useful to understand the main features of the \df\ and to calculate results for several different \oss\ scenarios. 

For applications, it is interesting to study the integrated fluxes (oscillated and unoscillated) above a threshold $E_{th}$: 
\be
\phi_w (E_{th})= \int^\infty_{E_{th}} \Phi_{w}(E) dE ~, \hskip 0.7truecm\phi^0_w (E_{th})= \int^\infty_{E_{th}} \Phi^0_{w}(E) dE ~.
\label{intflux}
\ee

\subsubsection{Dependence on the original \n\ spectrum}
\label{unoscill}

Here I describe how the \n\ fluxes  at production, $F^0_w$, influence the \df.   For simplicity, I consider the unoscillated fluxes $\Phi^0_w$ and $\phi^0_w$; as their features apply to the oscillated ones through the combination (\ref{practical}).

The integral in Eq. (\ref{phi0}) can not be calculated exactly, and so one can resort to certain approximations  or to numerical calculation. 
The following approximation  is valid at high  energy ($E \gta 15-20$  MeV) \cite{Lunardini:2006pd}:
\beq
\Phi^0_w& \simeq &  { R}_{{ SN}}(0)  \frac{c}{H_0} \frac{ L_w}
  {\Gamma (2+\alpha_w){\epsilon_w}^2}
~e^{-\frac{E}{\epsilon_w}} \sum^{\eta_w}_{k=0} \left[ \left(\frac{E}{\epsilon_w} \right)^{\alpha_w -1-k} \frac{\eta_w !}{(\eta_w -k)!}   \right]~,
\label{crudecompact}
\eeq
where $\Gamma$ stands for the Gamma function, $\eta_w \equiv \alpha_w+ \beta - 3 \Omega_m/2$  ($\beta$ is the power of growth of the \sfr\ and \snr\ with $z$, Eq. (\ref{sfrcurves})) and   $\epsilon=E_{0 w}/(1+\alpha_w)$.  The dependence on the parameters $\alpha$ and $\gamma$ of the \sfr\ is neglected.   The expression (\ref{crudecompact}) exceeds the exact result by up to 40\% above 19 MeV of energy.  It has also been observed  \cite{Malek:2003ki,Lunardini:2006pd} that a simple exponential form, $\Phi^0_w=\Phi^0_w(0) e^{-E/\langle E \rangle_w}$ (with $\langle E \rangle_w \sim \epsilon_w$) is adequate for realistic parameters and for $E \gg \langle E \rangle_w$.
In what follows, exact, numerically calculated results will be discussed. 

\begin{figure}[htbp]
  \centering
\includegraphics[width=0.47\textwidth]{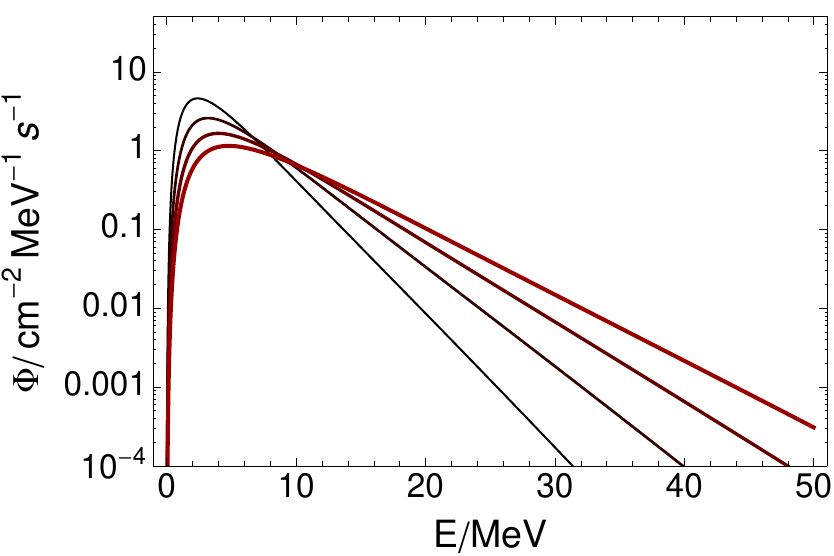}
\includegraphics[width=0.47\textwidth]{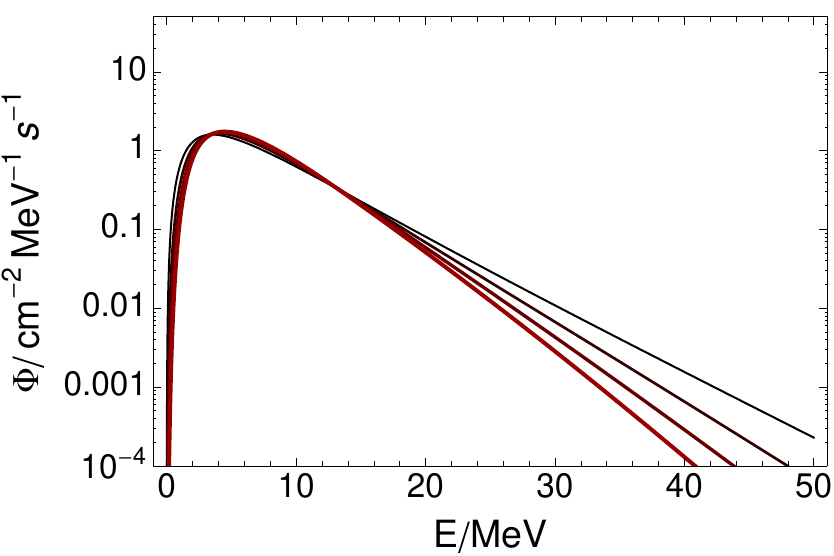}
\caption{Examples of unoscillated flux, $\Phi^0_w$ ($w=e,\bar e, x$) (Eq. (\ref{practical})), for different spectral parameters $E_{0 w}, \alpha_w$.  Left: the curves of increasing thickness (increasing color intensity) correspond to $E_{0 w}=9,12,15,18$ MeV, with $\alpha_w= 3$ .  Right:  the curves of increasing thickness (increasing color intensitiy) correspond to $\alpha_w= 2,3,4,5$ with $E_{0 w}=15$ MeV.  }
\label{dependences}
\end{figure}
Fig. \ref{dependences} shows $\Phi^0_w$ ($w=e,\bar e, x$) for different values of the the parameters $E_{0 w}, \alpha_w$ of the original flux $F^0_w$.  It appears that  $\Phi^0_w$ has a peak value of $\sim 1 - 5$ ${\rm  cm^{-2} s^{-1} MeV^{-1}}$  at 4-7 MeV, with an exponential decay at higher energy.  As expected from Eq. (\ref{crudecompact}), the decay is faster for smaller $\epsilon_w$, corresponding to lower average energy $E_{0 w}$ and/or larger $\alpha_w$.  For $\alpha_w=3$, varying $E_{0 w}$ between 9 and 18 MeV corresponds to a variation of the diffuse flux at 20 MeV by about one order of magnitude, while variations are more modest (a factor of 2 or so) when varying $\alpha_w$ between 2 and 5 with  $E_{0 w}$ fixed.   The dependence on the parameters becomes stronger with increasing \n\ energy.

\begin{figure}[htbp]
  \centering
\includegraphics[width=0.32\textwidth]{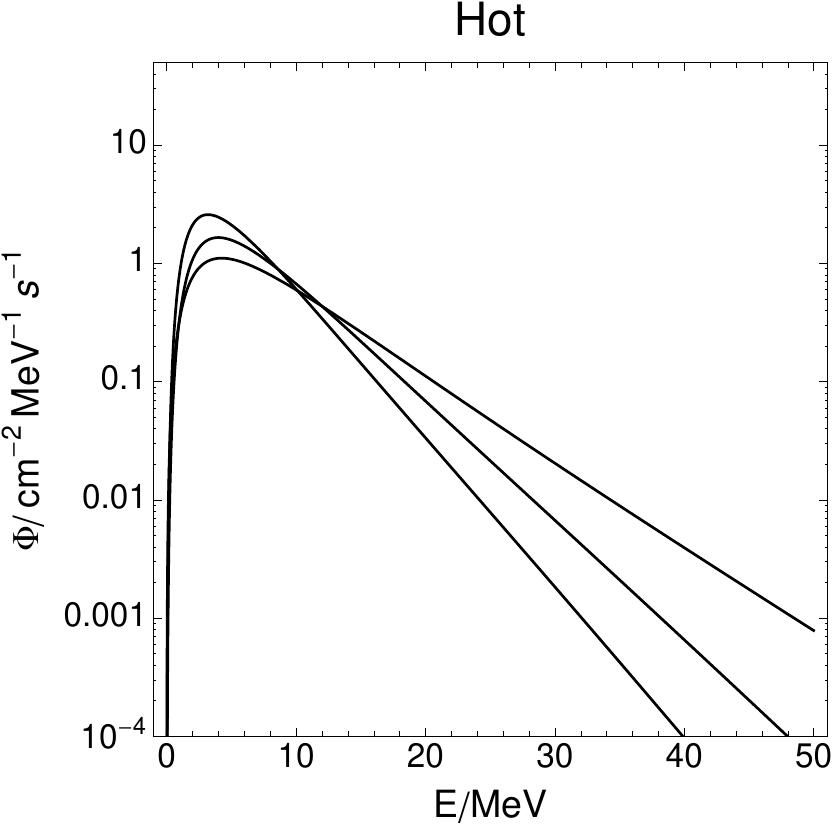}
\includegraphics[width=0.32\textwidth]{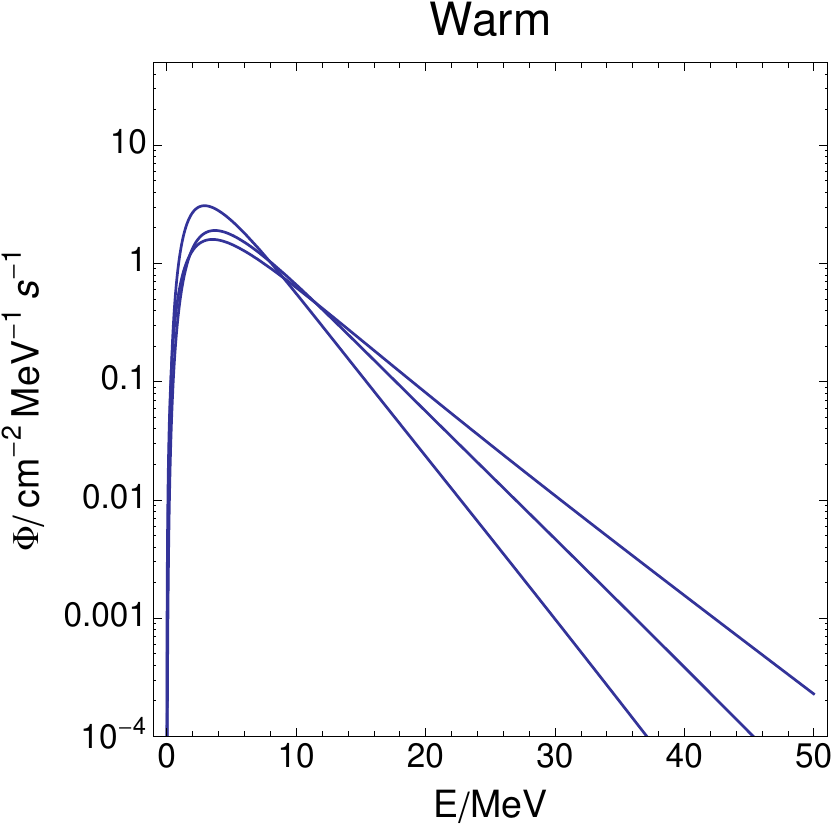}
\includegraphics[width=0.32\textwidth]{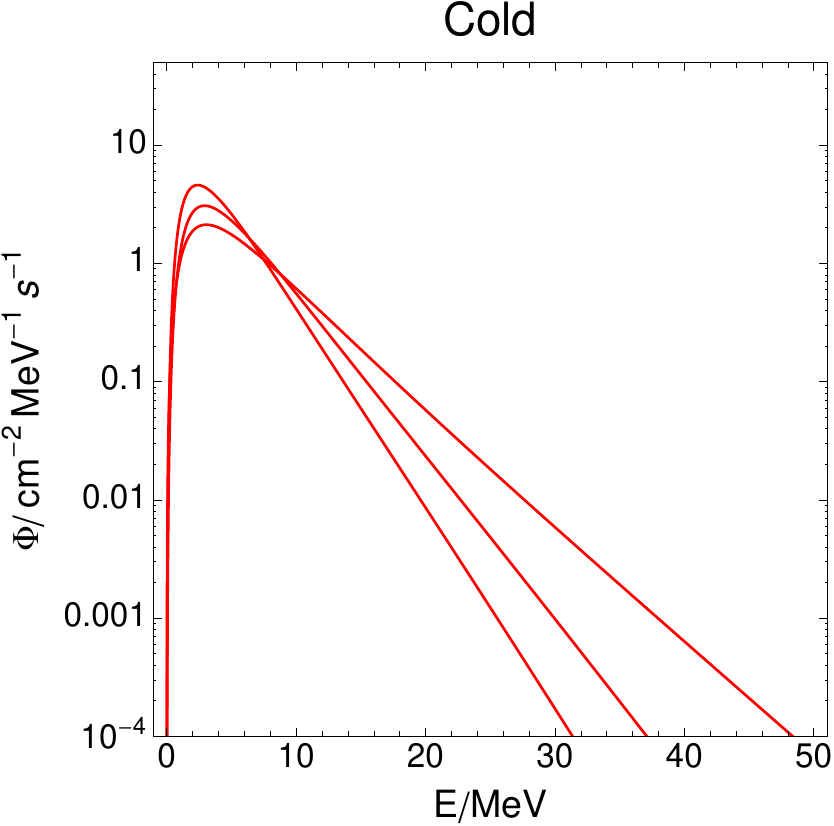}
\caption{The fluxes $ \Phi^0_e,\Phi^0_{\bar e},\Phi^0_x$ , defined in Eq. (\ref{practical}), for the three spectra examples in Table \ref{SNmodeltable}. For each of them, the colder to hotter spectra refer to $\nue,\barnue,\barnux$. }
\label{modeldiffuse}
\end{figure}
For further illustration, the fluxes  $ \Phi^0_e,\Phi^0_{\bar e},\Phi^0_x$ for the H, W and C spectra (Table \ref{SNmodeltable}) are presented in fig. \ref{modeldiffuse}.   Notice that the $\nux$ flux can easily be one order of magnitude larger than the $\nue$ one at 30-40 MeV.  This already gives an idea of how flavor conversion can strongly enhance the potential of detection of the electron flavor components above realistic thresholds (sec. \ref{diffearth}).

\begin{figure}[htbp]
  \centering
\includegraphics[width=0.47\textwidth]{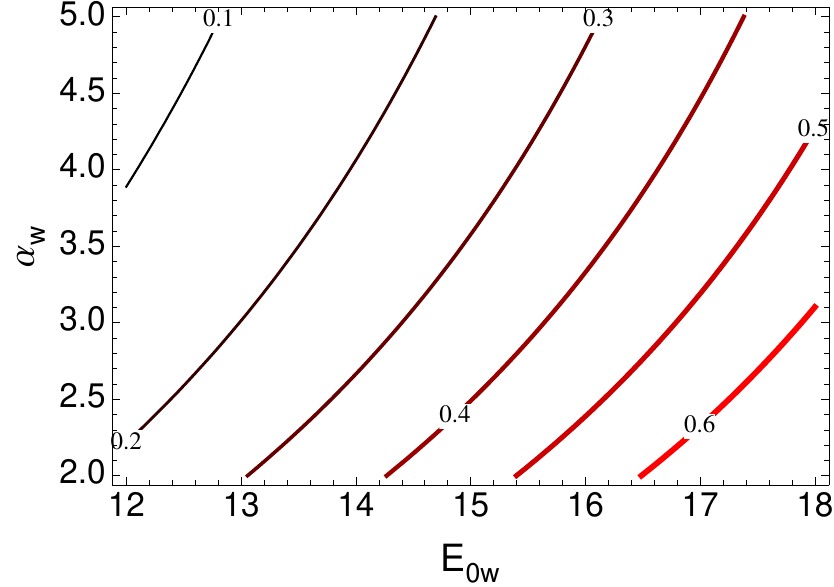}
\includegraphics[width=0.47\textwidth]{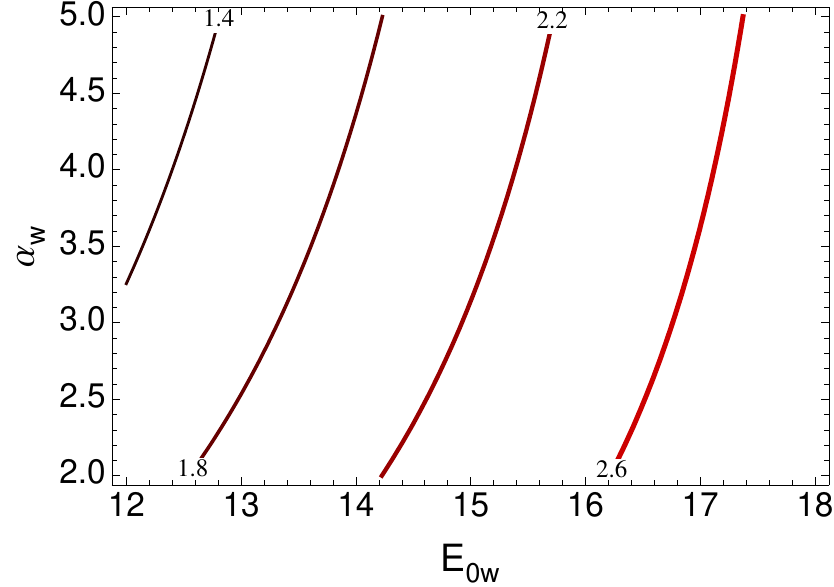}
\caption{Isocontours of the total (unoscillated) flux, $\phi^0_w$,  above the threshold energy $E_{th}$, in ${\rm  cm^{-2} s^{-1} }$, in the $E_{0 w}, \alpha_w$ plane.    Left: $E_{th}=19.3$ MeV.  Right: $E_{th}=11.3$ MeV.   }
\label{contours}
\end{figure}
The characteristics of the integrated flux $\phi^0_w$ reflect those already noted for the energy spectrum: for $E_{th}\gta 10 $ MeV,  $\phi^0_w$ increases with the increasing average energy $E_{0 w}$ and with decreasing values of $\alpha_w$. Fig. \ref{contours} show this for $E_{th}=19.3,11.3$ MeV, that are both relevant for water \ck\ detectors (sec. \ref{detection}). In the figure one notes the dramatic increase (up to one order of magnitude!) of the flux as the energy cut is lowered, thus approaching the peak of the spectrum.  

\begin{table}[htbp]
\newpage
\begin{center}
\begin{tabular}{| c |c |c |c |}
\hline
\hline
  &  \multicolumn{3}{|c|}{$ \phi^0_e,\phi^0_{\bar e},\phi^0_x$} \\
\hline
  & Hot  & Warm  & Cold   \\
\hline
\hline
total &  17.2, 13.8, 11.5  &  18.6, 14.8, 13.7  &   22.4, 18.6, 15.7   \\
\hline
$E>11.3$ MeV   &  1.44, 2.22, 2.90  & 1.17, 1.97, 2.37  &   0.64, 1.17, 1.91   \\
\hline
$E>19.3$ MeV    &  0.14, 0.35, 0.74   &   0.09, 0.27, 0.46  &  0.03, 0.09, 0.30  \\
\hline
\hline
\end{tabular}
\caption{The unoscillated fluxes at Earth integrated above energy thresholds of interest, in  ${\rm cm^{-2}~s^{-1}}$, for the  three cases in Table \ref{SNmodeltable}.  } 
\label{phi0table}
\end{center}
\end{table}
This is further illustrated in Table \ref{phi0table}, where the values of $\phi^0_e, \phi^0_{\bar e},\phi^0_x$ are given for the H, W and C spectra. The Table  evidences the large differences between the fluxes in the different flavors and in the different cases.    Among the total fluxes, integrated over all energies, the $\nue$ flux is larger: indeed, it has the same total energy than the other flavors, but each $\nue$ carries on average a lower energy.    When the integration is restricted to increasingly high energy, the $\nux$ unoscillated flux starts to dominate, being higher than the $\nue$ flux by a factor of 3-10.   For the H example, which has the most energetic $\nux$, $\phi^0_x$ is close (within a factor $\sim3-4$) to the current \sk\ limit, Eq. (\ref{sklim}), thus confirming that experiments are already probing the interesting region of the parameters. 
For example, with a \n\ spectrum equally or more energetic than the H one, one can constrain the \snr\ normalization \cite{Strigari:2005hu}.

\subsubsection{Dependence on the core collapse rate}
\label{ccdep}

\begin{figure}[htb]
      \includegraphics[width=0.7\textwidth]{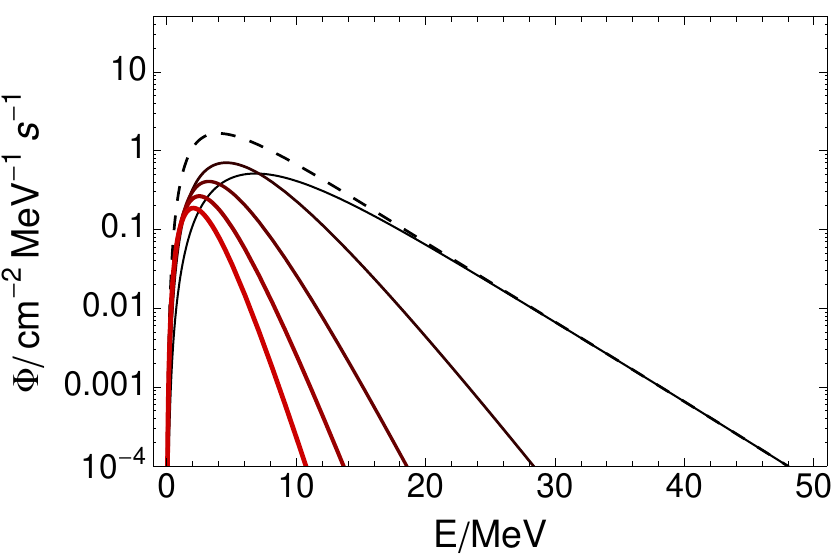}
    \caption{The contribution to the {\it unoscillated}  $\barnue$ flux of sources in bins of increasing redshift, for the best fit \snr\ parameter $\beta=3.28$ \cite{Hopkins:2006bw}.  The solid curves from thinner to thicker (darker to lighter  color) refer to the intervals: $z=0 - 1$, $z=1-2$, $z=2-3$, $z=3-4$ and $z=4-5$.  The dashed line is the total flux integrated over all redshifts.  The parameters of the H  case were used (Table \ref{SNmodeltable}).}
    \label{zdependence}
\end{figure}

It is interesting to study what fraction of the \df\ is due to sources at cosmological distances, i.e., $z \gta 1$. 
 Fig. \ref{zdependence}  addresses this question, showing the contributions to the flux of sources in bins of the form $\left[ z, z+1 \right]$, for the unoscillated $\barnue$ flux obtained with the H spectrum and the \snr\ in Eq. (\ref{snrpractical}).   It appears that the flux above $\sim 20$ MeV is practically all due to \sne\ at $z<1$, while the contribution of more distant sources becomes increasingly important at decreasing energy: sources at $z\sim 2$ should be included to reproduce the flux at 10 MeV, and at 2 MeV the dominant contribution is from sources at $z>2$.  This feature is explained with the larger redshift of the energy of \ns\ emitted at larger distances.  If we consider the total flux of \ns\ of all energies, the contributions of the first three redshift bins are about $\sim 40\%, 35\%,14\%$, while the remaining bins contribute for less than $\sim 10\%$.   See \cite{Ando:2004hc} for further details. 

\begin{figure}[htb]
      \includegraphics[width=0.7\textwidth]{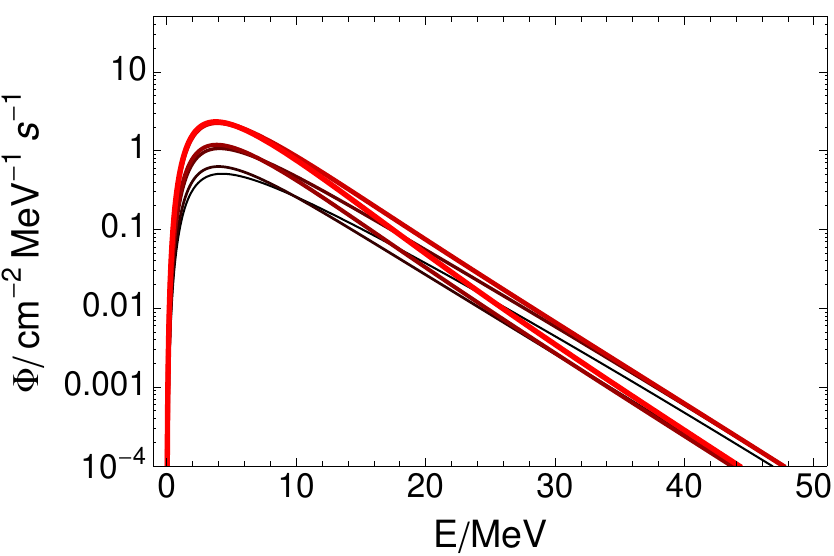}
    \caption{ 
    The  {\it unoscillated}  $\barnue$ flux calculated for different points in the largest allowed region of the parameters $\beta, R_{SN}(0)$ (fig. \ref{snrcontours}, outer dashed contour).  The curves from thinner to thicker (darker to lighter  color) refer to the points:  
  $(\beta, R_{SN}(0))= (1.6, 0.9),
(3.2, 0.4),
(4.6, 0.3),
(2.6, 1),
(5.6, 0.3), (4, 0.85)$, 
    where $R_{SN}(0)$ is in units of $10^{-4}~{\rm yr^{-1}~Mpc^{-3}}$. 
     The parameters of the H  spectrum were used (Table \ref{SNmodeltable}). 
     }
    \label{snrdependence}
\end{figure}
How does the \df\ depend on the \snr\ parameters, $\beta, R_{SN}(0)$\footnote{The dependence on $\alpha$ and $\gamma$ in Eq. (\ref{sfrcurves}) is weak, and negligible in first approximation.}?  This question is especially relevant if one considers the region allowed by \sn\ observations only (fig. \ref{snrcontours}), which is wider than that allowed by measurements of the star formation rate and exhibits a correlation between the normalization, $R_{SN}(0)$ and the power $\beta$.  Fig. \ref{snrdependence} shows the  unoscillated $\barnue$ flux obtained with the H spectrum and with a set of point in the space $(\beta, R_{SN}(0))$ that roughly map the 
95.4\% C.L. region in fig. \ref{snrcontours} (see caption of fig. \ref{snrdependence}  for details).   One can see that larger $\beta$ (faster growth of the \snr\ with $z$) corresponds to less energetic spectrum, reflecting the increased  contribution to the flux of the more redshifted \ns\ from cosmological sources. 
At certain points of the energy spectrum, the flux can vary by up to a factor of 5 as an effect of the variation of the \snr\ parameters, however the variation 
does not exceed a factor of 2
for the integrated fluxes above thresholds on interest.

\subsubsection{The $\barnue$ and $\nue$ fluxes in a detector}
\label{diffearth}

\begin{figure}[htbp]
  \centering
\includegraphics[width=0.45\textwidth]{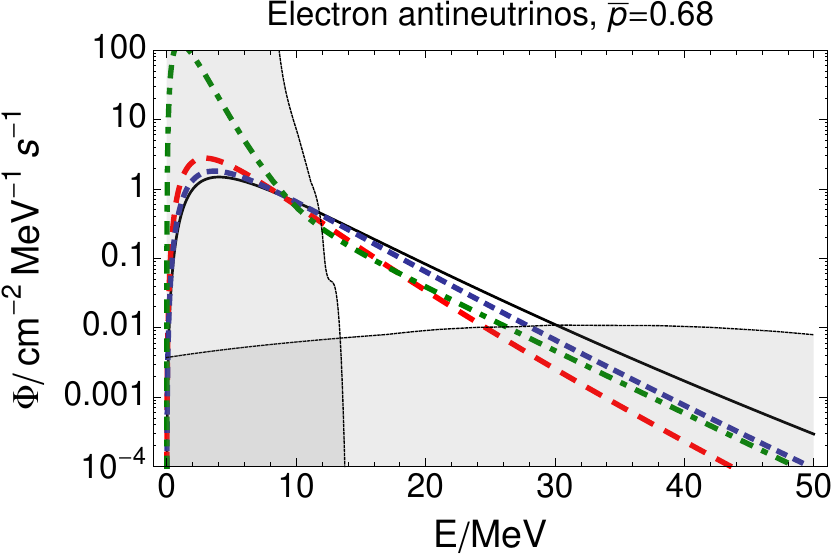}
\includegraphics[width=0.45\textwidth]{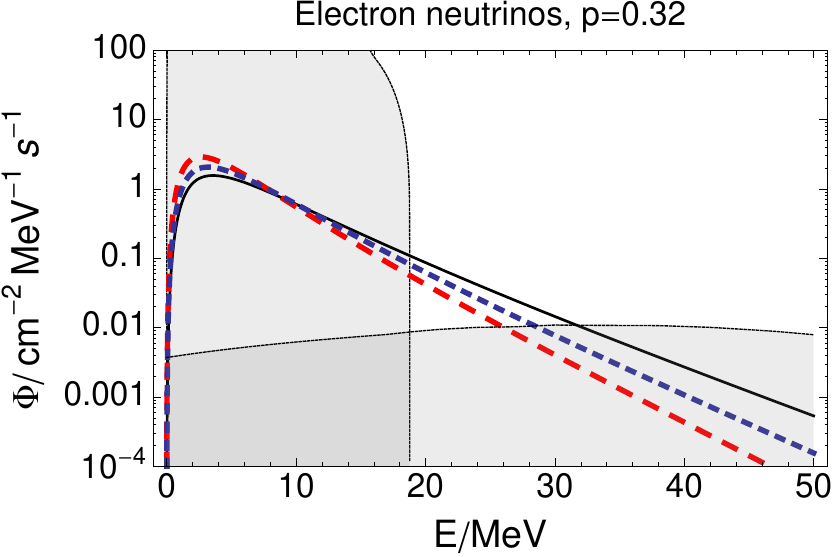}
\includegraphics[width=0.45\textwidth]{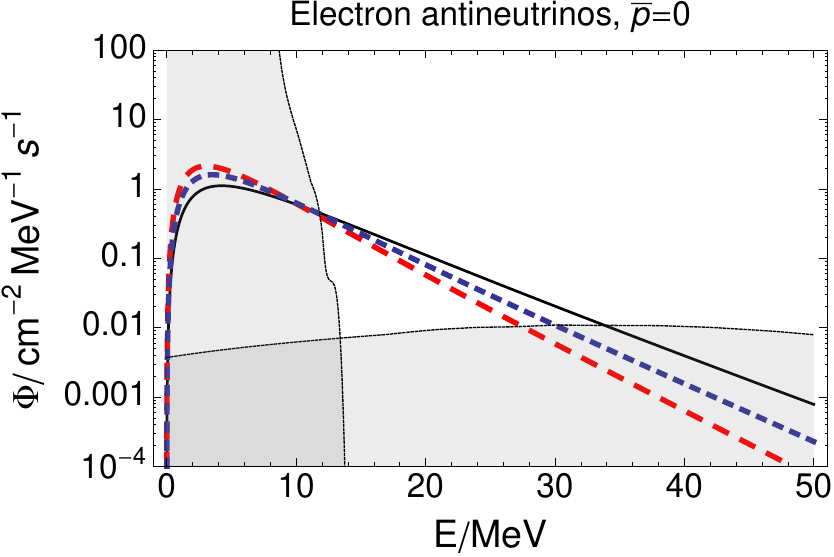}
\includegraphics[width=0.45\textwidth]{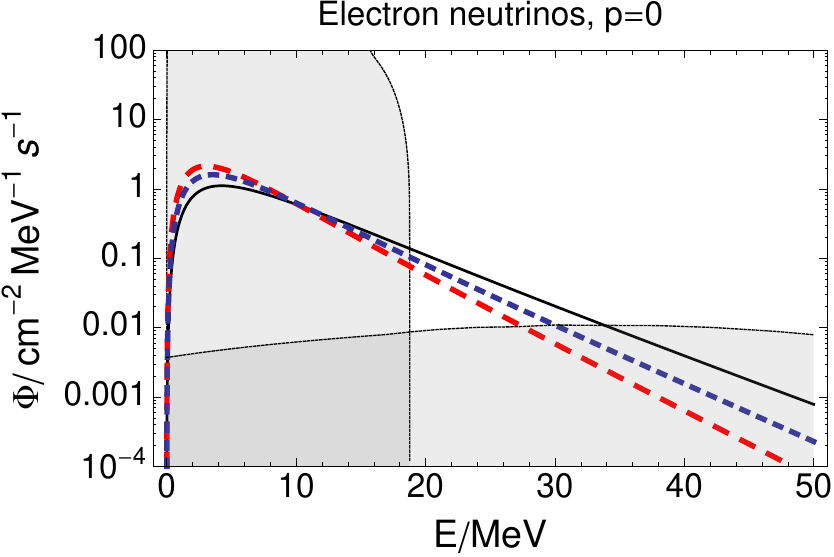}
\caption{Examples of energy spectra of the $\nue$ and $\barnue$ components of the
\df\ from the literature for the two extreme values of $p$ and $\bar p$.  The solid, short dashed and long dashed curves are for the 
Hot (H),  Warm (W) and Cold (C)  spectra (Table \ref{SNmodeltable}). The dashed-dotted line corresponds to the best fit of a multi-parameter statistical analysis of SN1987A data   \cite{Lunardini:2005jf} (Eq. \ref{87abest}).  All the curves were obtained for the piecewise \snr\ function (Eq. \ref{snrpractical}) with   $R_{-4}=1$.  The four panels also show backgrounds due to other \n\ sources (see sec. \ref{enwindow}): $\barnue$s from the atmosphere \cite{Battistoni:2005pd} and from reactors (below $\sim 11$ MeV, inclusive of oscillation effects, from \cite{Wurm:2007cy}), and $\nue$s from the atmosphere \cite{Battistoni:2005pd} and from the Sun (below $\sim$ 19 MeV) \cite{Bahcall:2004pz}.  All these background fluxes are for the Kamioka site. }
\label{spectradiff}
\end{figure}

\begin{figure}[htbp]
  \centering
\includegraphics[width=0.4\textwidth]{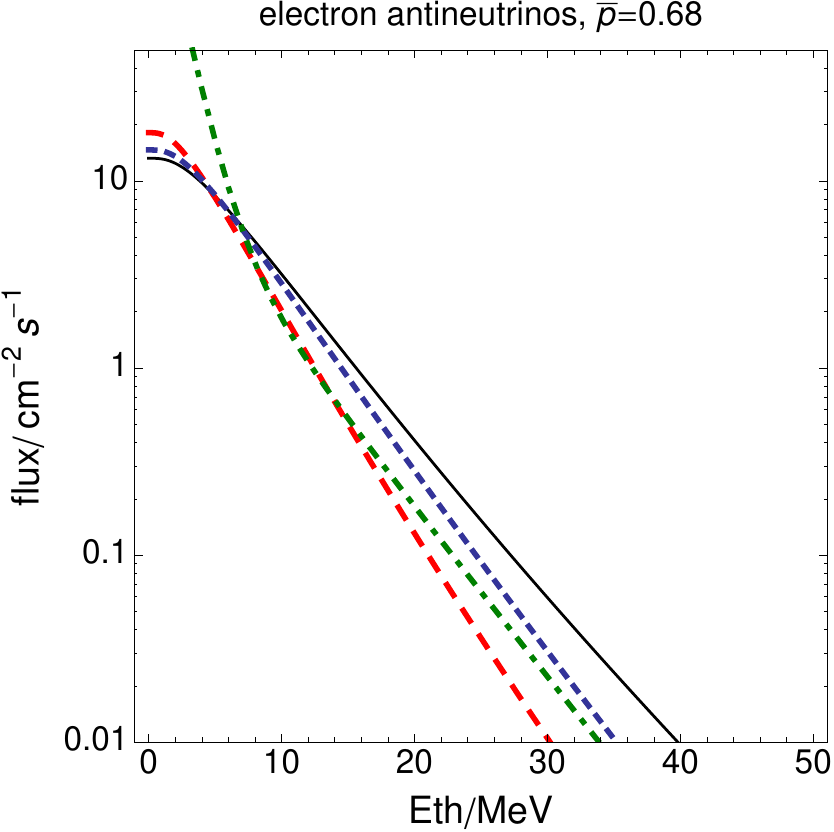}
\includegraphics[width=0.4\textwidth]{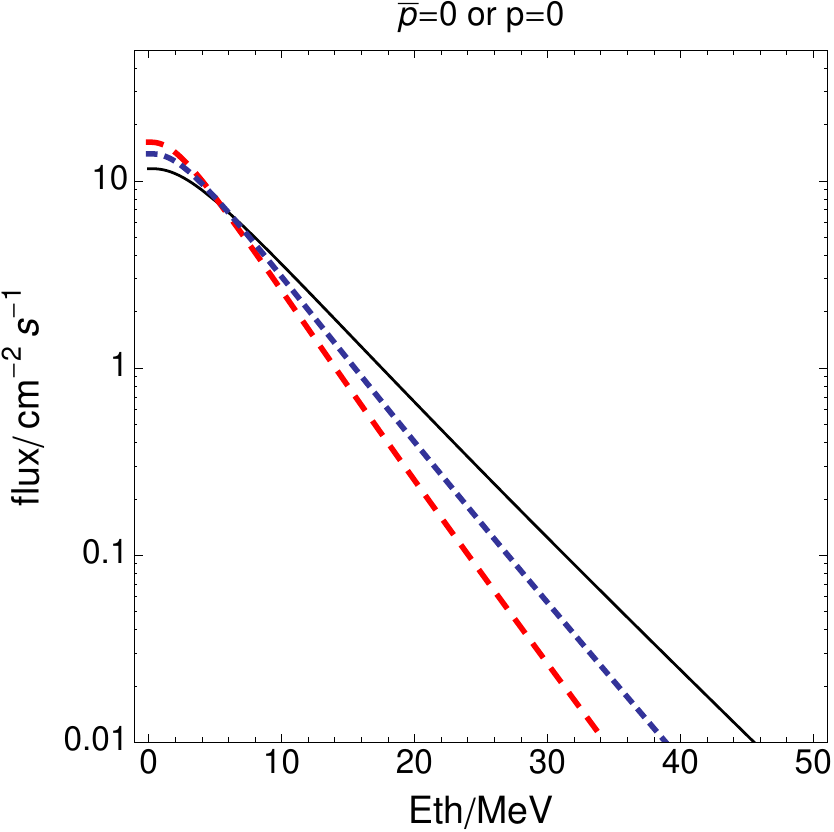}
\includegraphics[width=0.4\textwidth]{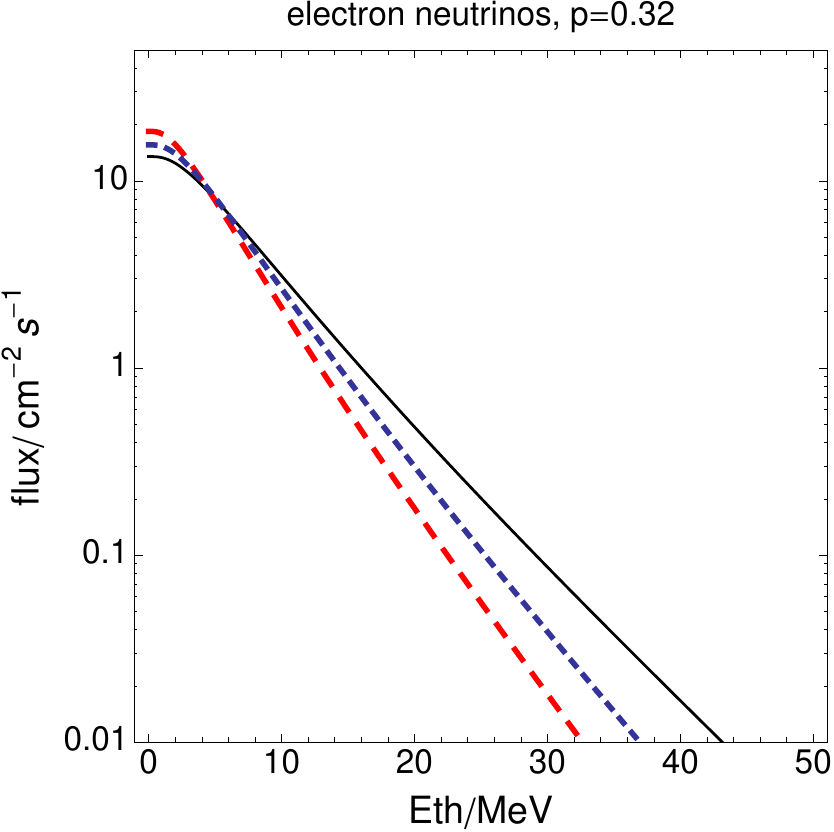}
\caption{  Diffuse $\nue$ and $\barnue$ fluxes integrated above a threshold energy $E_{th}$, as a function of $E_{th}$ for selected values of $p, \bar p$.  The middle panel refers to the case in which the $\barnue$ and $\nue$ fluxes both undergo complete flavor permutation into $\nux$, and therefore are equal.    The solid, short dashed and long dashed curves are for the 
Hot (H),  Warm (W) and Cold (C)  spectra  (Table \ref{SNmodeltable}).  The dashed-dotted line corresponds to the best fit of a multi-parameter statistical analysis of SN1987A data   \cite{Lunardini:2005jf} (Eq. (\ref{87abest})).  All other parameters are as in fig. \ref{spectradiff}.}
\label{diffint}
\end{figure}

When oscillation effects are included, the $\barnue$ and $\nue$ components of the \df\ receive a contribution from the original $\nux$ flux produced inside the star, eq. (\ref{practical}).  The oscillated $\nue$  and $\barnue$ fluxes are described  in figs. \ref{spectradiff} and \ref{diffint} and in Tables \ref{SNosctable} and \ref{SNosctablenue} for the H, W and C spectra, as well as for the flux that best fits the SN1987A data (Eq. (\ref{87abest})). 
All results refer to the extreme values of the survival probabilities $p, \bar p$, as they illustrate the maximum range of variation of the \df\ with the varying oscillations parameters.  Intermediate cases can be calculated from the unoscillated fluxes in the different flavors (sec. \ref{unoscill}, Table \ref{phi0table}). 

From the energy spectra  in fig. \ref{spectradiff} one can see that large variations (up to a factor of 3 or so) in the $\barnue$ flux can be expected for each model above 19.3 MeV as a consequence of the wide range of variation of $\bar p$.  The $\nue$ flux instead is always within 30-40\% from the unoscillated flux $\Phi^0_x$ as a consequence of the small component due to $\Phi^0_{e}$, $ p \leq 0.32$. 
Thanks to the larger contribution of the $\nux$ original flux,  the $\nue$ spectrum is always more energetic than the $\barnue$ one.   The $\nue$ and $\barnue$ diffuse fluxes can be equal in the limiting case $p=\bar p=0$. 
This is realized
for inverted (normal) hierarchy if collective oscillations are maximally effective for $\barnue$ ($\nue$) and completely suppressed for $\nue$ ($\barnue$) (see Table \ref{probtable}). 

Large variations in the flux are also seen for fixed $p,\bar p$ and varying spectral model, as already commented about the unoscillated fluxes.  Expectedly, at high energy the H spectrum gives a flux that is much larger (up to one order of magnitude at 50  MeV) that the other spectra. For the SN1987A best fit spectrum the \df\ has a high peak at 2 MeV, reflecting the very high $\barnue$ original flux, and exceeding the other predictions by two orders of magnitude.  For the energies of interest here, however, the flux is intermediate between the W and C spectra cases.  

\begin{table}[htb]
\begin{center}
\begin{tabular}{| c |c |c |c |c |}
\hline
\hline
  &  Hot  & Warm  & Cold  & SN1987A best fit (Eq. (\ref{87abest}))  \\
\hline
\hline
total &   13.2 & 14.6  & 18.1  & 305  \\
\hline
$E>11.3$ MeV   &    2.43 &  2.10 & 1.41  & 1.3  \\
\hline
$E>19.3$ MeV    &   0.47  &  0.33  & 0.16  &  0.21 \\
\hline
\hline
\end{tabular}
\caption{  Integrated $\barnue$ flux above thresholds of interest, in  ${\rm cm^{-2}~s^{-1}}$, for $\bar p=0.68$. All parameters are as in fig. \ref{spectradiff}.  
The results for the SN1987A best fit flux are from \cite{Lunardini:2005jf}, and are in agreement with those of \cite{Vissani:2011kx}, where a different method of analysis is used. 
} 
\label{SNosctable}
\end{center}
\end{table}
%
\begin{table}[htb]
\begin{center}
\begin{tabular}{| c |c |c |c |}
\hline
\hline
  &  Hot  & Warm  & Cold    \\
\hline
\hline
total &  13.31 &   15.3  & 17.8    \\
\hline
$E>11.3$ MeV   & 2.43 &    1.98  & 1.51   \\
\hline
$E>19.3$ MeV  &  0.55  &  0.35  &  0.21    \\
\hline
\hline
\end{tabular}
\caption{  Integrated $\nue$ flux above thresholds of interest, in  ${\rm cm^{-2}~s^{-1}}$, for $ p=0.32$. All parameters are as in fig. \ref{spectradiff}.} 
\label{SNosctablenue}
\end{center}
\end{table}

The integrated fluxes are given in  fig. \ref{diffint} and in Tables \ref{SNosctable} and \ref{SNosctablenue} for completeness. They reflect the features already noted for the unoscillated fluxes: the fast (exponential) decay of the flux with the increase of threshold energy $E_{th}$, and the variation by a factor of $\sim 2$ of the flux at high energy depending on the spectrum adopted.  Again, the  $\nue$ flux is generally larger above realistic thresholds due to the larger flavor permutation.

\subsubsection{Diffuse flux from failed \sne}
\label{diffailed}

Analogously to what done so far for \ns\ from \nts\ collapses, one can calculate the diffuse flux of \ns\ from \bh\ ones, $\Phi_{BH}$ \cite{Lunardini:2009ya,Lien:2010yb,Keehn:2010pn,Yuksel:2012zy,Nakazato:2013maa,Mathews:2014qba}, using the spectra in fig. \ref{BHspectra}  and considering that \nts\ (\bh )  collapses amount to a fraction    $f_{NS}= 0.78 - 0.91$ ($1- f_{NS}=0.09 -0.22$)  of the total (sec. \ref{snspectra}).  

\begin{figure}[htbp]
  \centering
 \includegraphics[width=0.45\textwidth]{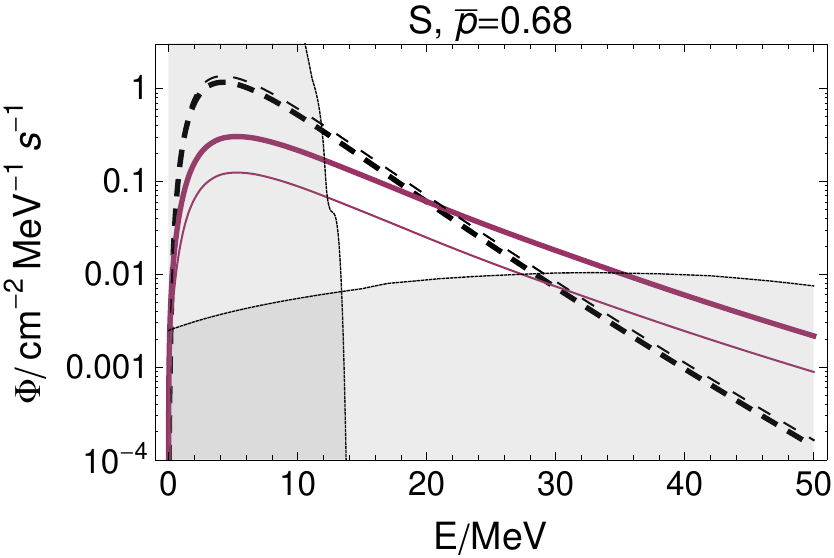}
 \includegraphics[width=0.45\textwidth]{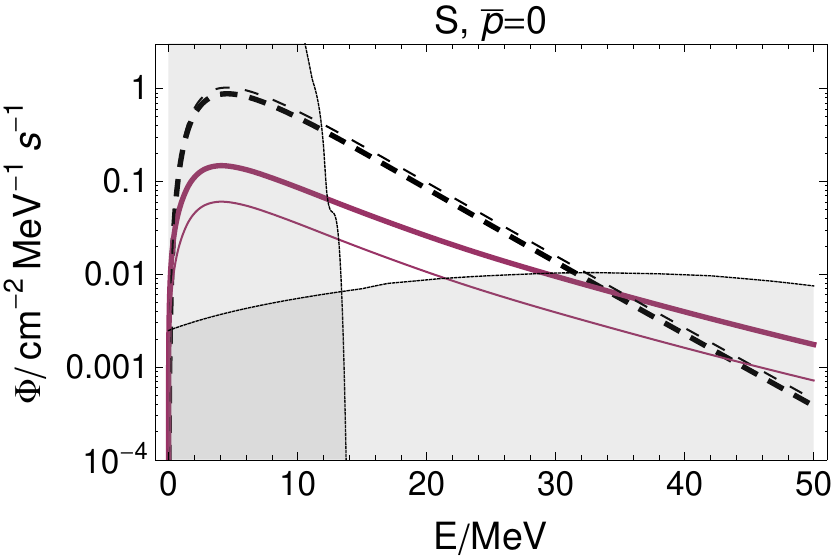}
 \includegraphics[width=0.45\textwidth]{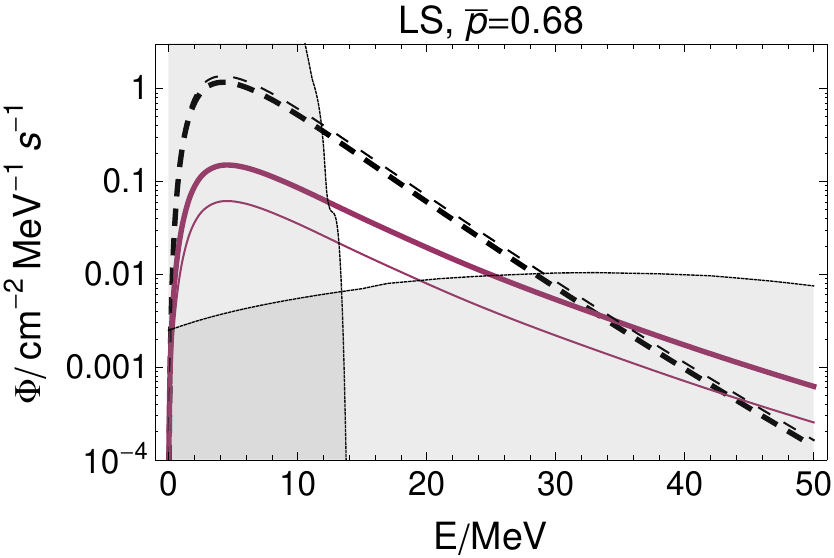}
 \includegraphics[width=0.45\textwidth]{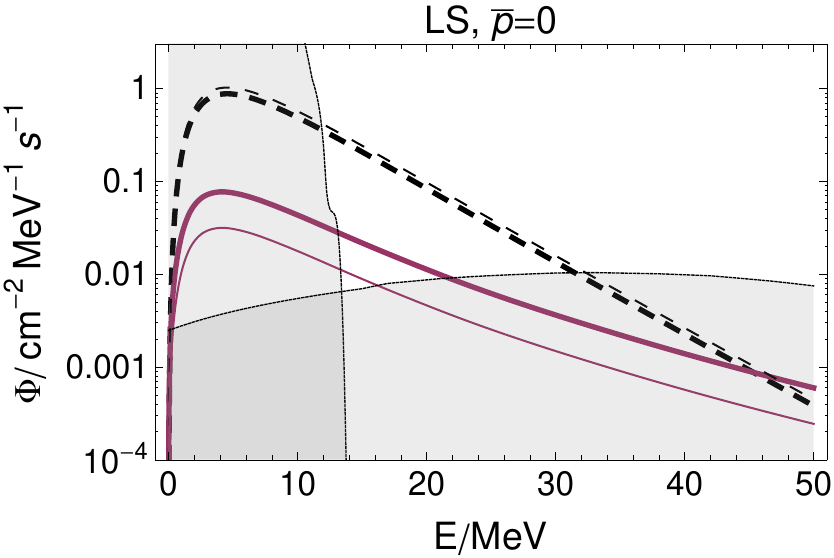}
   \caption{From \cite{Lunardini:2009ya,Keehn:2010pn}: the diffuse \f\ of $\barnue$ at Earth from direct \bh\ collapses, i.e. failed \sne\ (solid lines), for the Shen at al. (S) and Lattimer-Swesty (LS) equation of state 
  and different values of the survival probability $\bar p$.  The \f\ from \nts\
   collapses  is also plotted (dashed curves).
    Direct black hole -forming collapses
   are assumed to be 22\% (thick curves) or 9\% (thin curves) of the total.  The $\barnue$ atmospheric and reactor backgrounds are shown for comparison.  }
\label{spectradiffBH}
\end{figure}
Results are given in fig. \ref{spectradiffBH} 
 for the $\barnue$ 
 component of $\Phi_{BH}$ (similar conclusions hold for the $\nue$ component, see e.g., \cite{Keehn:2010pn})  
 and the extreme values of $f_{NS}$ and of $\bar p$, which is assumed to be the same for both types of collapses.  
    For comparison, examples of diffuse flux from \nts\ collapses are shown; they were calculated using the parameters  $E_{0 \bar e} = 15$ MeV, $E_{0
x} = 18$ MeV, $L_{\bar e}=L_x=5 \cdot 10^{52}$ ergs, $\alpha_{\bar
e}=3.5$ and $\alpha_x=2.5$, that is similar to the H spectrum case.
As expected,  $\Phi_{BH}$  has hotter
spectrum compared to the flux from \nts\ collapses, and thus it is increasingly important at higher
energy. 
Oppositely to the case of \ns\ from \nts\ collapses,
$\Phi_{BH}$ is larger for  minimal permutation ($\bar p=0.68$)
\cite{Nakazato:2008vj}, because of the especially luminous original fluxes in the electron flavor (fig. \ref{BHspectra}).
The dependence of the original fluxes on the EoS is evident in $\Phi_{BH}$.

Fig. \ref{spectradiffBH} evidences that $\Phi_{BH}$ might
dominate already at $E \sim 22$ MeV, implying a strong effect at
SuperKamiokande. For the most favorable parameters 
the total flux from both types of collapses
above 19.3 MeV is more than twice as large as the case of 
100\% \nts\ collapses. It reaches the value
$\phi_{BH}\simeq 0.89 ~{\rm cm^{-2} s^{-1}}$, tantalizingly close to the
current upper limit. 
It is more likely, however, that $\Phi_{BH}$ becomes dominant only
above 30-40 MeV, as the figure shows. If so, its effect would be below
the sensitivity of \sk\ -- which would therefore place limits on \ns\ from failed \sne\ \cite{Lien:2010yb}
 -- but might be visible at the more massive and more sensitive detectors of the next generation.

\section{Detection: the diffuse \n\ flux at \n\ \snew{observatories}}
\label{detection}

\subsection{The energy window }
\label{enwindow}

\begin{figure}[htbp]
 \centering
 \includegraphics[width=0.7\textwidth]{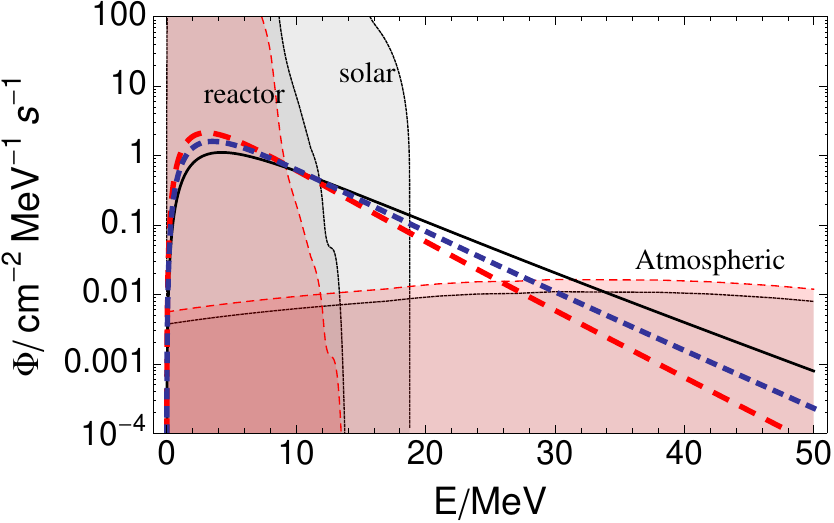}
\caption{   Backgrounds fluxes: $\barnue$s from reactors (taken from \cite{Wurm:2007cy}) and from the atmosphere \cite{Battistoni:2005pd}, and $\nue$ from the Sun \cite{Bahcall:2004pz} and from the atmosphere \cite{Battistoni:2005pd}.    These backgrounds are compared with the signal from the \df\ with $p=\bar p=0$ for different \n\ spectra, as in fig. \ref{spectradiff}.  The atmospheric and reactor fluxes are shown for the Kamioka (solid, gray) and Homestake (dashed, red) sites.  The atmospheric fluxes of $\nue$ and $\barnue$ are very similar, so only one of them is plotted.  The calculations of the background fluxes include oscillation effects, which are responsible for the visible modulation of the reactor spectrum. }
\label{allbackgrounds}
\end{figure}

A fortunate circumstance makes the \df\ detectable: the fact that part of it falls in a relatively quiet region of the \n\ spectrum, and precisely above the typical energies of \ns\ from nuclear processes  ($E_{nuc} \lta 18$ MeV) 
and below 
the bulk of the flux
 of \ns\ from cosmic rays (atmospheric \ns, $E_{atm} \gta 10^2$ MeV).
These \n\ fluxes are for the most part  ineliminable backgrounds, and thus place a natural limit to the sensitivity of any experiment to the \df.   Here I discuss the generalities of the $\nue$ and $\barnue$ components of these backgrounds, which dominate the relevant detection channels.  They are shown in fig. \ref{allbackgrounds}. 
Detector-specific backgrounds are discussed in  secs.  \ref{waterc}-\ref{lardetector}. 

\begin{table}[htdp]
\begin{center}
\begin{tabular}{|c|c|}
\hline
\hline
Detector location & energy window (MeV) \\ 
\hline
\hline
Kamioka (J) & 11.1 - 28.1 \\
Frejus (F)  & 10.8 - 26.4 \\
Kimballton (US) & 10.6 - 28.1 \\
Pyh¬asalmi (FIN) & 9.7 - 25.1 \\
Pylos (GR)&  9.4 - 28.1 \\
Homestake (US) & 9.0 - 26.4 \\
Henderson (US)&  8.9 - 27.2 \\
Hawaii (US) & 8.4 - 29.0 \\
Wellington (NZ) & 8.2 - 27.2 \\
\hline
\hline
\end{tabular}
\caption{\label{windows}  The location-dependent energy window for the detection of the $\barnue$ component of the \df. 
This window is defined (optimistically) 
as the interval where the signal exceeds the background fluxes of atmospheric and reactor \ns.    The table is adapted from \cite{Wurm:2007cy}, where the parameters used for the \df\ are $R_{-4}=0.87$, $E_{0\bar e}=15.4$ MeV,  $E_{0x}=15.7$ MeV, $\alpha_{\bar e}=4.2$, $\alpha_{x}=2.5$ 
\snew{
and $\bar p= 0.71$.
}
 }
\end{center}
\end{table}%

\begin{itemize}

\item {\it solar neutrinos.}  Electron, muon and tau \ns\ from the sun (with $\numu,\nutau$ generated by oscillations of originally produced $\nue$s) extend to $\sim 16$ MeV energy if they come from the $^8B$ chain, and even to $\sim 19$ MeV if produced through the $hep$ process.  Reaching $\sim 10^2 ~{\rm  cm^{-2} s^{-1} MeV^{-1}}$, the $hep$ flux dominates over the \df\ in the whole energy range nearly up to its endpoint.   Its $\nue$ component is shown in fig. \ref{allbackgrounds}, where the effect of  oscillations in the Sun has been included using the best fit oscillation parameters in  Eq. (\ref{values}).  

\item {\it reactor antineutrinos.}  Nuclear powerplants are a copious source of electron antineutrinos of energies up to $\sim 14 $ MeV. This reactor flux acquires a muon and tau component on the way to a detector due to oscillations \cite{Araki:2004mb}.    Very important as a signal for oscillation tests, the reactor antineutrino flux is a serious obstacle to the study of diffuse antineutrinos from \sne.  For example, at the Kamioka site the reactor flux is as high as $\sim 10^2 ~{\rm  cm^{-2} s^{-1} MeV^{-1}}$ at 8 MeV, and dominates over the \df\ below $\sim 12$ MeV (fig. \ref{allbackgrounds}).  The lower reactor flux in less nuclearized areas   would allow a slightly larger energy window for the \df, down to 9 MeV for Homestake (where the reactor flux is lower by a factor of $\sim 30$ compared to Kamioka)  and even to 8.2 MeV for a detector in New Zealand (table \ref{windows}).  Even in the best case, however, the reactor flux obscures the peak of the \df. 

\item {\it geoneutrinos.}  Even in the ideal scenario of complete absence of reactors, there is still another low energy background:  a flux of $\barnue$ from the natural radioactivity of the Earth. These geoneutrinos have been seen by KamLAND \cite{Araki:2005qa} and Borexino \cite{Bellini:2010hy}, and -- according to calculations \cite{Krauss:1983zn} --  they should dominate over the \df\ below $3.26$ MeV, where the spectrum of $\barnue$ from $^{214}Bi$ ends.  This energy is sufficiently low to expose the peak of the \df\ at 5-7 MeV;  
 however the possibility to reach such lower threshold is highly speculative,
 and is presented here only as motivation for minimizing the reactor flux. 

\item {\it atmospheric neutrinos.}   Collisions of cosmic rays with the Earth's atmosphere generate a shower of $\nue,\barnue,\numu,\barnumu$ (and $\nutau,\barnutau$ via oscillations) that appears in \n\ detectors primarily in the energy window $0.1 - 10^3$ GeV.  Still, a low energy tail of this atmospheric flux extends down to few MeV. The  spectrum of $\nue$s and $\barnue$s in this tail \cite{Battistoni:2005pd} is shown in fig. \ref{allbackgrounds}.  Oscillation effects have not been taken into account, since they are at the level of 5-10\%   at these energies \cite{Peres:2009xe}.    The figure shows that typically the atmospheric flux exceeds the \df\ above $\sim 30$ MeV or so depending on the intensity and spectrum of the \df. It should also be noticed that the atmospheric \n\ flux is location-dependent due to the dependence of the cosmic ray flux on the latitude, and increases as one moves towards the magnetic poles of our planet.  For example, the atmospheric flux is about $\sim 1.5$ times higher at Homestake than at Kamioka  \cite{Wurm:2007cy}, resulting in a smaller energy window for the \df.   This appears in fig. \ref{allbackgrounds} and in Table \ref{windows} \cite{Wurm:2007cy}, which  gives the energy windows for $\barnue$ detection in different locations. 

\end{itemize}

It is interesting to consider whether the  fluxes discussed here are truly ineliminable backgrounds or can in principle be distinguished from the \df. While the answer ultimately depends on  technology, 
here I observe that only the atmospheric neutrino flux is truly similar to the \df: it shares the same energy window, has similar flavor composition, and  has the same isotropic distribution in direction.  The other fluxes are not isotropic, and therefore can be distinguished {\it in principle}, while most likely not in practice. 

Besides the \n\ fluxes discussed here, one should consider a number of non-neutrino processes and various instrumental effects that also constitute background for a given detector and that generally restrict the energy window compared to what discussed here.   These backgrounds are described in sec. \ref{concepts}.

\subsection{Detection concepts}
\label{concepts}

\begin{table}[htdp]
\newpage
\begin{center}
\begin{tabular}{|c|c|c|c|c|c|}
\hline
\hline
Concept &  
 \begin{minipage}[h]{1.9truecm}
energy \\ window \\ (MeV)
 \end{minipage} & 
  \begin{minipage}[h]{2.5truecm}
detection \\ processes
 \end{minipage} &  
 \begin{minipage}[h]{1.9truecm}
experiment 
 \end{minipage} & 
  \begin{minipage}[h]{1.9truecm}
fiducial mass (kt)
 \end{minipage}&
   \begin{minipage}[h]{1.9truecm}
events\\
per year
 \end{minipage} 
  \\
\hline
\hline
 \begin{minipage}[h]{1.9truecm}
$H_2O$
 \end{minipage} &
       19.3 - 30  &
${\bf \barnue (p,n) e^+ }$  &
      {  \sk\  }   \cite{Bays:2011si}  &
            22.5  & 0.23 - 1.0  \\  \cline{4-6}
     &  [17.3 - 30]   &  $ \barnue (^{16}O,X ) e^+ $  &   Hyper-K  \cite{Abe:2011ts}  & 560   & 5.7 - 25.3  \\ \cline{4-6}
       &     & $ \nu_w  (e^-,e^-) \nu_w $   &   &  &    \\ 
              &     &  $ \nu_w  (p,p) \nu_w $  &   &    &  \\
                         &     &  $ \nu_w  (^{16}O,X) \nu_w $  &   &   &  \\
   \hline
     \hline
 \begin{minipage}[h]{1.9truecm}
$H_2O + Gd$
 \end{minipage} &
      11.3 - 30   &
      same as $H_2O$  &
        SuperK-Gd  \cite{Beacom:2003nk}   &
            22.5   &  0.93 -  2.3 \\  
                                                                       &     &    &    & &   \\

      \hline
        \hline
 \begin{minipage}[h]{1.9truecm}
Scintillator
 \end{minipage} &
      $\sim 11 - 30$   &
${\bf \barnue (p,n) e^+ }$  & JUNO \cite{An:2015jdp} &  17   &  0.8 - 1.9   \\ \cline{4-6}
 &     &  $ \nue (^{12}C,X) e^- $     & RENO-50  \cite{Kim:2014rfa} & 18   &  0.8 - 2.0   \\ \cline{4-6}

& &    $ \barnue (^{12}C,X ) e^+ $  &  &    &     \\ 
          &     & $ \nu_w  (e^-,e^-) \nu_w $    &   &   &  \\
                     &     &  $ \nu_w  (p,p) \nu_w $  &   &   &  \\
                         &     &  $ \nu_w  (^{12}C,X) \nu_w $  &   &   &  \\

         \hline
           \hline
    \begin{minipage}[h]{1.9truecm}
Argon
 \end{minipage} &
      $\sim 18 - 30$   &
${\bf \nue (^{40}Ar,X) e^- }$  &
DUNE  \cite{Adams:2013qkq} &  up to $40$  & up to $ 1.0 $\\ 
   &     &  ${ \barnue (^{40}Ar,X) e^+ }$   &   &   &  \\ 
            &     & $ \nu_w  (e^-,e^-) \nu_w $    &   &    &    \\
                         &     &  $ \nu_w  (^{40}Ar,X) \nu_w $  &   &   &  \\
\hline
\hline
\end{tabular} 
\caption{\label{tabconcepts}   Summary of running (\sk\ only) and near-future detectors for the \df\ having mass above 10 kt. The neutrino energy windows are indicative (as they depend on a number of factors, see e.g. sec. \ref{techwater}) and refer to the main detection process, which is highlighted in bold.   For water \ck\ I give the energy windows of both the first and second search of \sk\ (in brackets, see sec. \ref{techwater}); the rates are for the more conservative window.  Here $X$ stands for any final state and $\nu_w$ indicates a \n\ or antineutrino of any flavor.  The intervals of event rates include the different example spectra H, W and C (Table \ref{SNmodeltable}) and different oscillation scenarios.  
}
 
\end{center}
\end{table}

Neutrinos in a detector are studied through the products of their interaction with electrons and nuclei. For a charged current  process of the type $\nue + ^A_Z X \rightarrow ^A_{Z+1}Y  +  e^-$,  
the rate of  events  with electrons of {\it observed} kinetic energy $E_e$ is
\begin{eqnarray}
\frac{dN_{e} }{ dE_e} =
N_T\int_{-\infty}^{+\infty} dE_e' {\cal R}(E_e,E_e')
{\cal E}(E_e') \int dE \Phi_e (E)
\frac{d\sigma (E_e', E) }{ dE_e'}~,
\label{eq:dnem}
\end{eqnarray}
where  $E_e'$ is the {\it  true} energy of  the electron, $N_T$ is the number of target nuclei in the
fiducial volume and  ${\cal E}$ represents   the detection efficiency.
Here ${d\sigma (E_e', E)/dE_e'}$ is the differential cross section
of the detection reaction  and
${\cal R}(E_e, E_e')$ is the energy resolution  function.
An expression analogous to (\ref{eq:dnem}) holds for the events due to the 
$\barnue$ flux, $\Phi_{\bar e}$.

From what discussed so far, it appears that the detection of diffuse \sn\ \ns\ requires a detector with: (i) large mass (at least comparable with \sk, 50 kt), (ii) high detection cross section  (iii) good energy resolution (to identify and study the \df\ energy window) and (iv) excellent discrimination of the signal over the background. Currently \n\ \snew{observatories} are entering their mature phase, and several technologies are now available to achieve these requirements.  They are illustrated in Table \ref{tabconcepts}, and reviewed briefly in this section.  A more detailed discussion on specific designs is given in secs. \ref{waterc}-\ref{scintgad}.

\begin{itemize}

\item \underline{Water and water-based detectors.} Water \ck\ detection is probably the oldest and best known technology. It has already been used successfully to detect \sn\ \ns\ in the SN1987A event  \cite{Hirata:1987hu,Bionta:1987qt}. \sk\ has proven the feasibility of this technology for tens of kilotons mass, and -- thanks to its robustness and and contained cost -- masses up to 1 Mt are considered realistic (sec. \ref{waterc}). 

\begin{figure}[htbp]
 \centering
 \includegraphics[width=0.7\textwidth]{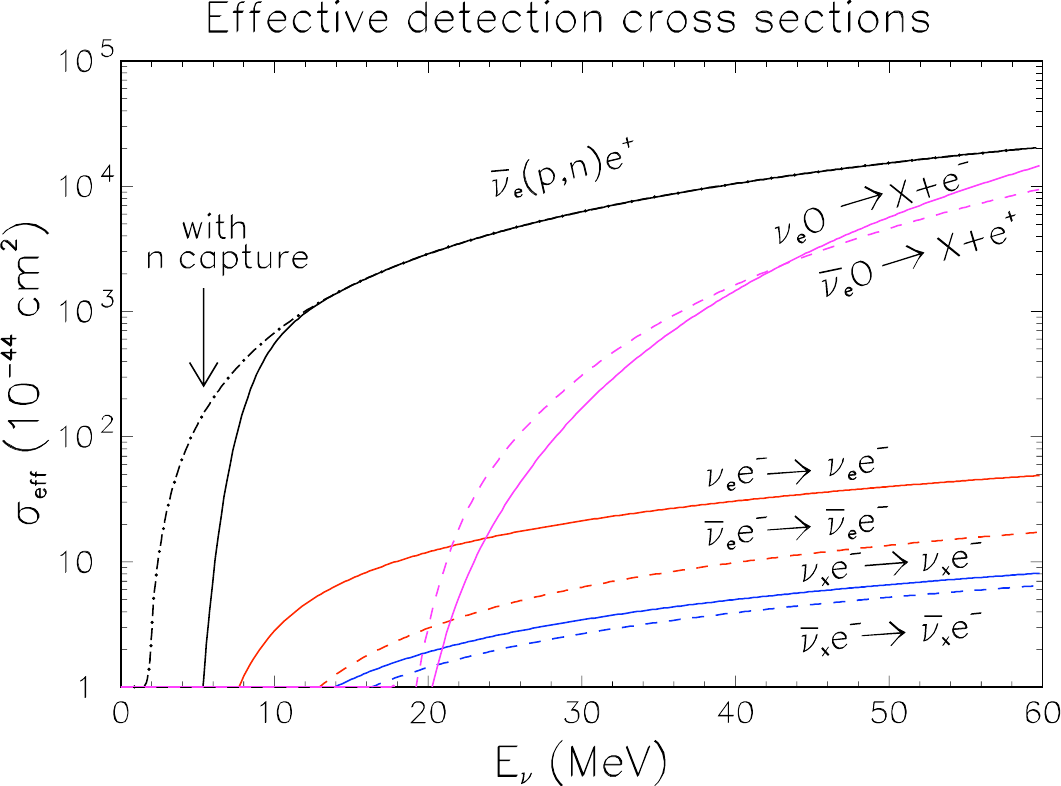}
\caption{Effective cross sections for \ns\ in water \cite{Strumia:2003zx}, including the \sk\ energy resolution and threshold effects. Figure from \cite{Fogli:2004ff}.  }
\label{water_xsec}
\end{figure}
In water, \sn\ \ns\ undergo several scattering processes with the production of an electron or positron (energies are sub-threshold of muon and tau production) that can be detected through its \ck\  cone. These processes (Table \ref{tabconcepts}) are elastic scattering of \ns\ of all flavors on electrons and charged-current interactions of $\nue$ and $\barnue$ on hydrogen and oxygen nuclei. Their cross sections are shown in fig. \ref{water_xsec}.  In realistic energy windows, inverse beta decay ($\barnue + p \rightarrow n + e^+$) exceeds all other channels by at least one order of magnitude in cross section, and therefore it
dominates a signal from the \df. All other channels can be neglected in first approximation.

The energy resolution $ {\cal R}(E_e,E_e')$  of a water \ck\ detector depends on photocatode coverage. For example, the resolution of \sk\  is well modeled by a Gaussian function of width $\Delta/{\rm MeV}\simeq (0.5 - 0.6) \times \sqrt{E_e/{\rm MeV}}$ \cite{suzukitalk,Fogli:2004ff}, corresponding to a resolution of $\sim 11 - 13\%$ at 20 MeV.   At energies above 5-7 MeV, the (hardware) detection efficiency  is very good, being close to 100\% at \sk\ \cite{smycomm}.   

The \df\ detection in water is background-dominated. 
Among backgrounds from other \n\ fluxes, events from solar \ns\ can be subtracted very effectively: they are due mostly to elastic scattering on electrons, which is directional (the emitted electron is nearly collinear with the incoming \n) and therefore pointing back to the Sun. The events from reactors and  atmospheric \ns\ instead are  mostly due to interactions with nuclei; they have little directional information, and thus are likely ineliminable backgrounds. Water detectors are also limited by other two  backgrounds: events from spallation and   events due to  atmospheric invisible muons\footnote{The wording ``invisible muon" denotes a muon whose only signature  in the detector is the track of the electron (or positron) produced by its decay. See e.g. \cite{Malek:2003ki}. } in  the detector. 
At \sk, most spallation events  are excluded  by a combination of an energy cut and of other techniques (see sec. \ref{techwater}), while invisible muons are included in data analyses (see sec. \ref{upperlimits}). 

A relatively recent chapter in the development of water-based detectors is the idea to use of a Gadolinium compound  dissolved in water for enhanced signal discrimination over background \cite{Beacom:2003nk}.   Gadolinium is a strong neutron capturer, already used in the past for the detection of reactor neutrinos (see e.g. \cite{Apollonio:1999ae}).  The capture of a neutron on Gd is followed by gamma ray emission from de-excitation; the detection in coincidence of the gamma ray and positron from inverse beta decay allows to distinguish this process from spallation and from invisible muons in most cases \footnote{A fraction of spallation and invisible muon events is accompanied by neutrons, and of these, some will be indistinguishable from inverse beta decay events in water with Gd.  Still the reduction in background is expected to be substantial \cite{smycomm}.}.  This results in a strong reduction of backgrounds, by a factor of $\sim 5$ for invisible muons and by at least an order of magnitude for spallation \cite{Beacom:2003nk,Fogli:2004ff}.
Therefore, the energy window would extend down to the barrier posed by reactor neutrinos.  Feasible at tens of kilotons scale, the water+Gadolinium technology might be impractical at larger masses.  Still, it is a very attractive idea for a fast and cost-effective enhancement of current detectors (see sec. \ref{scintgad}).

\item  \underline{Liquid scintillator}. Another well known technology, liquid scintillator detection has been used extensively in studies of \ns\ of all types, including \ns\ from galactic \sne\  \cite{Alekseev:1987ej,Dadykin:1987ek,Selvi:2007zz}.  Current liquid scintillator detectors reach 1 kt mass, and a scaling of about one order of magnitude is envisioned \cite{An:2015jdp,Kim:2014rfa}. 
Made of organic materials of the type $C_n H_{2n}$, scintillator can detect \sn\ \ns\ via elastic scattering on electrons and  scattering on hydrogen and carbon nuclei (Table \ref{concepts}). 
\begin{figure}[htbp]
 \centering
 \includegraphics[width=0.78\textwidth]{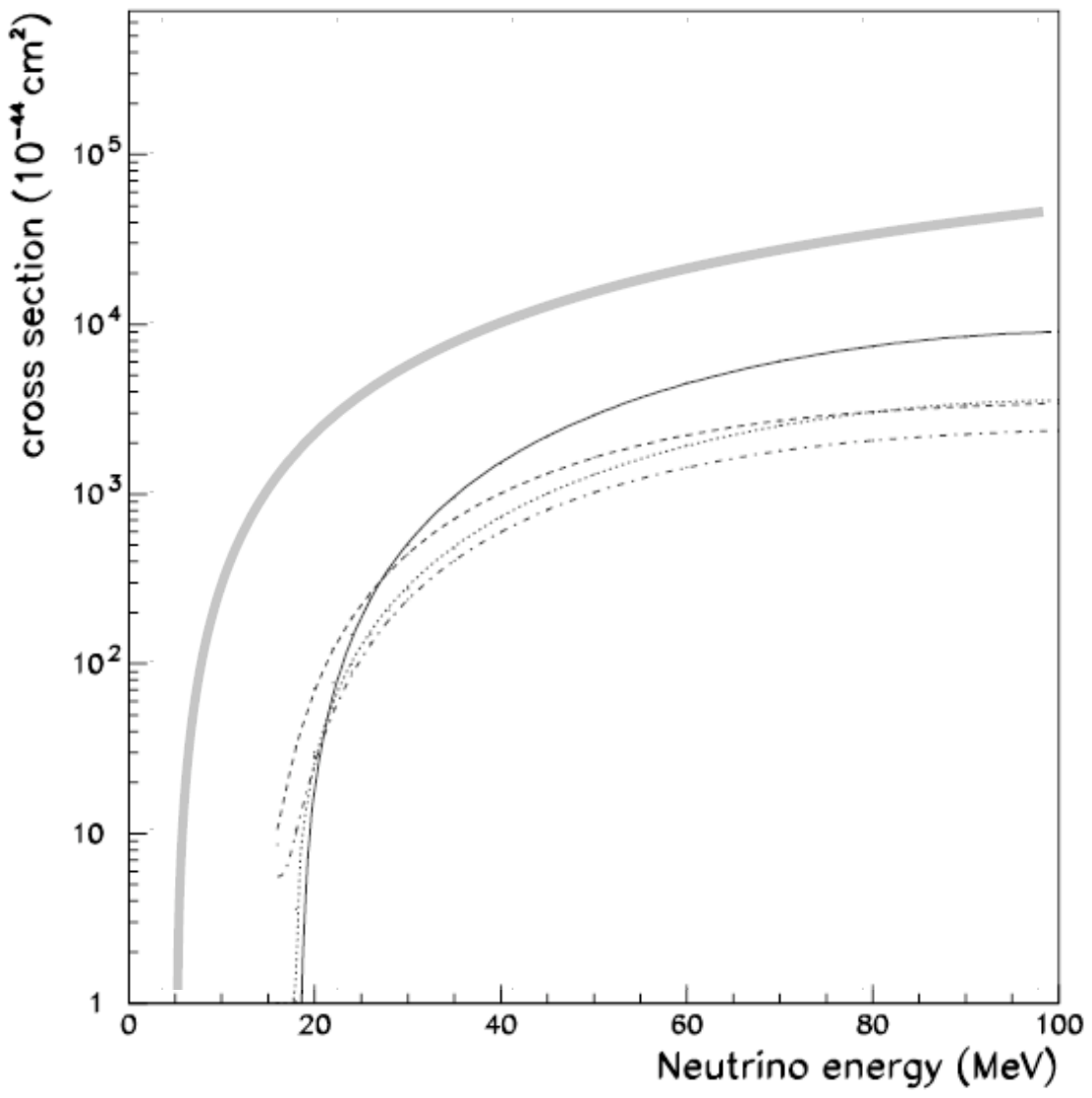}
\caption{Cross sections for \n\ scattering on nuclei relevant to a liquid scintillator experiment: inverse beta decay (thick solid), $\nue$ and $\barnue$ charged current scattering on $^{12}{\rm C}$  (thin solid and dashed), $\nue$ and $\barnue$ neutral current scattering on $^{12}{\rm C}$ (dotted and dot-dashed). Figure adapted from \cite{Agafonova:2006fz}.}
\label{scint_xsec}
\end{figure}
 Like in water, inverse beta decay dominates the event rates.  Compared to water, scintillator offers better background discrimination and better energy resolution.  Spallation and invisible muons \footnote{Backgrounds, of cosmogenic nature are present in scintillator, but reducible to low levels; see  \cite{Wurm:2007cy}.} are distinguished, for the most part, from inverse beta decay because the latter is accompanied by a gamma ray from the neutron capture on a free proton, $n(p,d)\gamma$. 
  The coincident positron can be detected with energy resolution up to 
  one order of magnitude better than water: $\Delta/{\rm MeV}= 0.03 \sqrt{E_e/{\rm MeV}}$, i.e., less than 1\% at 20 MeV \cite{An:2015jdp}.
 The same scaling to Mt mass as water detectors does not appear realistic, however.  Thus, scintillator detectors might be ultimately limited by low statistics (sec. \ref{scintgad}).

\begin{figure}[htbp]
 \centering
 \includegraphics[width=0.6\textwidth]{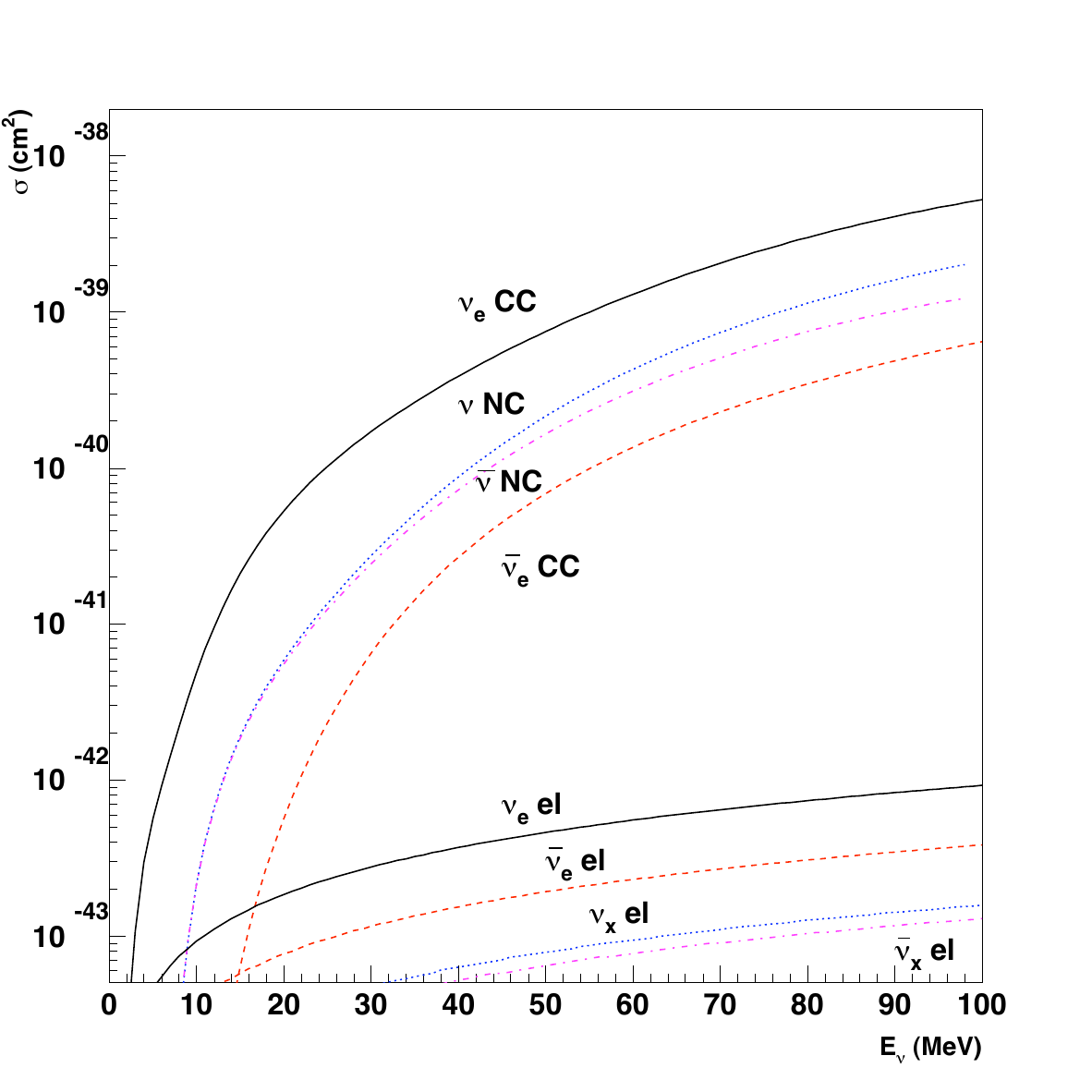}
\caption{Cross sections for \ns\ in liquid argon, from \cite{Cocco:2004ac}. CC (el) stands for Charged Current (elastic scattering). }
\label{lar_xsec}
\end{figure}
\item  \underline{Liquid Argon (\lar)}. A newcomer in the panorama of \n\ detection, the liquid argon technology has only been explored at the prototype level with the 600 tons ICARUS project \cite{Amerio:2004ze}. 
\snew{
The new generation project DUNE will reach ${\mathcal O}(10)$ kt mass \cite{p5report,Stahl:2012exa,cdr} (Table \ref{tabconcepts}). 
} 
A \lar\ detector operates by imaging the tracks left by charged, ionizing particles, using the Time Projection Chamber (TPC) method.  This imaging capability -- similar to that of a bubble chamber \cite{Cline:2006st,Rubbia:2009md} -- allows very good particle and process identification, as well as very good energy resolution.  
Of the several detection processes, given in Table \ref{tabconcepts}  (see fig. \ref{lar_xsec}), charged current scattering of $\nue$ on the $^{40}{\rm Ar}$ nucleus dominates, making liquid argon absolutely unique for its potential to detect the  $\nue$ component of the \df. Neutral current scatterings on $^{40}{\rm Ar}$  are not relevant for the detection of the \df\ because their signature -- the emission of  gamma rays below 11 MeV energy from the deexcitation of the daughter nucleus -- would be completely buried by solar neutrinos \cite{Cocco:2004ac}.  Scattering on electrons is subdominant due to its smaller cross section.

 The energy window for the \df\ at \lar\ is determined by atmospheric and solar neutrinos (sec. \ref{enwindow}), under the sensible assumption that other backgrounds can be effectively separated \footnote{Spallation products are expected in \lar\ detectors.  Their presence is likely to be irrelevant for the \df\ detection, since they lie below 20 MeV of energy, which is anyway precluded by solar \ns.  }.  
A full determination of backgrounds (including rare events, since the \df\ detection rate would be low) will 
  come from extensive R\&D studies that are currently ongoing.  Until then, the discussion of \df\ detection at \lar\ will necessarily be  indicative.

\end{itemize}

\subsection{Upper limits}
\label{upperlimits}

\subsubsection{Data and flux constraints}

So far, the \df\ has escaped detection.     Thanks to their larger volumes,  currently active detectors \cite{Malek:2002ns,Eguchi:2003gg,Aharmim:2004uf,Aharmim:2006wq,Collaboration:2011jza,Bays:2011si,Zhang:2013tua} have improved dramatically on the bounds set by the previous generation of experiments \cite{Zhang:1988tv,Aglietta:1992yk}.   Table
\ref{ex}  summarizes the most  stringent upper limits (see also fig. \ref{limitssummary} for a graphical representation).
\begin{table}[htdp]
\begin{center}
\begin{tabular}{|c|c|c|c|c|c|}
\hline
\hline
Species & Experiment & reference & energy  (MeV) &  limit (${\rm cm^{-2} s^{-1}}$) & C.L.\\
\hline
\hline
\multirow{4}{*}{$\nue$} & \sk\ & \cite{Lunardini:2008xd} & $>$ 19.3 & 73.3 -154 &90\% \\
 &  \sk\ (indirect) & \cite{Lunardini:2006sn}  & $>$19.3 &  5.5 & $\sim$ 98\%  \\
 &  SNO & \cite{Aharmim:2006wq} &   22.9 - 36.9   & 70  & \\
  &  \sk\  & \cite{Lunardini:2008xd} &   22.9 - 36.9   & 39 - 54  & 90\% \\
\hline
\multirow{2}{*}{$\barnue$} & \sk\  (2853 days) & \cite{Bays:2011si}  & $>$ 17.3 & 2.8 - 3.0 & 90\%  \\
 &  \sk\ (960 days) & \cite{Zhang:2013tua} &  13.3 - 31.3 &  $(1.48 - 9.9) \times 10^2$ & 90\%  \\
&  KamLAND & \cite{Collaboration:2011jza} &  8.3 - 31.8 &  $1.39 \times 10^2$  & 90\%  \\  
&  Borexino & \cite{Bellini:2010gn} &  1.8 - 17.8 &   $\sim 3 \times 10^5$  & 90\%  \\  
\hline
$\numu+ \nutau$ & \sk\ & \cite{Lunardini:2008xd}& $>$ 19.3 & (1.0 - 1.4) $\times 10^3$ & 90\% \\
\hline
$\barnumu+ \barnutau$ & \sk\   & \cite{Lunardini:2008xd} &$ >$19.3 & (1.3 - 1.8) $\times 10^3$ & 90\% \\
\hline
\hline
\end{tabular}
\caption{\label{ex} Summary of the most stringent bounds on the \df\ from current detectors, with their confidence level (C.L.).  %
Unless otherwise noted, the \sk\ bounds refer to the 1496 live days data published in \cite{Malek:2003ki}.
 The limit on the $\nue$ component labeled as ``indirect" proceeds from the \sk\ $\barnue$ limit with considerations of similarity of the $\nue$ and $\barnue$ fluxes at Earth due to neutrino oscillations in the star \cite{Lunardini:2006sn}.  Where applicable, intervals of limits are given, corresponding to the range of  neutrino spectra or models used in the analysis. The SNO result is also spectrum-dependent: the quoted bound is the median of several 90\% C.L.  limits found with different neutrino spectra. 
  The Borexino bound is an extrapolation from figure 2 in \cite{Bellini:2010gn}.
}
\end{center}
\end{table}%

For a long time, the strongest bounds on all \n\ flavors were those established in 2003 by the 1496 days search at \sk\ \cite{Malek:2002ns} for events with energy between the threshold of 18 MeV (lepton energy, established after a spallation cut) and 80 MeV.  
The data  
 were analyzed by the  \sk\ collaboration in the dominant detection channel, inverse beta decay induced by electron antineutrinos (Table \ref{tabconcepts}).  
The resulting bound was $\Phi_{\barnue}(E>19.3~ {\rm MeV})<1.2~{\rm cm^{-2} s^{-1}}$ at 90\% C.L., obtained by fitting the data with a single parameter, the flux normalization.  
A separate, multi-parameter, study of the same data \cite{Lunardini:2008xd} in the subdominant detection channels, gave limits on the $\nue$ component of the \df\ (from $\nue$ scattering on oxygen and on electrons, Tab. \ref{tabconcepts})  and on the non-electron flavors components as well (from scattering on electrons).  These limits, although loose, improved on the previous constraints from SNO \cite{Aharmim:2006wq} (heavy water, $\nue$ channel)
and LSD \cite{Aglietta:1992yk} (liquid scintillator, all flavors).

In 2012 the \sk\ collaboration updated the $\barnue$ bound, giving the result in Eq. (\ref{sklim}) \cite{Bays:2011si}.   This latter analysis uses 2853 days of data, is multi-parameter, and has a lower threshold, 16 MeV  of positron energy (corresponding to 17.3 MeV of neutrino energy). For comparison with earlier results, however, a limit with the 18 MeV threshold was also given for the model in \cite{Ando:2002ky}:  $\Phi_{\barnue}(E>19.3~ {\rm MeV})<2.0~{\rm cm^{-2} s^{-1}}$ at 90\% C.L..  The reasons for obtaining a looser bound, compared to the 2003 result, are improvements in the cross section and in the statistical method used, and a (non-significant) excess in the post-2003 data \cite{Bays:2011si}.

A different limit on the $\barnue$ flux was obtained in 2014 by a \sk\ search for inverse beta decay with tagging of neutron capture on hydrogen \cite{Zhang:2013tua}. The neutron tagging allowed for a lower energy threshold, 12 MeV of positron energy, however its efficiency is $\sim$17 - 18\% \cite{Zhang:2013tua}. Therefore the resulting \df\ limit is looser than that in eq. (\ref{sklim}), but it is nevertheless interesting due to the different energy region probed.

Complementary to water results, a limit on $\barnue$s at lower energy (threshold 7 MeV positron energy) was established by KamLAND \cite{Eguchi:2003gg,Collaboration:2011jza}, in a study originally designed mainly to constrain a possible $\barnue$ flux from the Sun. While limited by the small mass of KamLAND (1 kt), the result is an interesting test of the liquid scintillator technology, with its better signal discrimination, in anticipation of a next generation of ${\mathcal O} (10)$ kt mass experiments \cite{MarrodanUndagoitia:2006re}

Looking at the future, the \sk\ collaboration plans new, improved analyses, which could further lower the energy threshold by up to $\sim$2 MeV \cite{smycomm}.  However, major improvements will be possible only with new concepts like Gadolinium addition, currently under study (sec. \ref{scintgad}).

Aside from detailed data analyses, indirect upper bounds can be established on the basis of naturalness and theoretical considerations.
In particular, a constraint on diffuse $\nue$s follows from the constraint on the $\barnue$ component at  \sk\, by considering that the two components must be similar due to their common origin  in the non-electron neutrino flavors inside the star through neutrino oscillations \cite{Lunardini:2006sn}.  This is the ``indirect" limit given in Table
\ref{ex}, and, while considerably looser than the $\barnue$ bound, it is currently the strongest for the $\nue$ species.

\subsubsection{Implications of the \sk\ limit}

\begin{figure}[htbp]
 \centering
 \includegraphics[width=0.6\textwidth]{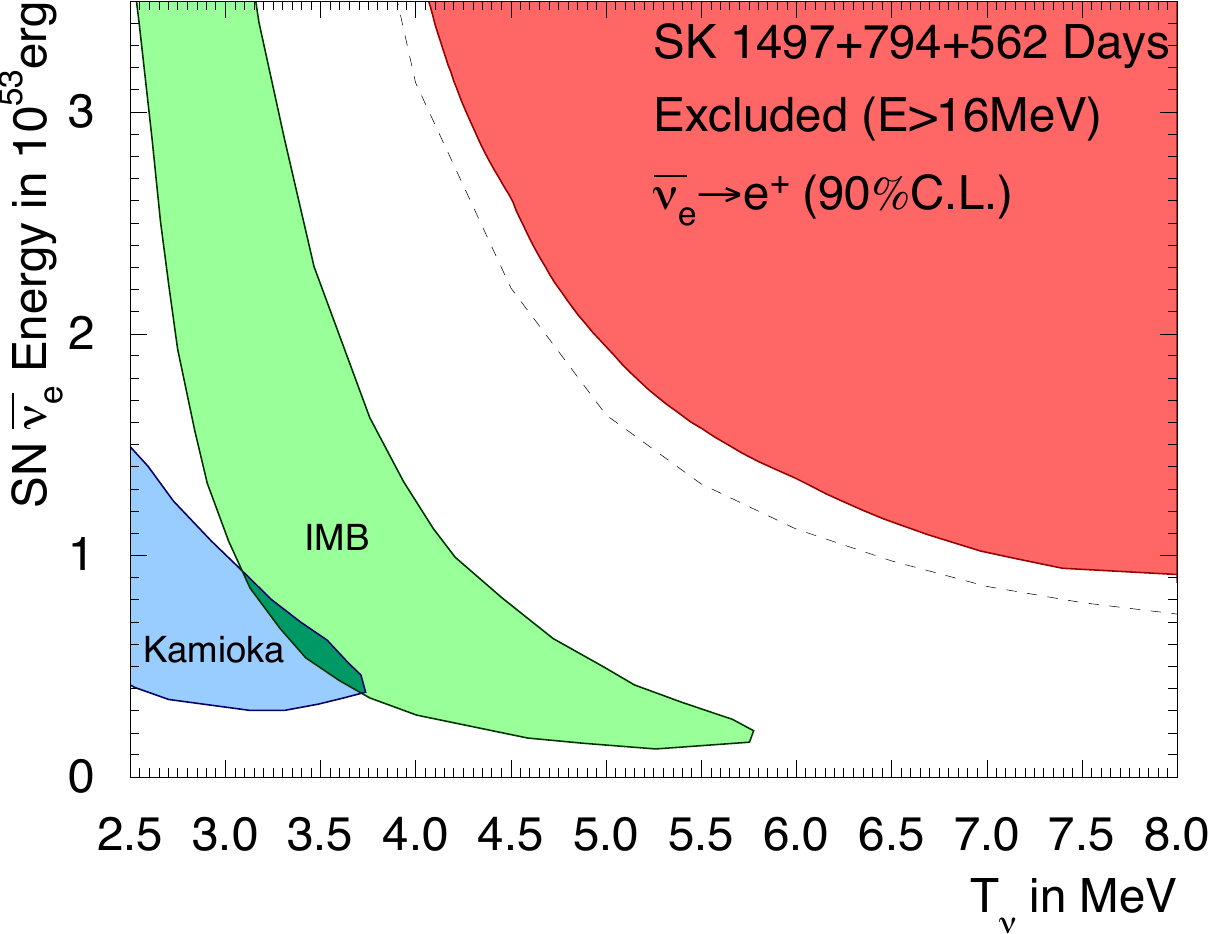}
\caption{
From \cite{Bays:2011si}: 90\% C.L. exclusion region from the 2011 \sk\ analysis (upper shaded area) in the space of the total and effective temperature of the $\barnue$ flux from an individual \sn.  A thermal \n\  spectrum is assumed,  
and the parameters $R_{-4}=1.25$, and $\beta = 3.4$ (from \cite{Beacom:2010kk,bayspriv}) were used to describe the  \snr\ (Eq. (\ref{sfrcurves})).  
Regions allowed by SN1987A data are shown as well (contours, originally from \cite{Jegerlehner:1996kx}; see also \cite{Yuksel:2005ae}). The dashed line refers to a one-dimensional analysis with the total energy $L_{\bar e}$ as free parameter and each value of the temperature taken as fixed. 
}
\label{bays_costraints}
\end{figure}

As already observed (fig. \ref{comparefig}), the \sk\ limit approaches the region of theoretical predictions of the \df, leaving most models unconstrained. 
Loose constraints can be obtained in the space of parameters that describe the flux, namely the \n\ spectra and total energies, and the normalization of the \snr.  
Only combinations of parameters giving the highest \df\ are excluded, and generally degeneracies imply only weak constraints on individual parameters.

\begin{figure}[htbp]
 \centering
 \includegraphics[width=0.7\textwidth]{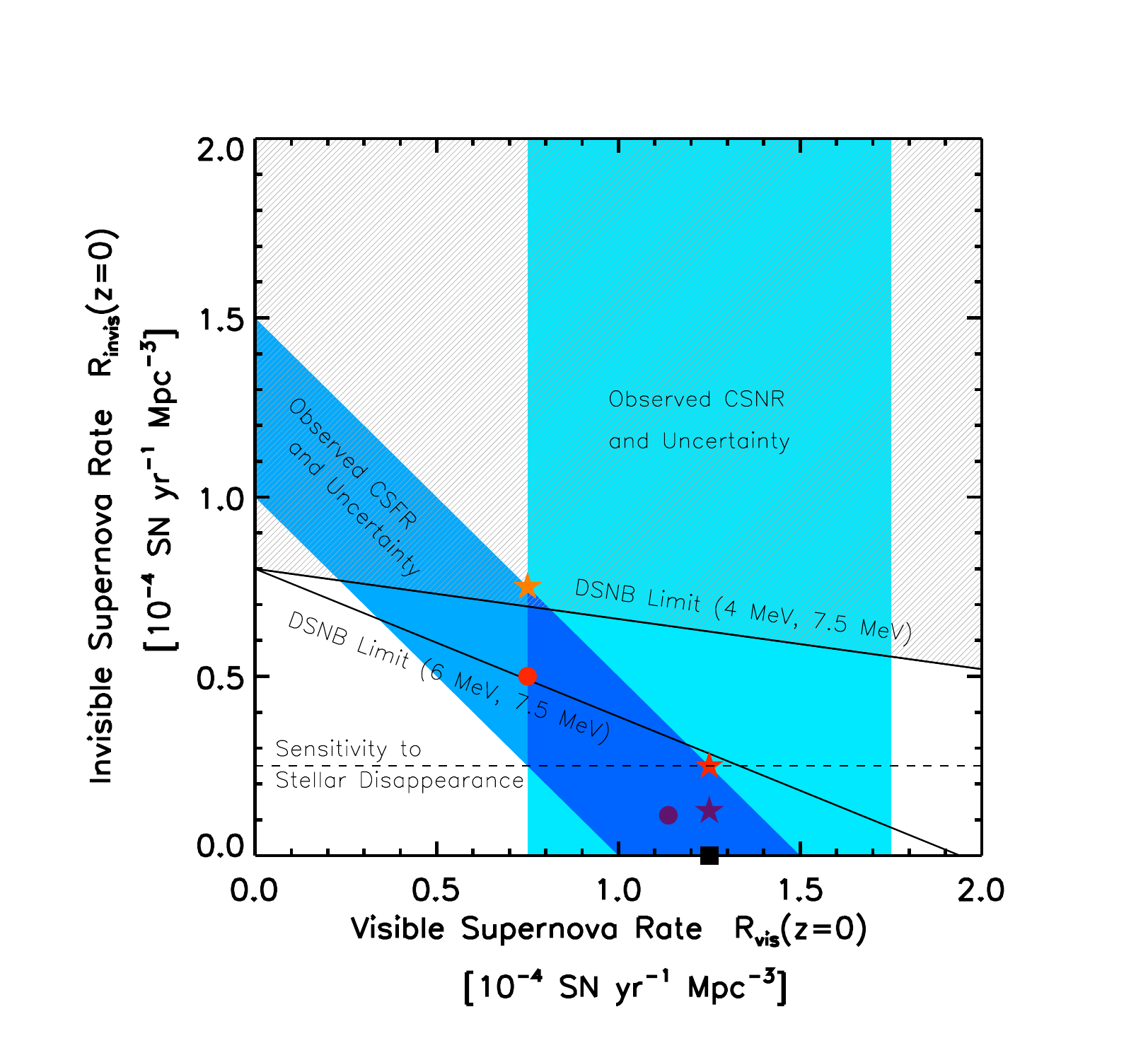}
\caption{From \cite{Lien:2010yb}:  exclusion region (shaded grey) from the 2003 \sk\ limit, in the space of the cosmic rates (at $z=0$)  of \nts\ (successful) and  \bh\ (failed) \sne.  For these two stellar populations, thermal \n\ spectra were used with temperatures  of 4 and 7.5 MeV respectively.  A second contour gives the constraint for a different set of temperatures (6 and 7.5 MeV).  %
The remaining shaded regions are those allowed by current measurements of the star formation rate (dark blue, diagonal) and of the supernova rate (light blue, vertical).  The dashed line represents the projected sensitivity of future searches for failed \sne.  Stars, circles and squares refer to the parameters used in \cite{Lien:2010yb} for flux predictions.}
\label{Lien_costraints}
\end{figure}
Two different sets of constraints are shown in figs. \ref{bays_costraints} and \ref{Lien_costraints} (from \cite{Bays:2011si,Lien:2010yb}).  
Fig.  \ref{bays_costraints} refers to fixed \snr\ (taken from \cite{Beacom:2010kk}) 
and shows the excluded region in the space of the total energy and 
effective temperature  of the $\barnue$ flux from an individual \sn.  
 Parameters compatible with SN1987A are allowed, and for the theoretically natural energy of $L_{\bar e}=5 \cdot 10^{52}$ ergs, temperatures as high as 8 MeV are allowed.    The allowed region is expected to be even wider if uncertainties on the \snr\ are included.

Fig.  \ref{Lien_costraints}  instead gives the exclusion of rates of \nts\ (successful) and  direct \bh\ (failed) \sne\ depending on the \n\ spectra for these two \sn\ populations, compared with astrophysical limits on the same quantities from star formation rate measurements and \sn\ observations.  Interestingly, the rate of failed \sne\ alone is constrained to be below $\sim 0.8 \cdot  10^{-4}{\rm  yr^{-1} Mpc^{-3}} $, which is however well above the rate expected for a  $\sim 20\%$ fraction of failed \sne\ 
(sec. \ref{snspectra}).

While detailed analyses of parameter constraints from the \sk\ limits have yet to be done, current results are sufficient to give a perspective of what can be expected from next generation searches (see sec. \ref{perspectives}).  With a factor of 2-4 improvement in flux sensitivity, the parameter constraints from the \df\ will become competitive with those from astronomy and from SN1987A \cite{Yuksel:2005ae}.  

\subsection{Water \ck\ detectors towards megaton scale}
\label{waterc}

\subsubsection{Number of events}

After the very successful experience of the 50 kt mass of \sk, water technology is now mature to expand to megaton scale. 
In this section I discuss the \df\ signal  
for a  representative setup  of a 
\snew{
0.5 Mt 
}
fiducial volume and 5 years running time, with energy resolution equal to that of \sk\ (sec. \ref{concepts}).  An efficiency of 93\% \cite{Hirata:1988ad,Bahcall:1996ha} is used as representative of a detector performance. However one must consider that the real efficiency depends on the details of the hardware and of the data analysis and therefore results will have to be rescaled once these details are known.
 I include only inverse beta decay events and neglect the subdominant channels (which contribute to less than $\sim 4$\% \cite{Volpe:2007qx})   and use backgrounds for the \sk\ location, taken from  \cite{Fogli:2004ff}.  
 To be conservative, all other specifications (like the energy threshold) will be taken to be the same as in the earliest \sk\ analysis of \cite{Malek:2003ki}.  
 This setup is necessarily indicative; some technical details will be discussed in sec. \ref{techwater}.  All results refer to the same flux parameters as in sec. \ref{diffearth}.

\begin{figure}[htbp]
\newpage
 \centering
 \includegraphics[width=0.6\textwidth]{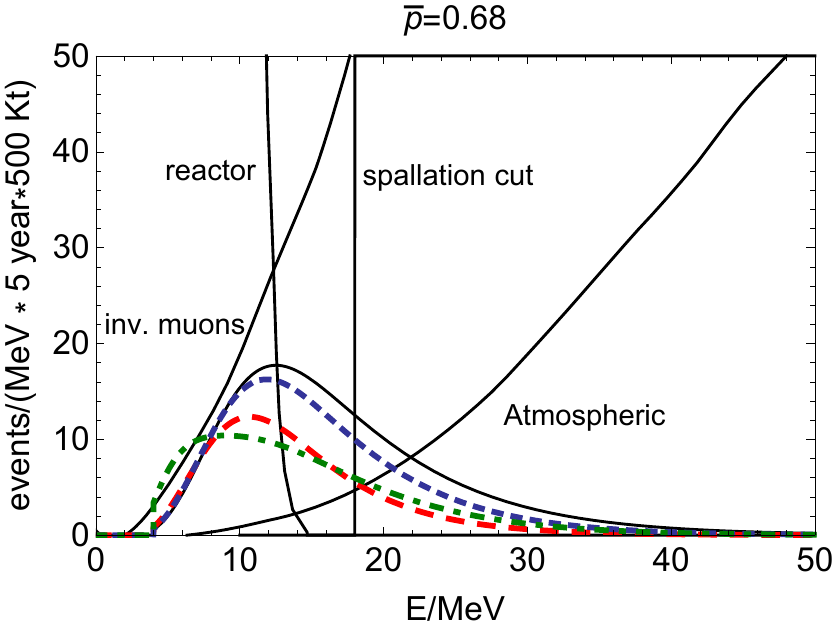}
 \vskip 0.4truecm
  \includegraphics[width=0.6\textwidth]{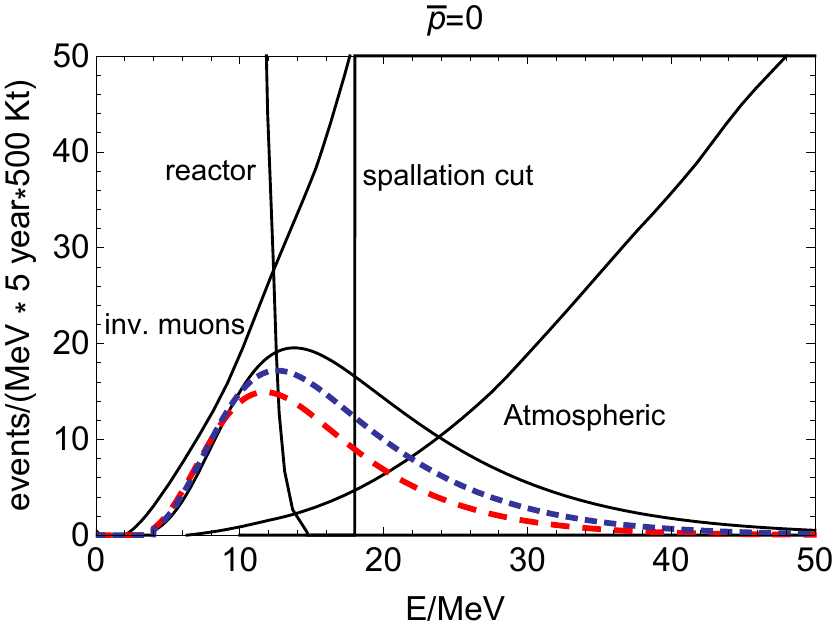}
\caption{  Energy distribution of positrons from inverse beta decay expected at a  water \ck\ detector with a 
\snew{
${\rm 2.5~Mt \cdot yr}$ 
}
exposure, located at Kamioka,  for the two extreme values of $\bar p$. The solid,  short dashed, long dashed and dash-dotted lines refer to the different \n\ spectra shown in fig.  \ref{spectradiff}, and specifically to the H, W and C spectra,  and the SN1987A best fit spectrum. 
The spectra of events from reactor \ns, atmospheric \ns\ and invisible muons  are shown for comparison (from \cite{Fogli:2004ff}).  A 18 MeV energy cut (due to spallation) is shown as well.  }
\label{waterkam}
\end{figure}
Fig. \ref{waterkam} shows the energy distribution of the signal and  background events.     It appears immediately that the signal to background ratio is always smaller than 0.2-0.3, and is maximum just above the threshold of 18 MeV. There the signal is comparable to or larger than the atmospheric background but is exceeded by the invisible muon one.  The atmospheric background starts to dominate at $\sim$25 MeV, thus closing the energy window (sec. \ref{enwindow}).  
It appears that even for the most energetic \n\ spectrum, the peak of the event distribution is below the spallation cut, and would require a better background discrimination to be observed.

Tables \ref{ratesmodel68} and \ref{ratesmodel0} give the numbers of  signal and background events in two intervals  of positron energy. While the smaller bin ($18<E_e<28$ MeV) is realistic and reflects the region where the signal/background ratio is maximal, the wider interval ($10<E_e<28$ MeV) might be relevant to a possible upgrade (for example water plus Gadolinium, sec. \ref{concepts}) where spallation is effectively removed and only reactor neutrinos remain as background at low energy.

\begin{table}[htbp]
\newpage
\begin{center}
\begin{tabular}{| c |c |c |c |c |c| c|}
\hline
\hline
  & \multicolumn{4}{|c|}{Signal  }  &   \multicolumn{2}{|c|}{Background  }  \\  \hline 
  &  Hot  & Warm   & Cold  &  SN1987A best fit  & atmosph. &  inv. $\mu$ \\
\hline
\hline
total & 275  & 227   &   150    & 166  &  &  \\
\hline
$10< E_e< 28$ MeV   & 203 &  167 &  103 &  102     & 115 & 1064    \\
\hline
$18< E_e< 28$ MeV   & 74.5 &   53.6 &  25.8 &  34.0    & 95 & 791  \\ 
\hline
\hline
\end{tabular}
\caption{  Numbers of  signal (from inverse beta decay) and background events from atmospheric \ns\ and invisible muons in energy intervals of interest at a detector of mass 0.5 Mt and livetime 5 years, for $\bar p=0.68$.  The total number of signal events over the whole spectrum is also given.  The \n\ fluxes in fig. \ref{spectradiff} were used; the backgrounds are for the Kamioka site \cite{Battistoni:2005pd}. }
\label{ratesmodel68}
\end{center}
\end{table}

\begin{table}[htbp]
\begin{center}
\begin{tabular}{| c |c |c |c |c |c|}
\hline
\hline
  & \multicolumn{3}{|c|}{Signal }  &   \multicolumn{2}{|c|}{Background  }  \\  \hline 
  &  Hot  & Warm   & Cold  &   atmosph. &  inv. $\mu$ \\
\hline
\hline
total &  358 &  268 & 207   &  &  \\
\hline
$10< E_e< 28$ MeV   & 259 &  199 &   151  & 115 & 1064    \\
\hline
$18< E_e< 28$ MeV   & 113 &  47.9 &  73.5  & 95 & 791   \\ 
\hline
\hline
\end{tabular}
\caption{ The same as Table \ref{ratesmodel68} for $\bar p=0$.  }
\label{ratesmodel0}
\end{center}
\end{table}
From the Tables it appears that the number of signal events above 18 MeV varies between $\sim$20 and $\sim$100 depending on the \n\ energy spectra and oscillation effects.    An even larger variation should be expected in consideration of the uncertainty in the normalization of the core collapse rate (sec. \ref{snrate}).  The $E^2$ dependence of the cross section magnifies the differences due to different energy spectra of the \n\ flux, so that the event rate for the  H spectrum  can easily be twice as large as in the other cases.  

\begin{figure}[htbp]
\newpage
  \centering
\includegraphics[width=0.6\textwidth]{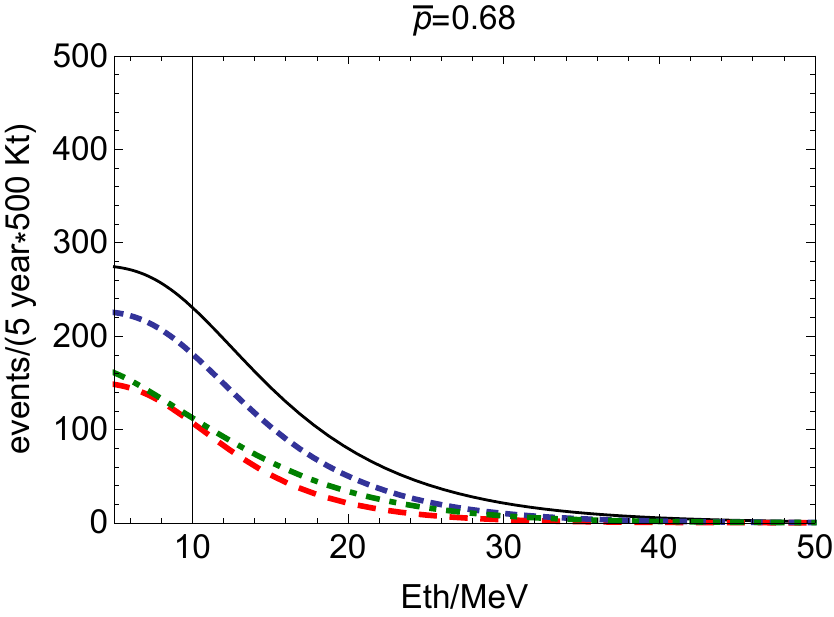}
 \vskip 0.4truecm
\includegraphics[width=0.6\textwidth]{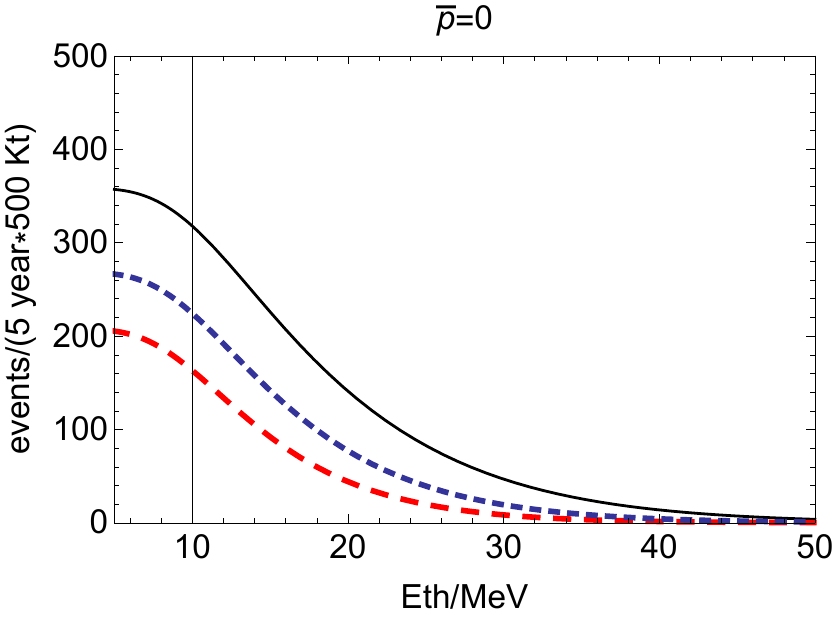}
\caption{   Inverse beta decay events above a threshold energy $E_{th}$, as a function of $E_{th}$ for the two extreme values of $\bar p$.  The solid, long dashed and short dashed curves correspond to the fluxes in fig.  \ref{spectradiff} (with the same coding of colors and dashings).  
 }
\label{diffposint}
\end{figure}
To answer the question of statistical significance, one should consider the $\sim 900$ background events in the 18-28 MeV bin, corresponding to a  $1 \sigma$ statistical error of about 30 events. It follows that, for the exposure considered here, 
\snew{
the signal could reach a  3$\sigma$ significance for the H spectrum.   In general, and depending on the \df\ normalization, a longer exposure would be necessary for a high significance observation. 
}
  In the worst case of $\sim$26 signal events, an exposure 12 times longer (for example, 20 years running time for a 1.5 Mt detector) would be required to reach a  $3\sigma$ significance\footnote{Notice that these considerations on statistical significance assume that the normalization of the background is well know independently, which is not the case at this time. If poorly constrained, this normalization will be included in the data analysis as a fit parameter, as is currently done by the \sk\ collaboration \cite{Malek:2003ki,Bays:2011si},  and this weakens the sensitivity to the signal.}. 

For an upgraded configuration with 
lower threshold and subtracted spallation background, the number of events from the signal can exceed 200. If the invisible muon background is subtracted as well,  the signal could be comparable to or larger than the background, thus ensuring excellent statistical significance even for the most unfavorable  parameters.  This is a very strong motivation for improvements in this direction.  A discussion of possibilities is given in sec. \ref{scintgad}. 

For generality, I also give numbers of signal events above a range of thresholds, in fig. \ref{diffposint}.  They can be used to calculate the event rates in other energy bins of interest.   For selected thresholds the figure also gives the 90\% C.L. interval predicted by imposing compatibility with SN1987A and with direct measurements of the \sn\ rate \cite{Lunardini:2005jf} (sec. \ref{snrate} and \ref{SN1987Acons}).
 One can see the substantial decline in the event rate with the increase of the threshold beyond the peak of the \df\ ($E \sim 6-7$ MeV, see sec. \ref{snspectra}). 
\begin{figure}[htbp]
\newpage
  \centering
  \includegraphics[width=0.65\textwidth]{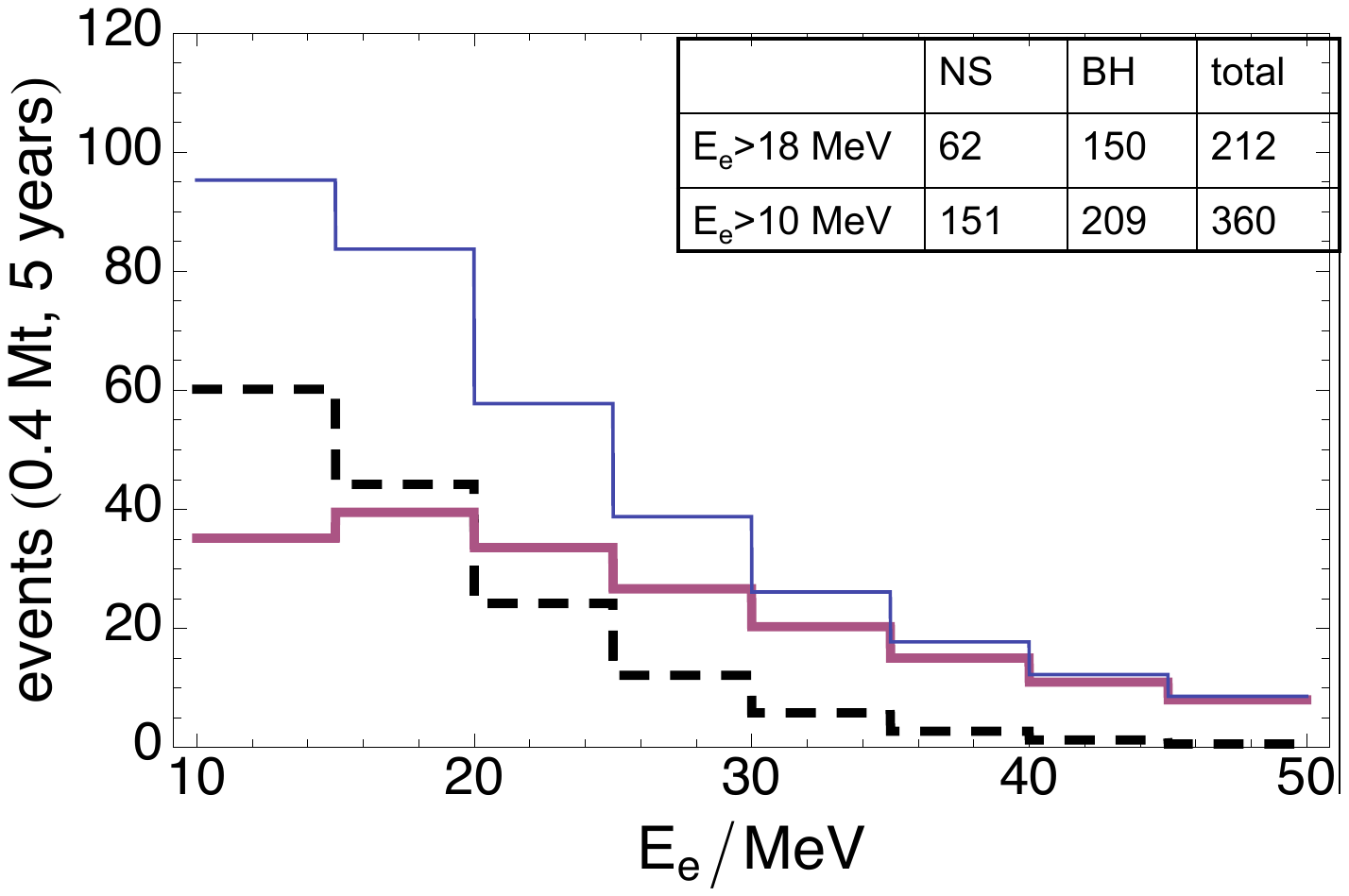}
   \caption{From \cite{Lunardini:2009ya}: events in water from direct \bh\ collapses (solid thick), from \nts\ collapses  (dashed) and the total of the two (thin line) for the best case scenario in fig. \ref{spectradiffBH}  (largest failed \sn\ flux, with the S EoS, $\bar p= 0.68$ and $f_{NS}=0.78$).  A 2 Mt$\cdot$yr exposure is used;   $E_e$ is the positron energy. The inset gives the number of events above selected thresholds.
}
\label{bhevents}
\end{figure}

The diffuse flux, and therefore the event rate, could be substantially enhanced if failed \sne\ are relatively numerous  ($\sim 20\%$ or so) in the universe, as discussed in sec. \ref{snspectra}.   Figure \ref{bhevents} gives the expected energy distribution and integrated rate of events expected in the best case scenario of fig. \ref{spectradiffBH}  (largest failed \sn\ flux, with S EoS, $\bar p= 0.68$ and $f_{NS}=0.79$).  It appears that above 18 MeV of positron energy the events due to failed \sne\ could amount to about 2/3 of the total, enhancing the rate up to about 200 events for an exposure of 2 Mt$\cdot$yr.  This enhancement could allow an earlier detection at 3$\sigma$ of the \df, already with half the exposure.  Thanks to the more energetic spectrum of failed \sne, lowering the energy threshold to 14 MeV or so would be sufficient to capture the bulk of the events from these objects and see the peak of their event distribution.  With a 10 MeV threshold the number of signal events can exceed 300. 

\subsubsection{Technical considerations}
\label{techwater}

Certainly the event rates given here are indicative, since they ultimately depend on the specific detector design.  For example,  the energy resolution and the fiducial volume (for a given total mass) are influenced by the detector's  geometry, photocathode coverage, electronics, etc..   As discussed in sec. \ref{enwindow}, location and depth influence the backgrounds importantly. 
Choosing the technical specifications of an experiment always involve balancing scientific, financial, technical and logistical requirements. Here I discuss some of the currently favored setups in their main physics aspects.

\begin{itemize}

\item \underline{geometry.  } The excavation of a cavern large enough to contain a single, undivided, $\sim$1 Mt mass of water is certainly challenging but possible \cite{Nakamura:2003hk}. 
However, current designs favor a modular approach, with the detector consisting of multiple separate volumes. An example is the proposed two-volume layout for Hyper-Kamiokande (fig. \ref{waterdesign}) \cite{Abe:2011ts} (see also \cite{Diwan:2006qf,Agostino:2012fd} for different concepts), where considerations of rock stability  suggest a quasi-cylindrical design for each module. 
Compared to a single volume, modularity would result in a smaller fiducial volume (for equal total mass), but is attractive because it breaks several challenges (excavation, instrumentation, maintenance, etc..) into more manageable parts that can be completed sequentially or in parallel depending on circumstances.  The modular design also shortens the waiting time, since it can start to deliver data as soon as the first module is completed, similarly to what happened for another supermassive project, IceCube \cite{Achterberg:2006md}.

For Mt masses,  the same photocatode coverage of \sk\  (40\%) may be unrealistic, and therefore a lower coverage (20-30\%)  is envisioned for  at least part of the detector \cite{Abe:2011ts}. This would result in a slight worsening of the low energy performance ($\sim 7$ MeV rather than $\sim 5$ MeV threshold, which is however inconsequential for the energy window of the \df), as well as a somewhat poorer energy resolution \cite{Autiero:2007zj}, with consequent poorer background discrimination and ultimately higher energy threshold for \df\ searches.   

\begin{figure}[htbp]
 \centering
 \includegraphics[width=0.8\textwidth]{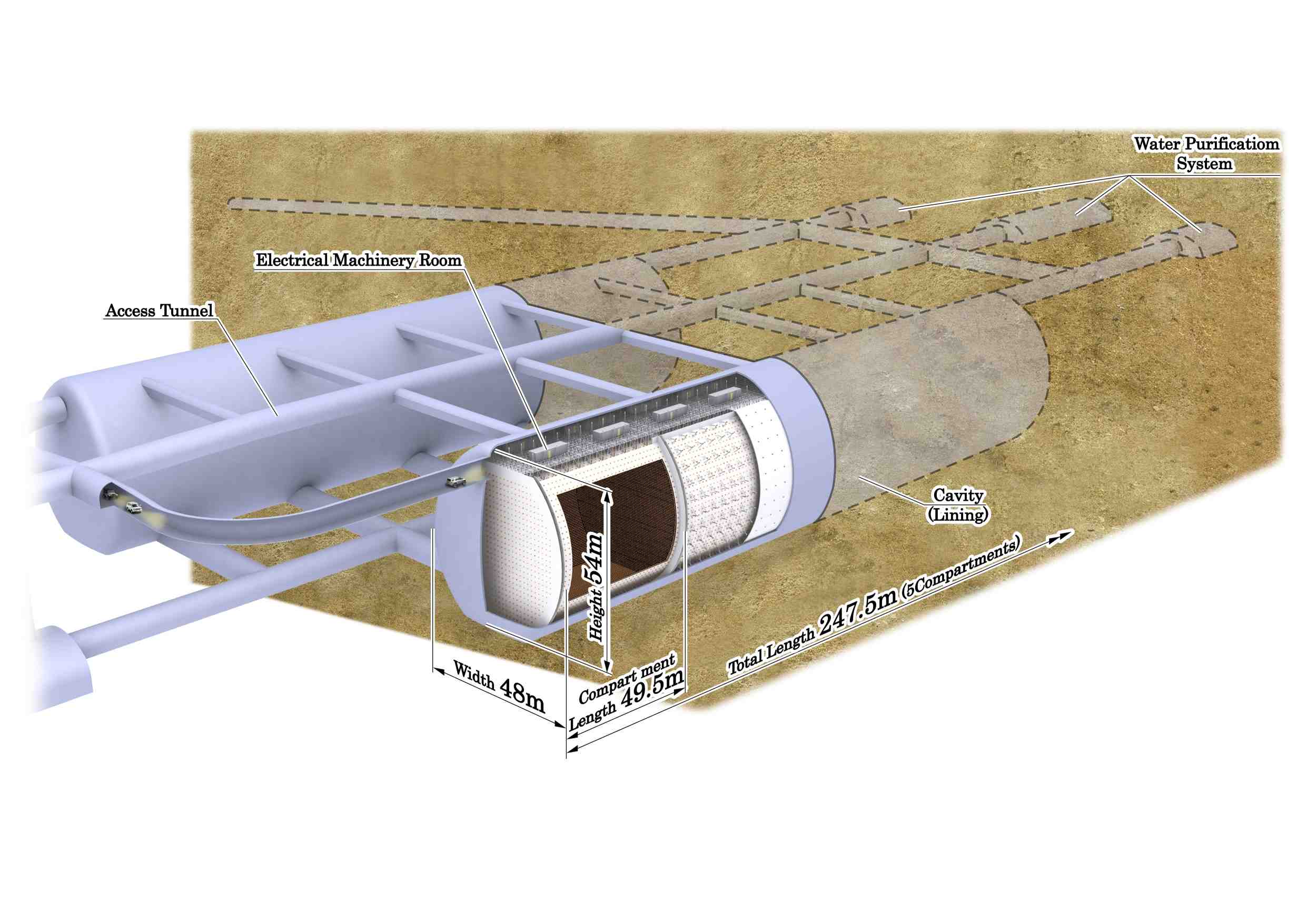}
\caption{
Conceptual design of Hyper-Kamiokande, \cite{Abe:2011ts}. 
}
\label{waterdesign}
\end{figure}

\item  \underline{ depth. }  Since the cosmic ray flux in a detector rapidly decreases with depth, even modest gains in depth result in substantially enhanced performance.  This is especially true for the \df, for which detection is dramatically limited  at low energy by spallation products of cosmic ray muons. A detailed study on depth requirements for a Mt \ck\ detector  \cite{Bernstein:2009ms}, gives the expected extension of the energy window with the increase of depth, relative to that of \sk\ (3300 ft, equivalent to 1005 m and to 2900 m water-equivalent).  Results are given in  Table \ref{tab:SN}.  For pure water, by going to 4850 ft depth the energy window can be pushed down to 15.5 MeV of positron energy, with an enhancement of the \df\ signal of about 40\%.  
\begin{table}[hbt]
\centering
\begin{tabular}{|l|l|l|l|}
\hline
rock depth & water equiv & Energy thres.  without  &  Signal rate without   (with) \\
 ft.   & km-w-e   &  (with) Gd (MeV)    &   Gd relative to 18 MeV    \\
\hline
4850  & 4.3 &  15.5 (12.0)  &   1.4   (2.0)       \\
3500  & 3.1  &  17.5 (15.0) &    1.1  (1.5  )      \\
3300  & 2.9  &  18.0 (15.5)  &     1.0  (1.4 )      \\
2000  & 1.8  &  20.5  (18.0) &      0.73 (1.0)       \\
300  &  0.27 &  25.0 (22.5)  &     0.36  (0.55)      \\
\hline
\end{tabular}
\caption{  Expected energy threshold for a water \ck\ detector with and without Gd addition as a
function of depth for detection of the \df, from \cite{Bernstein:2009ms}.  The \df\ spectrum used in \cite{Bernstein:2009ms} is close
\snew{
to the H spectrum (but with slightly hotter $\nux$, $E_{0x}=21.6$ MeV) with $\bar p=0.68$  
}
(see sec. \ref{snspectra}). }
\label{tab:SN}
\end{table}
This depth is also required to have sensitivity to the day-night asymmetry of solar neutrinos \cite{Renshaw:2013dzu}. 

\item  \underline{Location/latitude }  Even though latitude is important for the atmospheric \n\ background, considerations of depth always prevail in the choice of a particular underground location, together with considerations like access of people and technology to the site.   
\snew{
Although several sites have been considered recently for a Mt-scale detector (Table \ref{windows}), at this time efforts are concentrated on the Hyper-Kamiokande project in Japan \cite{Abe:2011ts}. 
}

\end{itemize}

\subsection{Reducing backgrounds with liquid scintillator and Gadolinium}
\label{scintgad}

\subsubsection{SuperK-Gd}

The idea of dissolving gadolinium in water was proposed by Beacom and Vagins \cite{Beacom:2003nk}, who discussed its cost-effectiveness and non-toxicity.  They envisioned applying the idea to the \sk\ detector, thus initiating a new phase of \sk\ called SuperK-Gd \footnote{\snew{In the original work by Beacom and Vagins \cite{Beacom:2003nk}, and until recently, this  upgrade was called GADZOOKS! ``Gadolinium Antineutrino Detector Zealously Outperforming Old Kamiokande, Super!".} }.  This initiative is especially attractive for its relatively low cost and short timeline; \snew{in July 2015, it has been officially approved by the \sk\ collaboration \cite{ikedatalk}. }

\begin{figure}[htbp]
 \centering
 \includegraphics[width=0.6\textwidth]{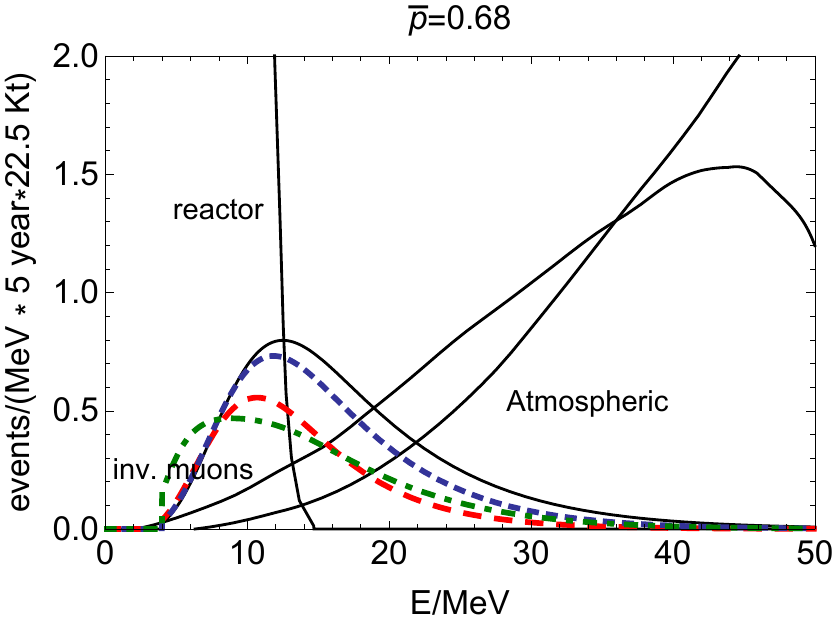}
  \vskip 0.4 truecm
  \includegraphics[width=0.6\textwidth]{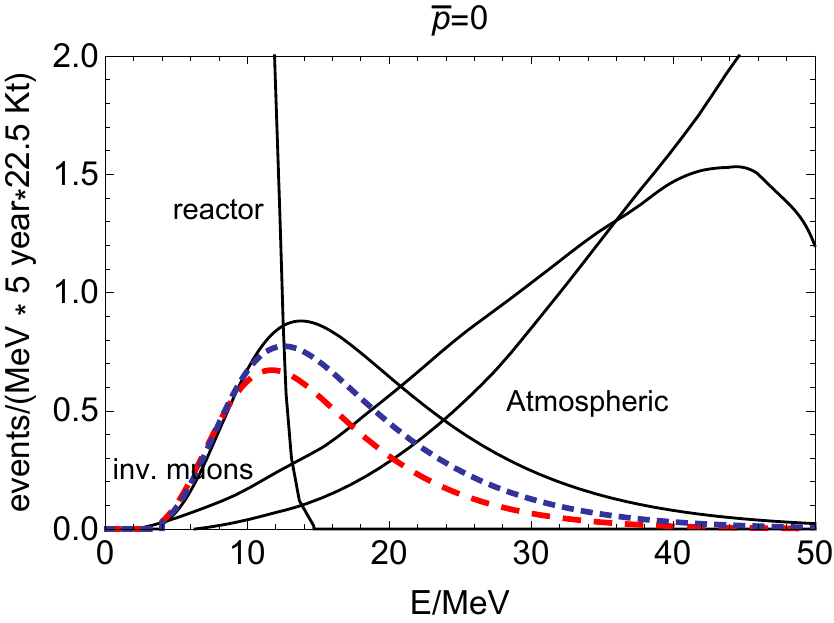}
\caption{  The same as fig. \ref{waterkam} for  SuperK-Gd (22.5 kt water with Gd). Notice the strong reduction of the invisible muon background. 
}
\label{waterkamGd}
\end{figure}
It is estimated that, with Gd, the background due to spallation will be subtracted almost completely and the one  due to invisible muons will be reduced by a factor of $\sim$5 \cite{Beacom:2003nk}. As a result, the lower end of the energy window 
for \df\ detection 
 is determined by reactor \ns, and would typically be 11-12 MeV in neutrino energy.  In the interval $\sim$12-20 MeV the signal dominates over the background.  This is illustrated in fig. \ref{waterkamGd}, where the enlarged energy window appears, possibly including (for the H spectrum ) the peak of the event energy distribution.

The number of events expected at SuperK-Gd can be evaluated by rescaling the event rates of Tables \ref{ratesmodel68} and \ref{ratesmodel0} by the appropriate volume factor.  
Assuming 10 years running time and  an indicative 22.5 kt fiducial volume  one expects 9 - 28 signal events and 29 background events  in the  window of 10 - 28 MeV of positron energy.  Thus, the excess due to the \df\ would be statistically significant for part of the parameter space, and the significance will likely be enhanced by the spectral analysis of the data. However, detailed studies of the signal properties would probably be beyond reach.

The recent intense R\&D work on SuperK-Gd focused on
 finding a Gd compound that meets  four basic criteria: (i) solubility, (ii) limited impact on water transparency, (iii) negligible reactions with the detector materials (e.g., corrosion, etc.)  and (iv) compatibility with a detector's filtration system \footnote{A detector like \sk\ requires constant purification of the water to avoid deterioration of its transparency. It is necessary to prove that a Gd compound can be effectively filtered and reinserted into the detector.  Another requirement is the feasibility to completely remove the Gd compound from the water to restore the detector to a pure water phase. }.

\snew{ Initially  ${\rm GdCl_3}$ was considered \cite{Watanabe:2008ru}, but was then discarded due to poor performance in water transparency,}  and since then a 0.2\% solution of Gd sulfate (${\rm Gd_2(SO_4)_4}$) has emerged as the favorite option, as it meets the criteria outlined above and is safe for stainless steel tanks \cite{Kibayashi:2009ih}.

\snew{This compound is now being tested further at EGADS (``Employing Gadolinium to Autonomously Detect Supernovas"), a dedicated 200 t prototype at Kamioka, where data-taking is in progress} \cite{Adams:2013ana,fernandeztalk} \footnote{Other important questions have to be addressed before SuperK-Gd can be realized: one of them is environmental concerns (real or perceived by the public opinion), related to the existence of a leak in the \sk\ tank.  I leave this aspect to more specialized literature (e.g., \cite{vaginstalk}).}.

\subsubsection{Liquid scintillator at the 20 kt scale}
\label{lenasec}

\snew{
Liquid scintillator detectors of tens of kt mass have been envisioned for at least a decade.  Initial studies focused on a 50 kt configuration, the european LENA (Low Energy Neutrino Astronomy)  \cite{MarrodanUndagoitia:2006re}, whose potential for the \df\ was assessed in dedicated studies \cite{Wurm:2007cy,Wurm:2011zn}.  Lately, upcoming projects have settled on a more realistic $\sim$20 kt mass, with the Asia-based JUNO (Jiangmen Underground Neutrino Observatory, in China) \cite{An:2015jdp}  and RENO-50 (Reactor Experiment for Neutrino Oscillation, in South Korea)  \cite{Kim:2014rfa}.   For definiteness, here I consider the parameters for JUNO :   17 kt fiducial volume, corresponding to $N_p=1.23 \cdot10^{33}$ protons \cite{An:2015jdp}.  When appropriate, a number of early results of the LENA collaboration will be adapted to the current context, as they are still the most detailed in many respects. 
Thanks to the detection in coincidence of the positron and neutron capture (sec. \ref{concepts}), the detection of diffuse $\barnue$s at JUNO will be similar to that at SuperK-Gd : the spallation and invisible muon backgrounds are effectively subtracted, so the energy window is determined by reactor and atmospheric neutrinos and extends from $\sim$10 MeV to $\sim$28 MeV. 
The signal event rate will be similar as well, considering that the number of protons in JUNO is only 22\% smaller than that of \sk/SuperK-Gd ($N_p=1.5 \cdot 10^{33}$).  Therefore about 8-24 events  are expected in 10 years at JUNO.  
}

The relevant backgrounds for liquid scintillator are currently under study.  Well understood backgrounds are the inverse beta decay events due to atmospheric and reactor neutrinos, and to a residual spallation product, fast neutrons  \cite{Wurm:2007cy}.   They contribute to $\sim$11 events per decade \cite{An:2015jdp}. 
Additional backgrounds are due to $\beta n$ emitters (mainly $^9$Li) \cite{Wurm:2007cy,Wurm:2011zn}, and to neutral current scattering of atmospheric neutrinos \cite{Wurm:2011zn}. The latter is especially dangerous, with a rate that exceeds the signal by about one order of magnitude \cite{efremenkotalk}.  However, subtraction techniques are being studied, and, preliminary results on pulse-shape discrimination indicate that it might be possible to achieve a signal-to-background ratio larger than 1 \cite{Wurm:2011zn,lenathesis,An:2015jdp}.

\begin{figure}[htbp]
 \centering
 \includegraphics[width=0.70\textwidth]{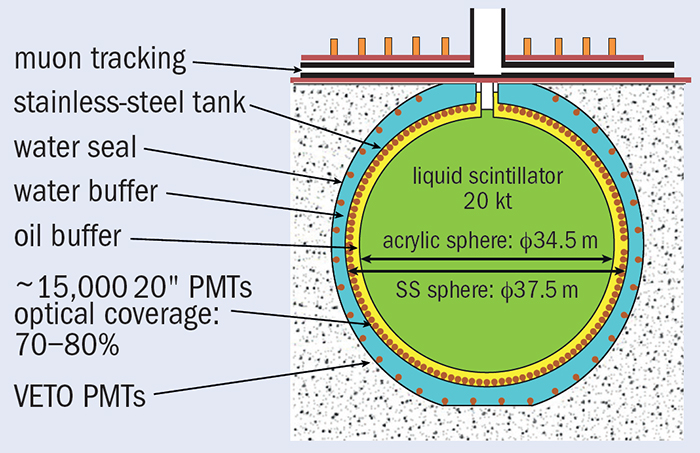}
\caption{  JUNO conceptual design, from \cite{cern}.  }
\label{junodesign}
\end{figure}
\snew{
The current design of JUNO is shown in fig. \ref{junodesign} \cite{An:2015jdp}.  It is characterized by a spherical geometry, with the inner detector surrounded by a water \ck\ volume acting as muon veto.  A 70-80\%  wall coverage of photomultipliers is envisioned.  The detector will be located in Jinji town, in the Guangdong province of China; its construction is now approved by the Chinese Academy of Sciences, and it is expected to start taking data in 2020  \cite{An:2015jdp}.  
}

\subsection{Liquid argon for $\nue$ detection}
\label{lardetector}

\subsubsection{Numbers of events}
\label{nevargon}

\snew{
From the intense scoping work of the past years it has been recognized that a  mass of  several tens of kilotons is necessary for non-beam physics with liquid argon.  This scale will be reached by the USA-EU joint project DUNE -- to be built in the Homestake mine, South Dakota, USA -- for which  40 kt fiducial mass is envisioned in its full configuration \cite{cdr}. 
 Here this mass is considered; and a running time of 5 years is chosen for illustration. 
 }
Here I include only the dominant process, charged current $\nue$ interaction with argon, (Table \ref{tabconcepts}). 
For this, the energy of the emitted electron differs from that of the incoming neutrino by $\sim$ 3-4 MeV depending on the nuclear transition taking place \cite{Cocco:2004ac}.  Since detailed information on the spectrum of these transitions is not available, however, here event rates will be discussed in terms of neutrino energy.  All results refer to the diffuse flux as in fig. \ref{spectradiff}.

\begin{figure}[htbp]
\newpage
 \centering
   \includegraphics[width=0.6\textwidth]{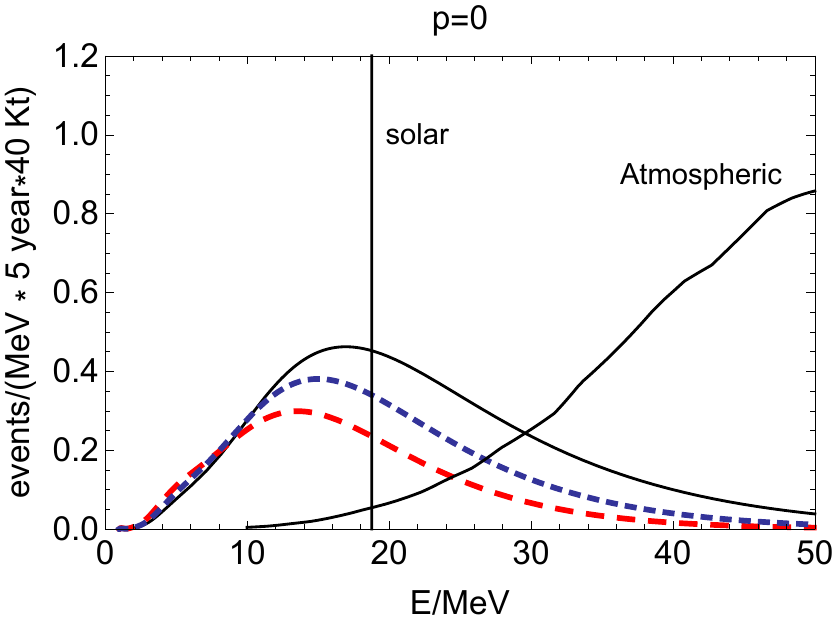}
   \vskip 0.4 truecm
 \includegraphics[width=0.6\textwidth]{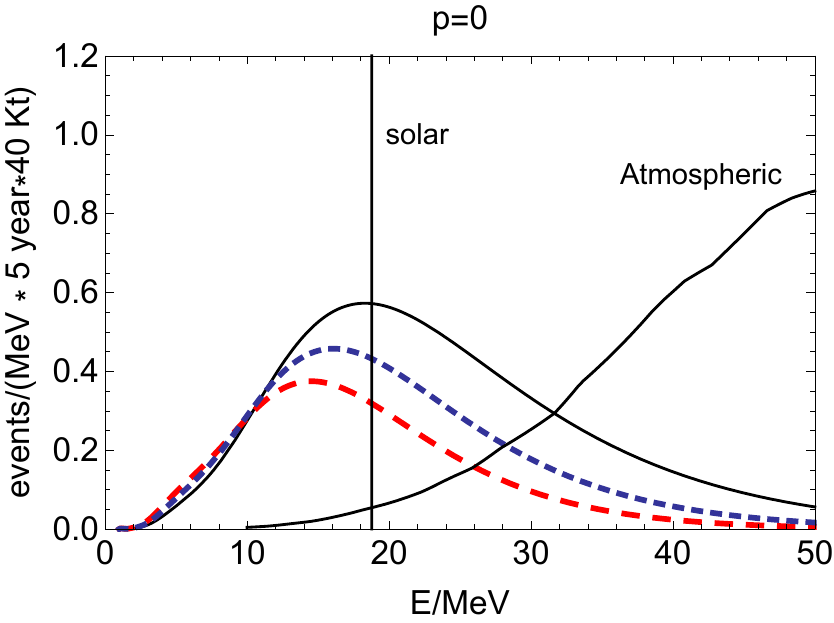}
\caption{  Distribution of charged current  $\nue$  events in liquid argon (Table \ref{tabconcepts}) in neutrino energy for a 40 kt LAr detector located at Homestake for the different models of $\nue$ diffuse flux given in  fig. \ref{spectradiff} (same dashings and color coding), and the two extreme values of $p$.  The spectra of events from the relevant backgrounds are shown for comparison. }
\label{LArevents}
\end{figure}
Fig. \ref{LArevents} shows the distribution of signal and backgrounds in neutrino energy, for the extreme values of the survival probability $p$. 
The signal peaks between 12 and 18 MeV depending on the oscillation scenario and on the original \n\ spectra. At the peak, $\sim 0.3 - 0.6$ events/MeV are expected. Backgrounds dominate below $\sim 19$ MeV and above 24-32 MeV depending on the parameters. This energy window is similar to that of a water detector (fig. \ref{waterkam}). Unlike the case of water, however, here 
\snew{it  may happen (for the most optimistic parameters)
that the peak of the signal might fall near or inside the energy window.  
}
This is thanks to the faster rise of the cross section with energy, compared to inverse beta decay.

\begin{table}[htbp]
\begin{center}
\begin{tabular}{| c |c |c |c |c|}
\hline
\hline
  &  Hot  & Warm  & Cold  & atmospheric \\
\hline
\hline
$19 < {\rm E/MeV} <29$   &  3.6 & 2.3 & 1.4   & 1.3    \\
\hline
 $19 < {\rm E/MeV} <39$   &   5.3 & 3.2 &  1.9   & 5.2  \\
 \hline
 $5 < {\rm E/MeV} <39$   & 9.9 & 7.3  &  5.3  & 5.4  \\
 \hline
\hline
\end{tabular}
\caption{  Rate of charged current $\nue$ interactions on ${\rm ^{40}Ar}$ (Tab. (\ref{tabconcepts})) in three energy windows of interest (given in terms of the neutrino energy, $E$) at a liquid argon detector of mass 40 kt and livetime 5 years, for $ p=0.32$.  All flux parameters are as in fig. \ref{spectradiff}. The atmospheric background is for the Homestake location, and is expected to be a factor of $\sim$1.5 lower at Kamioka, see fig. \ref{allbackgrounds}. } 
\label{ratesmodelLAr}
\end{center}
\end{table}
\begin{table}[htbp]
\begin{center}
\begin{tabular}{| c |c |c |c |c|}
\hline
\hline
  &  Hot  & Warm  & Cold  & atmospheric \\
\hline
\hline
$19 < {\rm E/MeV} <29$  &  4.8 &  3.1  &  2.0 &   1.3    \\
\hline
 $19 < {\rm E/MeV} <39$   & 7.3 &  4.3  & 2.6  & 5.2  \\
 \hline
 $5 < {\rm E/MeV} <39$   &  12.3  &   8.9 & 6.7 & 5.4  \\
 \hline
\hline

\end{tabular}
\caption{  Same as Table \ref{ratesmodelLAr} for $ p=0$.  } 
\label{ratesmodelLAr2}
\end{center}
\end{table}
Tables \ref{ratesmodelLAr} and \ref{ratesmodelLAr2} give the number of events from the signal and from the atmospheric background in energy intervals of interest.   In the 19 - 39 MeV window the detector may register between $\sim$3 and 5 events of signal and $\sim$5 events of atmospheric background.  The signal increases with the average energy of the $\nue$ flux entering the detector, and therefore is higher for the H spectrum  and for $p=0$ (complete swap between the original $\nue$ and $\nux$ fluxes).   
\snew{
Even in the most favourable case, a longer exposure (at least 20 years) would be required to reach a 3$\sigma$ statistical significance of the signal.  A similar result holds for a narrower  energy window, 19 - 29 MeV, where the increase in the signal to background ratio is roughly compensated by the decrease in statistics. 
  One should remember, however, that the overall normalizations of the \df\ and of the atmospheric neutrino flux are highly uncertain, and therefore the signal to background ratio could very well be a factor of a few larger for all models, thus leading to more encouraging results. 
  }

Tables \ref{ratesmodelLAr} and \ref{ratesmodelLAr2} also give numbers of events in a larger energy window, 5 - 39 MeV, which might be applicable if a method is found to subtract events from solar neutrinos, probably on the basis of directional information from kinematics reconstruction. This possibility has not been investigated so far, and so, while speculative, it remains open.  The Tables show that accessing lower energies might enhance the event rate by up to a factor of 3 for the models with lower \n\ average energy.  

\begin{figure}[htbp]
  \centering
\includegraphics[width=0.6\textwidth]{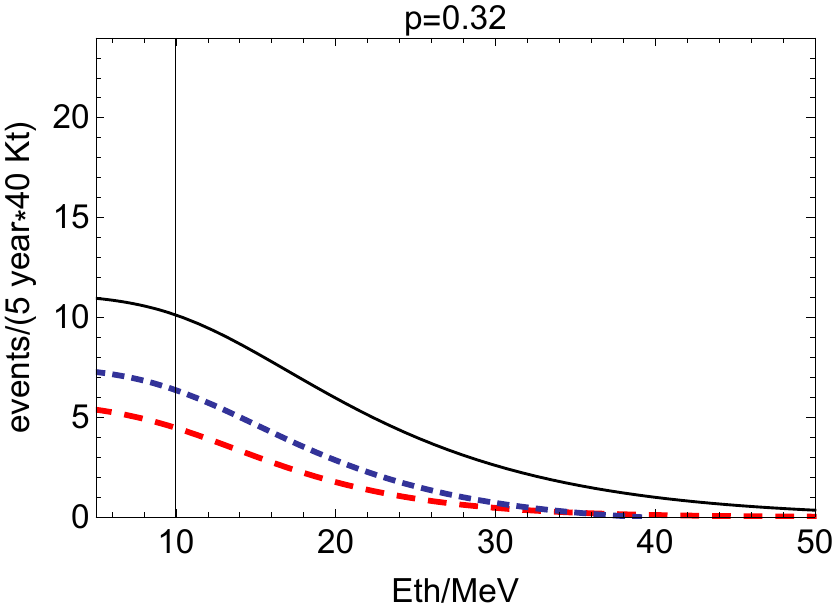}
\includegraphics[width=0.6\textwidth]{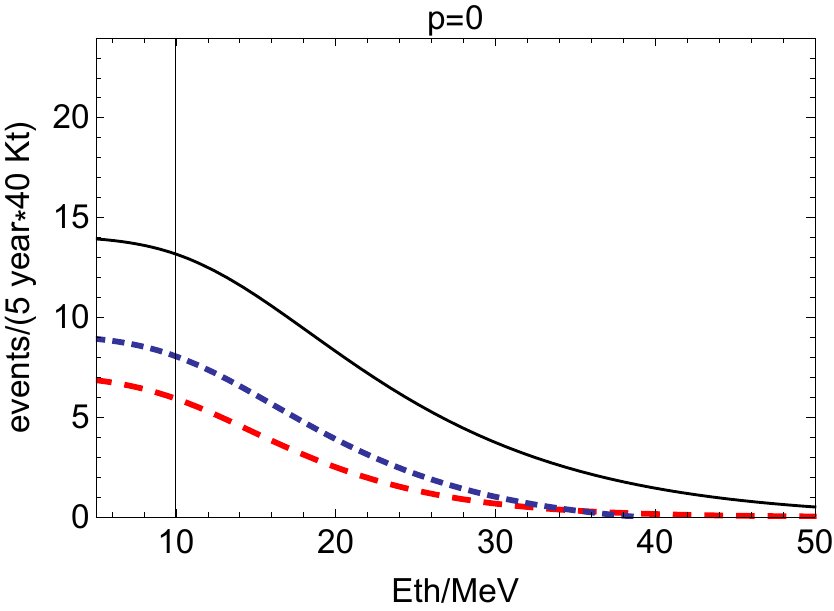}
\caption{  $\nue$ CC  events in liquid argon above a certain neutrino energy $E_{th}$, as a function of $E_{th}$ for the two extreme values of $p$.  The solid, short dashed and long dashed curves are from the 
H, W and C spectra examples (see Table \ref{SNmodeltable}).   All other parameters are as in fig. \ref{LArevents}.}
\label{totalAr}
\end{figure}
For completeness, fig. \ref{totalAr} gives the numbers of \df\ events above a certain neutrino energy $E_{th}$, as a function of $E_{th}$.

\begin{figure}[htbp]
  \centering
  \includegraphics[width=0.6\textwidth]{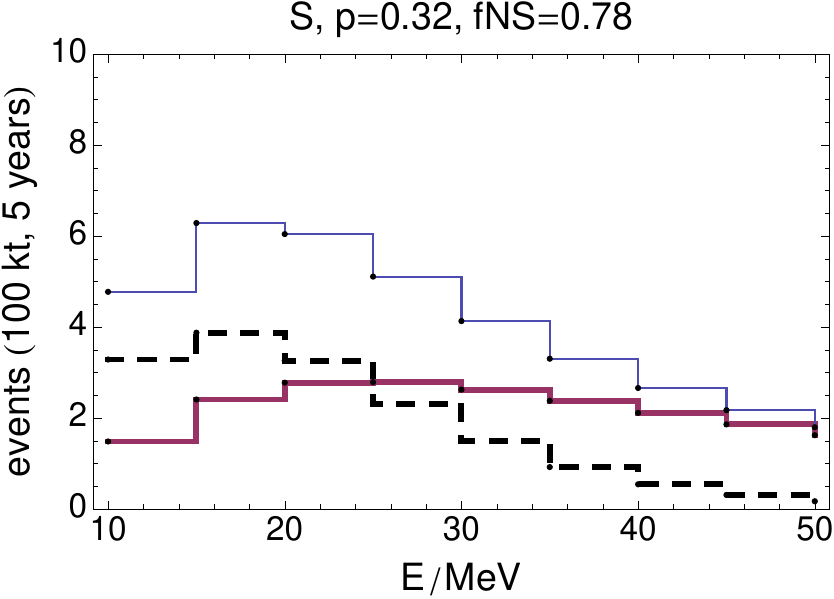}
   \caption{From \cite{Keehn:2010pn}: events in \lar\ from \bh\ collapses (solid thick), from \nts\ collapses  (dashed) and the total of the two (thin line) for the parameters that maximize the failed \sn\ flux (S EoS, $ p= 0.32$ and $f_{NS}=0.78$, see sec. \ref{diffailed}).  Events are plotted in bins of neutrino energy.  
  \snew{Note the long exposure: 500 kt$\cdot$yr, equivalent to 12.5 years of operation for DUNE. 
    }
}
\label{bheventslar}
\end{figure}
Similarly to water detectors, also for \lar\ the event rate could be enhanced by failed \sne.  Fig. \ref{bheventslar} illustrates this, by showing the number and energy distribution of events from normal and failed \sne\ for the most optimistic parameters  (largest failed \sn\ flux, with the S EoS,  $ p= 0.32$ and $f_{NS}=0.78$, sec. \ref{diffailed}).   It appears that 
the contribution of failed \sne\ peaks in the energy window and enhances the event rate by a factor of $\sim 2$ in the same energy interval. This implies a higher chance that a \lar\ experiment might see an indication of signal in its earliest phase of operation.  

\subsubsection{technical considerations}
\label{techargon}

\begin{figure}[htbp]
 \centering
  \includegraphics[width=0.75\textwidth]{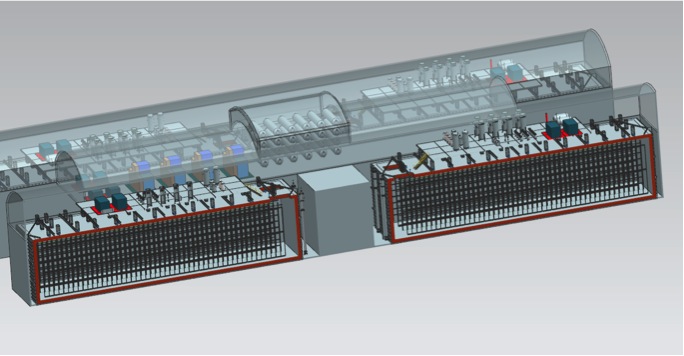}
\caption{
\snew{
Conceptual design of  a 10 kt module of the US-based DUNE detector \cite{cdr}. This is the proposed single phase solution; an alternative dual phase design is also considered, see \cite{cdr} for details.  The full DUNE detector will be made of four modules.
 }
 }
\label{LArexp}
\end{figure}

The technical realization of the liquid argon concept at 40 kt mass is still a subject of discussion.   Current research is focused on testing the performance of liquid argon in lower mass projects, on  designing the necessary electronics,  and on the engineering of the final, full scale, structure and its accommodation  underground.  Essential criteria are the scalability (in mass) of the  design and safety in operations of handling and containing the argon under pressure. 

Current low mass efforts that are specific for \n\ physics include ICARUS \cite{Amerio:2004ze,Antonello:2013ypa,Raselli:2013oxa}, MicroBOONE \cite{Chen:2007ae,Soderberg:2009rz} (now under construction),  and ArgoNEUT  \cite{Soderberg:2009qt,TingjunYang:2013vva} (running).

\snew{
Detailed designs for DUNE already exist; 
fig. \ref{LArexp} illustrates a project for a 10 kt \lar\ module, functioning as a time-projection chamber (TPC), where the readout anode is composed of wire planes inside the LAr volume.  An alternative concept is characterized by a dual phase, with the ionization charges being detected in a layer of gaseous Argon above the liquid phase. See \cite{cdr} for more details.
}

\section{Perspectives: what can be learned and how?}
\label{perspectives}


The question of what can be learned from data on the \df\ has been explored only partially so far.  
It is reasonable to expect that, as data start to appear, there will be a first phase of analysis that will constrain the basic ingredients of the \df\ at the coarse level, with errors of the order of 50-100\% due to backgrounds and parameter degeneracies.  What these basic ingredients are emerges from the  previous sections: the neutrino spectra in the different flavors, conversion/survival probabilities, and normalization parameters like the total energy emitted in \ns.  \snew{These are discussed in sec. \ref{spectesting} and \ref{osctesting}. Exotic physics and subdominant effects are briefly discussed in sec. \ref{exotictesting} and \ref{smalleffects}; directions for improvements and final considerations then follow to conclude this section. }

\subsection{Testing the physics of supernovae: neutrino spectra}
\label{spectesting}

Naturally, the sensitivity of an experiment to the \df\ spectrum depends critically on its energy window.   Limited sensitivity is expected  for pure water and \lar, where the window is only  10-15 MeV wide, while liquid scintillator and SuperK-Gd-type (water+Gd) designs should perform better thanks to the larger window. 
Detailed studies have been done for  water, water+Gd, and liquid scintillator.  Most of them discussed  the  dependence of event rates  on the \n\ spectra (e.g, \cite{Fogli:2004ff,Chakraborty:2008zp,Volpe:2007qx}), while the question of parameter degeneracy has been addressed only in part \cite{Lunardini:2006pd,Lunardini:2012ne,Nakazato:2013maa,Mathews:2014qba}.

One way to eliminate the degeneracy with normalization parameters ($L_{\bar e}$, $L_{\bar x}$ and $R_{-4}$), is to study ratios of events, for example the 
ratio $r$ of the number of events in the first and second energy bin \cite{Lunardini:2006pd}\footnote{Including higher energy bins enhances the backgrounds more than the signal and therefore it worsens the sensitivity to the \n\ spectrum \cite{Lunardini:2006pd}. }.
Taking bins of 5 MeV width,  this ratio is:
\be
r \equiv \frac{N(18\leq E_{e }/{\rm MeV}<23)}{N(23\leq E_{e }/{\rm MeV}<28 )}~,
\label{ratio}
\ee
for the water only case, and 
\be
r \equiv \frac{N(10\leq E_{e }/{\rm MeV}<15)}{N(15\leq E_{e }/{\rm MeV}<20)}~,
\label{ratio_gd}
\ee for water+Gd, in terms of the positron energy $E_{e }$. 

 \begin{figure}[htbp]
  \centering
\includegraphics[width=0.60\textwidth]{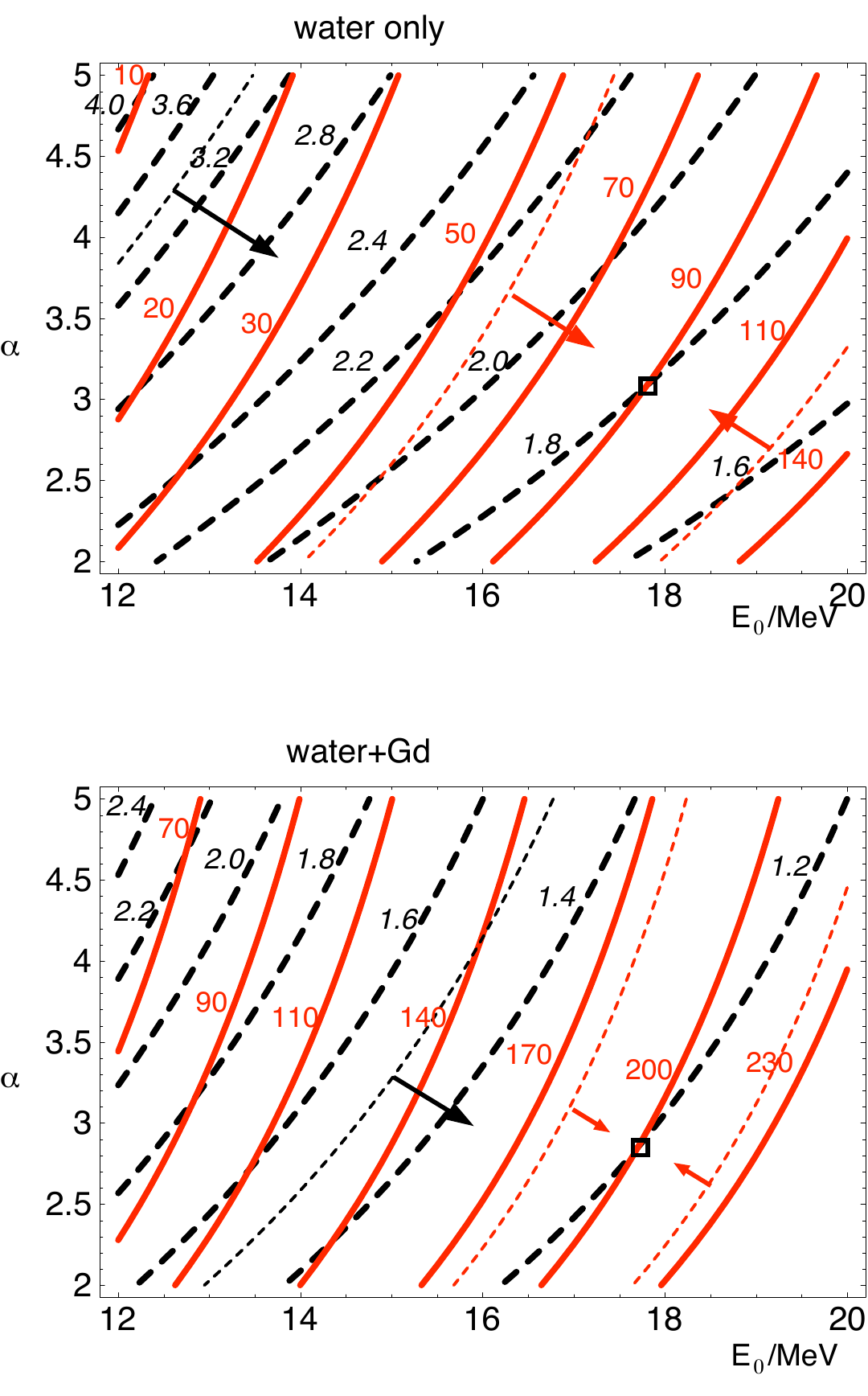}
\caption{From \cite{Lunardini:2006pd}: isocontours of the numbers of events $N$ (solid red lines), and of the ratio $r$ of events in the first two energy bins  (Eqs. (\ref{ratio}) and (\ref{ratio_gd}), long dashed black lines), at a water \ck\ detector with 1.8 Mt$\times$ yr exposure, in the space of the spectral parameters $\alpha_w - E_{0 w}$ (in the assumption that the spectral form in Eq. (\ref{nuspec}) applies to the $\barnue$ flux at Earth).  In each panel  a possible measurement of $N$ and $r$ is shown, with central values (diamond) and 1$\sigma$ statistical errors (regions within the short dashed lines, marked with arrows; the wider region is the error on $r$). 
The \snr\ parameters and total energy emitted in \ns\ are as in fig. \ref{spectradiff}.}
\label{curvessensitivity}
\end{figure}

\snew{
Fig. \ref{curvessensitivity} \cite{Lunardini:2006pd} shows isocontours of $r$ in the space of the spectral parameters $\alpha_w - E_{0 w}$, for fixed \snr\ and in the assumption that the spectral form in Eq. (\ref{nuspec}) applies to the $\barnue$ flux at Earth.
}

 For pure water   $r$ varies in the range $r \simeq 1.5 - 4.3 $.  For a 
 \snew{
 2 Mt$\times$ yr exposure, 
 }
 the error on $r$ is larger than
$\sim 100\%$ due to the high background rate \cite{Lunardini:2006pd}; this confirms the conclusion that the sensitivity to the
neutrino spectrum is limited for pure water.  
For the  configuration with extended energy window (water+Gd) one gets $r \simeq 1 - 2.5 $. Thanks to the better background reduction, the error is down to
$\sim 20 - 30 \%$, meaning that at least the extreme cases should be
distinguishable, for 
\snew{
typical values of the normalization parameters. 
}

If in the future the normalizations and the background fluxes become known precisely, the number of signal events $N$ in the energy window 
can be used jointly with the ratio $r$ to constrain the spectral parameters. Isocontours of $N$ are given in fig.  \ref{curvessensitivity}; they intersect the isocontours of $r$, suggesting that a joint measurement of  $N$ and $r$ could be   
more constraining than the two separately.

In essence, the results in fig. \ref{curvessensitivity} tell us that data will allow us to reconstruct the spectrum of the \n\ emitted by a \sn\  in terms of two {\emph effective} parameters, an average energy and a shape parameter.  The  next step of the analysis would be to reconstruct the more fundamental quantities that determine the effective ones, such as the \n\ spectra in the different flavors at the production point, and the  \n\  mixings  and mass hierarchy.   This second level of detail is certainly more challenging, and is likely to require a combination of precision-phase experiments and improved theoretical modeling.  
\snew{
An example of this is given in a detailed study  \cite{Wurm:2007cy} for a liquid scintillator detector with an exposure of $\sim$400 kt$\times$yr (equivalent to about 24 years of running for JUNO). Considering the number of events in two different energy bins, and including \n\ oscillations, it was found that different spectrum models can be discriminated with $\sim 2 - 2.5\sigma$ confidence level. This conclusion is only moderately encouraging, and may become weaker when it is updated with a full inclusion of all the backgrounds (see sec. \ref{lenasec}).
}

\subsection{Testing \n\ flavor conversion}
\label{osctesting}

The sensitivity of the \df\ to \n\ oscillation effects,
and therefore to the mass hierarchy and to the phenomenology of collective oscillations (sec. \ref{flavconv}), 
remains an open question.  Many studies have included conversion effects in predictions of the \df, examining their impact on the rates and energy distributions of events at detectors (sec. \ref{detection}). 
Effects are strong for both quantities, however, what exactly can be concluded on oscillation parameters is not clear, in consideration of the many degeneracies.  Specifically, event rates have a degeneracy with the 
\snew{
normalization parameters,
}
while energy distributions  are meaningful only with priors (or a measurement from a galactic \sn) on  the originally emitted \n\ spectra in the different flavors.   A detailed reconstruction of spectral shapes might resolve this degeneracy to some extent, but only for the larger energy windows of SuperK-Gd and of a liquid scintillator detector, provided that they run for at least a decade to accumulate sufficient statistics. 

Some of the degeneracies can be resolved by the combination of several complementary datasets. For example,  a set of data from $\nue$ and another from $\barnue$ would help to eliminate degeneracies with the normalization parameters and would allow to probe the permutation parameters $p$ and $\bar p$.  This 
can lead to the discrimination 
of the mass hierarchy, as already observed for  an individual \sn\ (e.g., \cite{Lunardini:2003eh,Skadhauge:2006su,Minakata:2008nc}).  For the \df, this step of connecting measured probabilities with the fundamental oscillation parameters might be complicated by the integration over the diverse \sn\ population; 
\snew{
therefore it might be advisable to focus on probability determination, 
}
with the understanding that the measured probabilities should be intended as effective/averaged parameters. The connection with the swap probabilities $P_c$, $\bar P_c$, and with the mass hierarchy\footnote{Rigorously, the mass hierarchy is the only independent parameter that is still unknown. In principle, $P_c$ and $\bar P_c$ are completely determined by neutrino masses and mixings, as well as by \n\ flavor spectra and total energies at production. However, they are also very sensitive to subtle and poorly known effects of non-linearity and symmetry breaking. For this reason,  it may be practical to consider them as independent parameters in a data analysis. } may be attempted if  the formulae in Table \ref{probtable} are assumed to apply for the \df.  

\subsection{Sensitivity to \n\ exotica}
\label{exotictesting}

Analogously to what happened with SN1987A, the data from the diffuse \sn\ \ns\ will test innumerable new physics hypotheses --typically the existence of new particles and/or new forces   --  from searches of unexpected features in the \n\ signal.   

The sensitivity of the \df\ to these exotica has been studied only minimally.  In view of the substantial cosmological contribution to it (see sec. \ref{ccdep}), one expects the \df\ to be particularly constraining, compared to SN1987A, of phenomena that require large propagation distances, such as effects of decay or absorption of the \ns\ during their propagation across the universe.

\begin{figure}[htbp]
 \centering
 \includegraphics[width=0.5\textwidth]{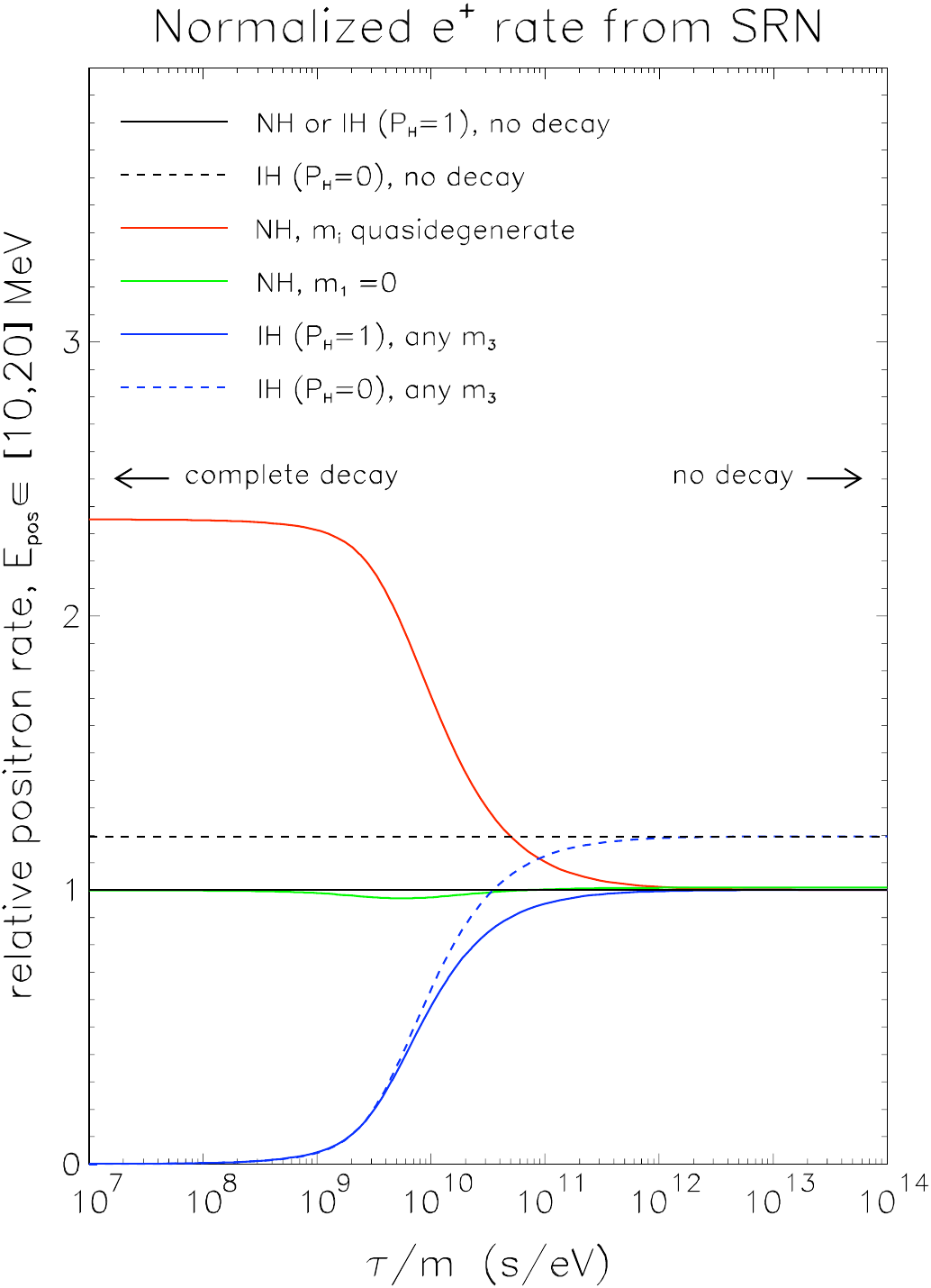}
\caption{From \cite{Fogli:2004gy}: positron event
rates in water in the energy range $[10,20]$~MeV for various decay
scenarios, relative to the case of no decay with $\bar p\simeq 0.68$ (solid horizontal line).   }
\label{foglidecay}
\end{figure}

Effects of \n\ decay on the \df\ were studied by  \cite{Ando:2003ie,Fogli:2004gy}.
Depending on the \n\ mass spectrum, the decay can enhance or suppress the electron flavor component, in a way that could be observable for ratios of  
the neutrino lifetime over mass as high as  $\tau/m \simeq 
10^{10}~{\rm s/eV}$ (fig. \ref{foglidecay}).  One should be mindful, however, of the possible degeneracy with the normalizations of the \n\ flux and of the \snr. 

Exotic absorption effects could be caused by a high rate of 
neutrino-antineutrino annihilation in the cosmological relic neutrino background, mediated by new light gauge bosons \cite{Goldberg:2005yw}, or by resonant \n-dark matter scattering \cite{Farzan:2014gza}.

About non-propagation (between source and detector) effects, it has been shown that resonant spin flavor conversion inside the star influences the expected event rates \cite{Volpe:2007qx}.
It has also been observed that the \df\ could in principle provide a test of dark energy,
complementary to astrophysical measurements \cite{Hall:2006br}. 

Besides  affecting the \df, physics beyond the Standard Model could also mimic it under certain circumstances.    These could be the annihilation  or decay of light dark matter into \ns\ of energy comparable to the \df\ (see e.g. \cite{PalomaresRuiz:2007eu,PalomaresRuiz:2007ry,Bernal:2012qh}).    A flux similar to the \df\ could be made of solar antineutrinos, if their production is enhanced by resonant spin-flavor conversion \cite{Raffelt:2009mm}.    Ultimately, when detectors reach the precision phase, it will be possible to distinguish the \df\ from other, non-supernova fluxes, on the basis of spectral and/or directional information.  For the first detection phase, however, it is likely that a complete disambiguation might not be possible, so that a first detection of a diffuse flux will have multiple interpretations.  

\subsection{Subdominant effects}
\label{smalleffects}

\snew{For completeness, it is worth mentioning other phenomena that will be difficult to study with the \df, either due to their modest ($\sim 10\%$ effect or less) impact, or due to background. They might become of interest for an advanced precision phase of the \df\ in the more distant future. }

\begin{itemize}

\item regeneration of $\nue$ and $\barnue$ due to oscillations inside the Earth.  In the \df\ these effects are below $\sim 5$\% \cite{Lunardini:2006sn} and therefore statistically insignificant.

\item the  swap in the \n\ energy spectrum due to \n-\n\ refraction (sec. \ref{snspectra}).  The spectral step caused by the swap for a single supernova will be smoothed out by the integration over the redshift and by the presence of individual differences between \sn\ progenitors \cite{Chakraboty:2010sz,Lunardini:2012ne}.  Besides, 
for part of the parameter space, 
the swap occurs below 10 MeV of energy, and thus is likely to fall outside the energy window for \df\ detection. 

\item the \snr\ at large redshift, $z \gta 1$.   This will be difficult to probe because \n\ produced at large redshifts accumulate at low energy, where backgrounds dominate (see sec. \ref{ccdep}). 

\item the normalization of the supernova rate.  This is degenerate with the \n\ flux normalization.   Probably, it will be better constrained by astronomical surveys of \sne. 

\item shockwave effects on the \n\ spectrum due to oscillations. First pointed out in \cite{Schirato:2002tg}, these effects are very interesting probes of oscillation parameters as well as of the physics of the shock.    Once integrated over the duration of a \n\ burst, and over a diverse population of stars at different redshifts, they reduce to a $\sim 10\%$ size or less \cite{Ando:2002zj,Galais:2009wi}, with no distinctive spectral features in the \df.  Therefore, they are likely to be indistinguishable from many other effects of similar magnitude.

\item the time structure of \n\ emission from a \sn, including the neutronization burst, the accretion phase and the cooling phase, with their characteristic time scales \cite{Lunardini:2012ne,Nakazato:2013maa}.  The \df\ is a sum of many time-integrated  \sn\ bursts, which causes loss of information.  

\end{itemize}

\subsection{Directions for improvement}

\subsubsection{Phenomenology of the \df}

The directions in which phenomenology can develop are many. 
Here I outline some that I consider most important to advance the field.

\begin{itemize}

\item  {\it Contribution of different \sn\ types.}  Perhaps, the truly unique value of the \df\ -- in comparison to \ns\ from individual \sne\ -- is its sensitivity to the diverse stellar population that contributes to it, which spans decades of progenitor masses and different stellar properties (magnetic fields, metallicities, etc.).    A comprehensive study of how this diversity is reflected in the \df\ would be extremely important to fully explore the physics potential of this flux.  It would require a systematic modeling of the \n\ emission for a large variety of \sn\ progenitors, a goal that is certainly ambitious but realistic, in principle. 
 At this time, numerical simulations exist for selected values of the progenitor mass, and some  \cite{Fischer:2009af,Nakazato:2012qf}) have been used for initial studies on the \df\ \cite{Lunardini:2012ne,Nakazato:2013maa} . Results indicate that the effect of the progenitor mass distribution is of the order of 10\% or less.

\item  {\it from one \sn\ to many: what does really matter?}  As commented previously, when summing over many sources and including redshift effects, some loss of information is unavoidable.  It would be desirable to have a systematic study of the features and size of each different  effect  (e.g., oscillations, propagation effects, time structures, etc.).
An initial study obtained with a specific set of numerical flux predictions \cite{Lunardini:2012ne} shows that the MSW-driven flavor conversion may dominate over other effects like spectrum cooling and the dependence on the progenitor mass.   However, it is possible that, in certain circumstances, several small effects may sum to a sizable contribution, with no single one of them dominating.   Other literature discusses the dependence on the stellar dynamics; the effect of the time of shock revival was found to be of the order of tens of per cent \cite{Nakazato:2013maa}.
 A description at such level of detail might be premature at this time, but will be needed when the precision phase begins.  

\item  {\it What fundamental physics will we learn?} Related to the previous point is the question of how much information will be possible to extract from the \df\ data on specific quantities of interest, like the \snr, \n\ spectra, oscillation parameters, etc.    This question is closely connected to that of experimental capability, and requires a careful study of how data can be processed for best parameter extraction. It will be of immediate relevance after the \df\ is first detected. 

\end{itemize}

In addition to these directions, it is expected that studies of the  \df\ will be updated over time, following the developments in the physics of \ns\ and of \sne.   New, more precise determinations of the \snr\ will be used after a new phase of \sn\ surveys, some of which are already taking data e.g., \cite{snls,des,lsst,jwst,elt}.  A more detailed description of oscillation effects will have to be adopted once the phenomenology  of \n\ self-interactions  (sec. \ref{flavconv}) is completely  understood. 
Progress in the determination of oscillation parameters -- especially if the mass hierarchy is established -- will help to converge towards a uniquely determined oscillation scenario.
 Finally, as the modeling of \n\ emission and transport in \sne\ progresses, updated \n\ fluxes will be available with a reduced uncertainty on the \n\ flavor spectra and total energies. 

If a galactic \sn\ is observed in the next decade or so, the study of the \df\ will take a different turn.  The abundance of data from the single nearby \sn\ will eliminate or greatly reduce the uncertainty on the original \n\ fluxes and on oscillations.  This will allow more precise predictions for the \df, with more focus on its sensitivity to the \sn\ population of the universe and a number of cosmological and astrophysical effects. 

\subsubsection{Experimental searches}

From what was discussed so far,  one can see that the experimental study of the \df\ has two strong limiting factors: the detector mass and the backgrounds.   

Increasing the mass of a detector is of course desirable for a higher statistical significance of a  \df\ signal.   At the moment, any expansion beyond $\sim$ 1 Mt ($\sim$ 50 kt) mass for water \ck\ (LAr) seems speculative.  However, 
\snew{
a conceptual study has been presented
}
for a 5 Mt mass water \ck\  detector to be built under sea, off the coast of Japan \cite{Suzuki:2001rb}.  The sensitivity of this design to \sn\ \ns\  has been studied at a basic level \cite{Kistler:2008us}, but many questions of technical nature remain open. 
  In particular, the larger volume is likely to come at the price of worse performance in terms of energy threshold, background, etc., in a way that remains to be assessed. 
Similarly, a $\sim 10$ Mt under-ice detector has been suggested as an upgrade of the IceCube detector \cite{Boser:2013oaa}. Although a rough similarity with an under-sea concept is expected, for this proposal the sensitivity to the \df\ has not been determined yet.

For currently planned detectors, efforts to better discriminate the backgrounds at low energy, and therefore expand the energy window, are in progress (sec. \ref{detection}).
Even though it might be possible to gain 2-3 MeV by purely technical upgrades, a new approach might be needed to discriminate backgrounds that are orders of magnitude larger than the signal, like solar \ns\ for \lar\ detectors or reactor antineutrinos for water and liquid scintillator.  In the absence of concrete ideas, here I outline some possible directions for further thinking, some of which are purely speculative at the moment.

\begin{itemize}

\item {\it directional detection.}  Reconstructing the direction of  arrival of neutrinos and antineutrinos in a detector would be a very effective way to discriminate the solar and reactor   backgrounds.  Unfortunately,  the relevant detection processes have weak correlation between the direction of the \n\ and that of the daughter lepton at the energies of interest (see e.g., \cite{Strumia:2003zx}). Therefore the complete kinematics of the process would need to be reconstructed for directional sensitivity.  This seems unrealistic for water and \lar\ detectors \cite{rubbiacomm}.  Studies in the context of geoneutrinos show that some directional information can be extracted by adding  $^6$Li, $^{10}$B or Gadolinium to liquid scintillator \cite{Hochmuth:2005nh,Terashima:2008zz,tanaka},  with  $^6$Li emerging as the best candidate \cite{Terashima:2008zz,tanaka}.   It is not clear, however, if the improvement in directionality would be sufficient to separate reactor \ns\ at the level needed to study the \df.

\item {\it turning off or optimizing nuclear reactors.}  Considering  the current efforts in developing renewable energy, it is conceivable that in the far future  fission reactors will be abandoned.  The absence of reactor \ns\ would allow a sensitivity to the $\barnue$ component of the \df\ down to at least 4.5 MeV, where the flux of geoneutrinos terminates \cite{Krauss:1983zn}\footnote{\snew{In fact, such situation has already been realized, with the temporary general shutdown of nuclear reactors in Japan, following the Fukushima nuclear accident in 2011. This allowed a background-free study of geoneutrinos \cite{Gando:2013nba}.}}. This implies the possibility to study the peak of the \df, which would allow a detailed reconstruction of the \n\ spectrum and would provide precious information on \sne\ at large redshift ($z>1$, see sec. \ref{ccdep}).  
If a new generation of fission reactors is built, instead, its flux might be reduced by negotiating down times with the reactor managements, or even by negotiating a reactor-free zone within a certain distance from major detectors. Negotiations could prove unreliable, however.   Interestingly, one could minimize the reactor flux also by arranging the detector-reactor distance to maximize the suppression effect of oscillations, which can be as large as $(1-\sin^2 2\theta_{12}) \simeq 0.13$.  This could be done either by planned construction of detectors and reactors, or with a movable detector \cite{Learned:2007zz}. 

\item {\it extraterrestrial detectors.} In the light of a renewed interest in manned exploration of space, one can dream that a few centuries from now underground laboratories might exist on the moon and/or on Mars or other planets of the solar system. Even without a colonization of them, scientific labs may flourish there the same way they do in Antarctica today, even though with the new challenges of extraterrestrial conditions.  
A  large \n\ detector on Mars, for example, 
would benefit from the lower solar flux and from  the suppressed atmospheric \n\ flux reflecting the thinner atmosphere. Reactor \ns\ might be absent there if power is provided by alternative forms of energy (fusion, solar, etc.). Of course,  investing in such detector would require a strongly motivated and diverse scientific agenda, which might include studying the radioactivity  of the planet with geoneutrinos.

\end{itemize}

\subsection{Final considerations}

At the opening of a new phase of \n\ \snew{observatories}, the possibility to observe diffuse \sn\ \ns\ has a quiet appeal of its own, that well complements the thrill of playing the lottery of galactic \sne.  As new detectors turn on, the \df\ could be the first new signal to emerge, nicely sandwiched between higher precision data from known sources at low and high energy.   This is a guaranteed flux that must be there -- even though with a large normalization uncertainty -- and once detected, it will mark the transition of \sn\ \ns\ from rare events (SN1987A) to everyday physics, like solar and atmospheric \n\ have been in the past several years.  

Over decades, the \df\ will go through the phases of discovery, maturity and precision, delivering unique information on the population of core collapse \sne\ all the way to cosmological distances, and thus complementing information from individual \sne\ on the physics of core collapse and on the properties of the \n\ as a particle. 
In addition, there might be surprises: unexpected turns of which \n\ physics has been rich so far.

\subsection*{Acknowledgments} 

I am indebted to  Alessandra Tonazzo, Michael Smy, Mark Vagins and Michael Wurm for important comments and feedback; many thanks to Shunsaku Horiuchi, James Keehn, Orlando L.~G. Peres and Lili Yang for useful discussions. I acknowledge the support of the NSF under Grant No. PHY-0854827 and PHY-1205745.  

%
%


\newpage

\bibliographystyle{elsarticle-num}
\bibliography{masterbiblio}

\end{document}